\documentclass[11pt]{article} 

\PassOptionsToPackage{bbgreekl}{mathbbol}

\usepackage[]{ajtex}

\usepackage{physics}
\usepackage{simpler-wick}

\title{Schwinger-Keldysh effective field theory for stable and causal relativistic hydrodynamics}

\author[1,2]{Akash Jain}\email{ajain@uva.nl}
\author[3]{Pavel Kovtun}\email{pkovtun@uvic.ca}

\affiliation[1]{Institute for Theoretical Physics, University of Amsterdam, 1090 GL Amsterdam, The Netherlands}

\affiliation[2]{Dutch Institute for Emergent Phenomena, 1090 GL Amsterdam, The Netherlands}

\affiliation[3]{Department of Physics \& Astronomy, University of Victoria, 3800 Finnerty Road, Victoria,
British Columbia V8P 5C2, Canada}

\abstract{We construct stable and causal effective field theories (EFTs) for describing statistical fluctuations in relativistic diffusion and relativistic hydrodynamics. These EFTs are fully non-linear, including couplings to background sources, and enable us to compute $n$-point time-ordered correlation functions including the effects of statistical fluctuations. The EFTs we construct are inspired by the Maxwell-Cattaneo model of relativistic diffusion and M\"uller-Israel-Stewart model of relativistic hydrodynamics respectively, and have been derived using both the Martin-Siggia-Rose and Schwinger-Keldysh formalisms. The EFTs non-linearly realise the dynamical Kubo-Martin-Schwinger (KMS) symmetry, which ensures that $n$-point correlation functions and interactions in the theory satisfy the appropriate fluctuation-dissipation theorems. Since these EFTs typically admit ultraviolet sectors that are not fixed by the low-energy infrared symmetries, we find that they simultaneously admit multiple realisations of the dynamical KMS symmetry. We also comment on certain obstructions to including statistical fluctuations in the recently-proposed stable and causal Bemfica-Disconzi-Noronha-Kovtun model of relativistic hydrodynamics.}

\newcommand\bbg{\mathbb g}
\newcommand\ssfS{{\mathsmaller\sfS}}
\newcommand\ssfT{{\mathsmaller\sfT}}

\setlist[itemize]{itemsep=0pt,topsep=5pt}
\setlist[enumerate]{itemsep=0pt,topsep=5pt}

\usepackage{tikz}
\usetikzlibrary{decorations.markings,decorations.pathmorphing,arrows.meta}

\tikzset{aux/.style={decorate,decoration={snake,segment length=1.5mm,
      amplitude=0.4mm}}}
\tikzset{left/.style={arrows={Stealth[scale length=0.5, scale width=1.5][sep=2pt]-}}}
\tikzset{right/.style={arrows={-[sep=-2pt]Stealth[scale length=0.5, scale width=1.5]}}}
\tikzset{lleft/.style={arrows={Stealth[scale length=0.5, scale width=1.5][sep=2pt]-}}}
\tikzset{right/.style={arrows={-[sep=-2pt]Stealth[scale length=0.5, scale width=1.5]}}}

\begin{document}

\maketitle

\section{Introduction}

Hydrodynamics has been an immensely successful framework for describing near-equilibrium dissipative many-body systems, without relying on the precise intricate knowledge of their microscopic constituents and mutual interactions. The fundamental requirement for hydrodynamics to apply is a separation of ``microscopic'' vs. ``macroscopic'' scales, i.e.\ any characteristic spacetime scales of the microscopic theory, e.g. the mean free path or mean free time, must be much smaller than the spacetime scales at which the system is being probed. If so, most excitations that operate at the microscopic scales, dubbed the ``fast'' degrees of freedom, decay by the time we reach the macroscopic scales. The only remaining ``slow'' degrees of freedom are the collective excitations of conserved charges that are protected by the global symmetries of the system under consideration and thus cannot decay locally. In this hydrodynamic regime, we expect that the effective low-energy description of the system can be universally captured by its conserved charges and the respective conservation equations. All the intricate microscopic information about the system coalesces into the hydrodynamic constitutive relations: how the fluxes of conserved charges are related to the conserved densities and their derivatives. 

Like every theoretical model in physics, the validity of the assertions above crucially depends on the type and precision of the questions being asked. Even with a clear separation of scales,  interactions between the ``slow'' and the ignored ``fast'' degrees of freedom can add up over macroscopic scales and lead to significant and qualitative deviations from the naive hydrodynamic predictions~\cite{1963AnPhy..24..419K, Martin:1973zz, Pomeau:1974hg, Hohenberg:1977ym, DeDominicis:1977fw, Khalatnikov:1983ak}. As an example, $n$-point retarded correlation functions or response functions of hydrodynamic observables (i.e. conserved densities and respective fluxes), which measure how the system responds to external perturbations, are known to admit ``long-time tails'' as a consequence of stochastic interactions that are not predicted by classical hydrodynamics~\cite{Forster:1976zz, Jain:2020zhu}.
Furthermore, $n$-point symmetric correlation functions, which measure how the system is correlated across space and time, cannot be accessed using the classical theory of hydrodynamics at all and require a systematic effective theory including stochastic noise. 

The state-of-the-art solution to these problems is presented by the Martin-Siggia-Rose (MSR) formalism~\cite{Martin:1973zz}, where the hydrodynamic constitutive relations are supplemented with random stochastic noise and physical observables are computed by averaging over all possible noise configurations sampled from a Gaussian distribution. The ambiguity in the spread of the Gaussian distribution is fixed by invoking the fluctuation-dissipation theorem (FDT) from thermal field theory~\cite{Callen:1951vq}. It requires that the Fourier-space symmetric and retarded 2-point correlation functions of arbitrary observables $\cO$ and $\cO'$, computed in a thermal state with inverse temperature $\beta_0 = 1/T_0$, must satisfy
\begin{equation}
    G^S_{\cO\cO'}(\omega,k)
    = \frac{2}{\beta_0\omega}
    \Im G^R_{\cO\cO'}(\omega,k),
    \label{eq:fdt}
\end{equation}
where $\omega$ and $k_i$ denote the frequency and wavevector in Fourier-space. Similar FDTs also exist for higher-point time-ordered (i.e. symmetric, retarded, and partially retarded) correlation functions~\cite{Wang:1998wg}, which can be used to fix how the spread of the Gaussian distribution  functionally depends on the hydrodynamic fields. In recent years, a more systematic Schwinger-Keldysh (SK) effective field theory (EFT) framework has been developed~\cite{Grozdanov:2013dba, Harder:2015nxa, Crossley:2015evo, Haehl:2015uoc, Haehl:2018lcu, Jensen:2017kzi}, where one can start from symmetry principles to directly construct an effective action for stochastic hydrodynamics and use this to obtain $n$-point time-ordered correlation functions of hydrodynamic observables. In particular, the FDT requirement in \cref{eq:fdt} and its higher-point generalisations are not imposed by hand in this formalism, instead they are conveniently realised via a dynamical Kubo-Martin-Schwinger (KMS) symmetry of the effective action. SK-EFT enables us to compute stochastic, as well as classical non-linear, corrections to $n$-point correlation functions perturbatively as ``loop corrections'' coming from non-quadratic interactions in the effective action.  As it turns out, assuming the distribution of stochastic noise to be Gaussian,\footnote{Gaussian noise here means that every term in the effective action is at most quadratic in noise fields. Interactions involving one or two noise fields and multiple physical hydrodynamic fields can be described by the MSR formalism, as we illustrate in the forthcoming discussion. However, interactions involving three or more noise fields require the full SK-EFT formalism.} the MSR and SK formalisms yield the same EFTs for stochastic hydrodynamics up to redefinitions of the stochastic noise fields. We will thoroughly explore this connection during the course of this work.

In this work, we are particularly interested in the EFT framework for relativistic hydrodynamics. In fact, the SK-EFT formalism itself was originally developed for relativistic hydrodynamics~\cite{Glorioso:2018wxw}, thanks to the enormous control provided the spacetime Poincar\'e symmetries, and was only later generalised to non-relativistic contexts~\cite{Jain:2020vgc, Armas:2020mpr}. A systematic study of stochastic noise in relativistic hydrodynamics can help us better describe a number of physical systems where stochastic effects are naturally pronounced. Potential applications include relativistic heavy-ion collisions~\cite{Romatschke:2017ejr, Gale:2013da, Murase:2013tma, Young:2014pka} and the putative critical point in Quantum Chromodynamics~\cite{Hohenberg:1977ym, Son:2004iv, An:2019csj, Martinez:2019bsn}.
While developing SK-EFT for relativistic hydrodynamics, one is forced to confront the unfortunate fact that the conventional textbook formulation of dissipative relativistic hydrodynamics is unstable and acausal~\cite{Hiscock:1985zz, Hiscock:1987zz}. In the linear regime, stability and causality can be investigated through the linearised mode spectrum of the theory, $\{\omega(k)\}$, defined as poles of the Fourier-space retarded 2-point correlation function $G^R_{\cO\cO'}(\omega,k)$. We must have
\begin{equation}
    \Im\omega(k) \leq 0, \qquad 
    \left|\Re\frac{\omega(k)}{k}\right| \leq 1\,,
    \label{eq:stability-causality}
\end{equation}
at real $k$.
The first condition ensures that a small external perturbation does not cause an exponential growth in the hydrodynamic response. It also ensures the hierarchy between cause and effect: the retarded correlation function vanishes when the measurement precedes the perturbation. The second condition ensures that an initial perturbation does not induce a response faster than the speed of light.
Both the conditions in \cref{eq:stability-causality} are violated in the textbook formulation of relativistic hydrodynamics, even without loop corrections, when viewed by a boosted inertial observer.

The problems typically arise at timescales $t\sim\tau_*$ set by the magnitude of dissipative transport coefficients, like conductivities and viscosities. Since hydrodynamics is only supposed to be reliable at spacetime scales much longer than all other scales in the system, the stability and causality issues lie well beyond the hydrodynamic regime of applicability and do not really signal an inconsistency of the hydrodynamic framework itself, see e.g.~\cite{Geroch:1995bx}. Nonetheless, they render the theory inadequate for any practical application that involves actually solving a Cauchy problem for the relativistic hydrodynamic equations to determine the fluid flow. To circumvent these issues, we need to appropriately modify the hydrodynamic equations near or before the timescales of instabilities $\tau_*$ in a way that the physics in the hydrodynamic regime remains unchanged. There are two popular proposals for this in the literature: the M\"uller-Israel-Stewart (MIS) formalism~\cite{Muller:1967zza, Israel:1976tn, 1976PhLA...58..213I, Israel:1979wp, 1979RSPSA.365...43I, Hiscock:1983zz, Baier:2007ix, Denicol:2012cn} and more recently the Bemfica-Disconzi-Noronha-Kovtun (BDNK) formalism~\cite{Bemfica:2017wps, Kovtun:2019hdm, Bemfica:2019knx, Hoult:2020eho}. 
The MIS formalism allows the fluxes of conserved charges to evolve independently of the respective densities, such that they relax to their hydrodynamic values after some characteristic relaxation timescale $t\sim\tau$. By appropriately tuning $\tau\gtrsim\tau_*$, we can mend
the stability and causality issues that would otherwise appear near $\tau_*$.
Closely related to these are the so-called ``divergence-type theories'' where the evolution equations for fluxes take the form of sourced total-divergence equations~\cite{1986AnPhy.169..191L}; see the review by~\cite{Geroch:1990bw}.
On the other hand, the BDNK formalism uses clever redefinitions of hydrodynamic fields to introduce the relaxation timescale $\tau$ into the hydrodynamic equations, so that they are perfectly causal and stable when expressed in terms of the new variables.

The above procedure of modifying the physics at short timescales has a natural analogue in the language of EFTs. Physical observables in an EFT are obtained by averaging over all off-shell field configurations with arbitrary frequencies. Any low-energy EFT, which is not intended to describe high-energy ultraviolet (UV) phenomena, needs an appropriate UV-regularisation prescription to handle high-frequency field configurations that lie beyond its regime of applicability. In this sense, the MIS and BDNK formalisms outlined above can be thought of as potential stable and causal UV-regularisations of relativistic hydrodynamics. In the context of SK-EFTs, an important requirement for a regularisation prescription is that the regularised $n$-point correlation functions must also satisfy the respective FDTs. As shown in~\cite{Gao:2018bxz}, this ensures that stochastic loop corrections do not alter the analyticity properties of $n$-point correlation functions.\footnote{If one wishes to use a regularisation prescription that is not compatible with FDT, such as dimensional-regularisation or zeta-function regularisation, one is forced to add additional BRST ghost fields into the framework to preserve the analyticity properties of $n$-point correlation functions~\cite{Crossley:2015evo,Gao:2017bqf}.} In particular, provided that retarded 2-point correlation functions have no unstable or acausal poles at tree-level (i.e. in the classical linear theory), an FDT-compatible UV-regularisation prescription ensures that stability and causality remains intact perturbatively at arbitrary loop orders in the SK-EFT.

The goal of this work is to develop UV-regularised SK-EFTs for relativistic hydrodynamics that are stable and causal by construction, at least perturbatively in interactions, and realise the dynamical KMS symmetry to ensure compatibility with FDTs. In particular, since the UV sector of these SK-EFTs is not fixed by any global symmetry requirements, we find multiple consistent realisations of dynamical KMS symmetry that leave the effective action invariant simultaneously.
We will first consider a toy model with single diffusive conserved current $J^\mu$ in \cref{sec:MIS-diffusion}, the MIS-like version of which is known as the Maxwell-Cattaneo (MC) model of relativistic diffusion~\cite{1989RvMP...61...41J}. We will construct a stable and causal SK-EFT for relativistic diffusion using the MSR formalism and then present a first-principle derivation of the same using the SK formalism. In \cref{sec:full-hydro}, we will generalise these results to construct a stable and causal SK-EFT for relativistic hydrodynamics based on the MIS formalism. In particular, we shall see that the right choice of variables is quintessential for non-linearly realising the dynamical KMS symmetry in the effective theory. Working with stress as a relaxed degree of freedom directly, as is done in the conventional treatments of the MIS-hydrodynamics~\cite{Muller:1967zza, Israel:1976tn, 1976PhLA...58..213I, Israel:1979wp, 1979RSPSA.365...43I}, is not compatible with the dynamical KMS symmetry. We will end the main part of the paper with some discussion and future directions in \cref{sec:discussion}. 

As it turns out, the BDNK formalism of relativistic hydrodynamics, while appropriate for making the classical hydrodynamic equations stable and causal beyond their regime of applicability, does not generalise to a consistent SK-EFT.\footnote{We thank J.\,Noronha and M.\,Hippert for bringing these issues to our attention and for various useful discussions in this regard.} We comment on these issues in \cref{sec:bdnk}. In the same appendix, we also compare with the recent work of~\cite{Abbasi:2022aao}, where authors used a BDNK-like EFT for diffusion to compute UV-regulated one-loop stochastic corrections to 2-point correlation functions of conserved charge. Though this EFT agrees with 2-point FDT at tree-level, we argue that compatibility with higher-point FDTs or with 2-point FDT at loop-levels requires the dissipative coefficients, i.e. charge conductivity and relaxation time, to be constants and not depend on the conserved charge. In \cref{sec:3pt-functions}, we outline a perturbative diagrammatic procedure for computing various time-ordered correlation functions in the UV-regulated EFT for diffusion. We explicitly compute the tree-level 2- and 3-point correlation functions and investigate their consistency with FDTs. The generalisations of this appendix to include stochastic loop corrections and extensions to the complete theory of relativistic hydrodynamics are left for future work. Finally, in \cref{app:alt}, we discuss alternate prescriptions of dynamical KMS symmetry in the SK-EFTs for MC-diffusion and MIS-hydrodynamics.

\noindent
\textsc{Notation:} We work in $d+1$ spacetime dimensions. The spacetime indices ``$\mu,\nu,\ldots$'' collectively run over the time coordinate ``$t$'' and the spatial indices ``$i,j,\ldots$''. The spacetime indices are raised/lowered using the background spacetime metric $g_{\mu\nu}$, which in flat spacetime is given by the Minkowski metric $\eta_{\mu\nu} = \diag(-1,1,1,\ldots)$. The frequency $\omega$ and wavevector $k_i$ in Fourier space is collectively denoted by $p_\mu = (-\omega,k_i)$. We will assume natural units where the speed of light $c$ and the Boltzmann constant $k_B$ are set to 1, however we will keep the reduced Planck's constant $\hbar$ explicit.

\noindent
\textsc{Note:} While this paper was nearing completion, we received a preliminary draft of~\cite{Mullins:new} that offers a different perspective on the problems explored here and appears on the same day on arXiv.

\section{Maxwell-Cattaneo model of diffusion}
\label{sec:MIS-diffusion}

In order to transparently highlight the conceptual aspects of our construction, let us start with a simple diffusion model with single relativistic conserved charge. We will revisit the Maxwell-Cattaneo (MC) model of relativistic diffusion from the perspective of the second law of thermodynamics and review how it circumvents the issues with causality and stability that arise in the ordinary theory of relativistic diffusion; see e.g.~\cite{1989RvMP...61...41J, Kostadt:2000ty}. We will explore how to incorporate stochastic fluctuations in the theory of MC-diffusion using the MSR framework and make our way to the associated SK-EFT. We will pay close attention to the dynamical KMS symmetry in the EFT, which ensures the conformity of stochastic fluctuations with FDTs. The goal of this section is to setup the groundwork for the construction of a causal and stable EFT of charged relativistic hydrodynamics, which we will undertake in \cref{sec:full-hydro}.

\subsection{Theory of diffusion}

Consider a physical system invariant under certain internal U(1) global symmetry, with the associated Noether conserved current $J^\mu$ satisfying the conservation equation
\begin{equation}
    \dow_\mu J^\mu = 0.
    \label{eq:conservation}
\end{equation}
We will also find it helpful to introduce a background gauge field $A_\mu$ coupled to the current $J^\mu$. The conservation of $J^\mu$ means that the theory must be invariant under background gauge transformations of the gauge field $A_\mu \to A_\mu + \dow_\mu\Lambda$. Given an inertial relativistic observer with constant four-velocity $u^\mu_0$ (normalised as $u^\mu_0 u_\mu^0 = -1$), the current $J^\mu$ can generically be parametrised as
\begin{equation}
    J^\mu = n u^\mu_0 + \cJ^\mu,
    \label{eq:param-J}
\end{equation}
where $n$ and $\cJ^\mu$ (normalised as $u^\mu_0 \cJ_\mu = 0$) respectively denote the charge density and flux in the local rest frame of the inertial observer. The charge conservation \eqref{eq:conservation} is a scalar equation and, given the flux $\cJ^\mu$, can be thought of as determining the dynamics of the density $n$. In the long-distance late-time regime, where all other non-conserved ``fast'' degrees of freedom have relaxed, the constitutive relations for the $\cJ^\mu$ can effectively be expressed in terms of $n$, $A_\mu$, and their spacetime derivatives, known as the \emph{constitutive relations}. 
To leading order in derivatives, we have the well-known Fick's law for diffusion
\begin{equation}
    \cJ^\mu = - \sigma\,\Delta^{\mu\nu}
    \Big( \dow_\nu \mu - F_{\nu\rho} u^\rho_0 \Big),
    \label{eq:Ji-consti}
\end{equation}
where $\Delta^{\mu\nu} = \eta^{\mu\nu} + u_0^\mu u_0^\nu$ is the spatial projection operator, $\mu(n)$ is the chemical potential thermodynamically conjugate to $n$, $\sigma(n)$ is the conductivity transport coefficient, and $F_{\mu\nu} = \dow_\mu A_\nu -\dow_\nu A_\mu$ is the external electromagnetic field. 

Note that the response of charge propagation to electric fields $F_{\mu\nu}u^\nu_0$ and gradients of chemical potential $\dow_\mu\mu$ is controlled by the same transport coefficient $\sigma$. A physical implication of this is that the ``hydrostatic equilibrium configuration'' $\mu = \mu_0 + u^\mu_0 A_\mu$ identically solves the diffusion equation when coupled to background gauge fields that are time-independent in the rest frame of the inertial observer $u^\mu_0\dow_\mu A_\nu = 0$.

The particular form of the constitutive relations can also be understood as a consequence of the local second law of thermodynamics, i.e. there exists an entropy current $S^\mu$ whose divergence is non-negative for all solutions of the conservation equation. To see this, let us start with the internal energy density $\epsilon(s,n)$ as a function of the thermodynamic entropy density $s$ and charge density $n$, defining the thermodynamic equation of state of the system. The first law of thermodynamics can be parametrised as
\begin{equation}
     \df\epsilon 
     = T_0 \df s 
     + \mu \df n,
     \label{eq:firstlaw}
\end{equation}
which defines the chemical potential $\mu$. 
The thermodynamic temperature $T_0$ is taken to be constant; to allow for a spacetime-varying profile of temperature, we will need to bring in another equation for the conservation of energy; see \cref{sec:full-hydro}. Since we are only considering systems with conserved charge, we take the rate of change of energy density to be given by the associated Lorentz-power term $\dow_\mu (\epsilon u^\mu_0) = J^\mu F_{\mu\nu} u^\nu_0$. Using these, we can obtain the expression for entropy production
\begin{equation}
    \dow_\mu S^\mu
    = - \frac{1}{T_0}\cJ^\mu
    \Big(\dow_\mu\mu  - F_{\mu\nu} u^\nu_0 \Big)
    \geq 0,
    \label{eq:second-law}
\end{equation}
where $S^\mu = s u^\mu_0 -\mu/T_0\,\cJ^\mu$. 
We require that the constitutive relations for $\cJ^\mu$ must be such that entropy is strictly produced for all onshell configurations of $\mu$. This allows us to identify the constitutive relations in \cref{eq:Ji-consti} for an arbitrary non-negative conductivity $\sigma\geq 0$. Having done this, the entropy production rate is given as
\begin{equation}
    \dow_\mu S^\mu
    = \frac{1}{T_0\sigma} \cJ_\mu \cJ^\mu \geq 0.
    \label{eq:ec-net-rate}
\end{equation}

The utility of coupling the theory to background gauge fields $A_\mu$ is that we can use these to obtain the response functions (retarded correlation functions) of the conserved current $J^\mu$. With the forthcoming generalisation to hydrodynamics in mind, let us use a collective notation for the operators $\cO=(J^\mu)$ and the respective sources $\phi = (A_\mu)$. We define the classical expectation values of operators in the presence of sources
\begin{equation}
    \langle {\cal O}\rangle
    = {\cal O}\,\big|_{\text{onshell}},
\end{equation}
formally evaluated on the solutions of the conservation equation \eqref{eq:conservation}. The retarded 2-point retarded correlation functions can then be defined as simply
\begin{equation}
    G^R_{\cal O\cal O'} = \frac{\delta\langle \cO\rangle}{\delta\phi'}\bigg|_{\phi_0},
    \label{eq:retarded-var}
\end{equation}
where $\phi_0 = (A_\mu = 0)$ denote flat equilibrium sources. We can similarly obtain the higher-point retarded correlation functions by taking higher variations with respect to $\phi$. In Fourier-space, with four-momentum $p_\mu = (-\omega,k_i)$, and assuming the inertial observer to be at rest $u^\mu_0 = \delta^\mu_t$, we find that the retarded 2-point functions of the charge density $J^t$ and the transverse flux $J^i_\perp = k^i_j J^j$ (where $k_{ij} = \delta_{ij} - k_{i}k_j/k^2$) take the form
\begin{align}
    G^R_{J^tJ^t}(\omega,k)
    &= \frac{- \sigma k^2}
    {i\omega - D k^2}, \nn\\
    G^R_{J^i_\perp J^j_\perp}(\omega,k)
    &= i\omega\sigma\,k^{ij},
    \label{eq:diff-response}
\end{align}
where $D=\sigma/\chi$ is the diffusion constant with $\chi = \dow n/\dow\mu$ being the charge susceptibility. The response functions involving the longitudinal flux $J_\| = J^i k_i/k$ can be obtained using the conservation law: $-i\omega J^t + ik J_\| = 
0$. The pole structure of the response functions yield the mode spectrum of the theory. There is a single diffusive mode carried by the conserved charge fluctuations, with the dispersion relations
\begin{equation}
    \omega = -i D k^2 + \ldots,
    \label{eq:diff-mode}
\end{equation}
where ellipses denote the higher-$k$ corrections arising from the possible higher-derivative generalisations of the diffusion model. Thermodynamic stability of the equilibrium state requires that $\chi\geq 0$. Combined with the second law requirement $\sigma\geq 0$, this implies that $D\geq 0$, ensuring the stability of the diffusive mode.

We can obtain the retarded correlation functions and mode spectrum of relativistic diffusion in a boosted frame of reference $u^\mu_0 = (1,v^i_0)/\sqrt{1-v_0^2}$ by performing a Lorentz boost on the dispersion relations in the local rest frame, i.e.
\begin{gather}
    \omega \to \frac{1}{\sqrt{1-v_0^2}}
    \lb\omega - kv_0\cos\theta\rb, \qquad 
    k_i \to 
    \frac{1}{\sqrt{1-v_0^2}} \lb k \cos\theta
    - \omega v_0 \rb 
    \frac{v^0_i}{v_0} + k_j\!\lb \delta^j_i - \frac{v^0_iv_0^j}{v^2_0} \rb, \nn\\
    k^2 \to 
    \frac{1}{1-v_0^2} \lb k \cos\theta
    - \omega v_0 \rb^2
    + k^2\sin^2\theta,
    \label{eq:boost}
\end{gather}
where $k_i v^i_0 = kv_0\cos\theta$.
In a boosted reference frame, the dispersion relations are determined by a quadratic polynomial in $\omega$, giving rise to an additional mode. We find a boosted diffusive mode and a gapped mode
\begin{equation}
    \omega = k v_0 \cos\theta
    - iD k^2\sqrt{1-v_0^2}
    \lb 1 - v^2_0\cos^2\theta \rb + \ldots, \qquad 
    \omega = i\frac{\sqrt{1-v^2_0}}{v^2_0 D} + \ldots.
    \label{eq:boosted-diff-modes}
\end{equation}
While the diffusive mode is still stable, disturbingly, the new gapped mode has a positive imaginary part and hence is unstable. This is a generic feature of relativistic dissipative hydrodynamic theories: the rest-frame dispersion relations have higher powers of $k$ than $\omega$ due to the presence of dissipation, whereas after a boost transformation, we generically have as many powers of $\omega$ as $k$. This generates new unphysical unstable gapped modes in the boosted frame of reference, making the theory unusable beyond the perturbative regime. The theory also features modes propagating faster than the speed of light for large wavevectors and small boost parameter
\begin{equation}
    \omega = \lb \frac{\cos\theta}{v_0}
    \pm i\sqrt{\frac{1-v_0^2}{v_0^2}\sin^2\theta}
    + \cO(1/k) \rb k,
\end{equation}
whose real part is clearly not bounded from above.

These are not really conceptual problems with the relativistic diffusion model itself: the formalism is only strictly valid in the low-frequency regime. Since the unphysical modes typically show up at scales $|\omega|\gtrsim 1/\tau_* \equiv 1/D$, they lie outside the regime of applicability of the model. However, as noted in the introduction, they do pose practical problems when solving the Cauchy problem in a relativistic context, in particular the diffusion equation can not be coupled to Einstein's equations of General Relativity. To overcome these issues, one popular solution is to introduce additional degrees of freedom in the hydrodynamic theory. The extra degrees of freedom are gapped, so that the small-$k$ prediction in \cref{eq:diff-mode} remains untouched; their purpose is to serve as a consistent UV regularisation of the relativistic diffusion model.

\subsection{Maxwell-Cattaneo model}

While setting up the model of diffusion above, we assumed that all non-conserved ``fast'' degrees of freedom in the system have relaxed and thus the flux of conserved charge $\cJ^\mu$ is entirely given in terms of the density $n$. We can go slightly beyond this ``hydrodynamic regime'' by allowing the flux $\cJ^\mu$ to evolve independently of the density, with certain characteristic relaxation timescale $\tau$. The resulting theory is sometimes referred to as the Maxwell-Cattaneo (MC) model of diffusion \cite{1989RvMP...61...41J}. The inclusion of extra degrees of freedom can be thought of as conservatively including the ``slowest'' gapped non-hydrodynamic modes (one each in longitudinal and transverse sectors) into the framework. More formally, we should think of this procedure as introducing an ultraviolet regularisation to the diffusion model at the frequency scales $1/\tau$. If we tune $\tau\gtrsim \tau_*$, the new flux degrees of freedom will modify the physics before/near the onset of instabilities in \cref{eq:boosted-diff-modes} in frequency space, and will enable us to construct a diffusion model that is well-behaved for all Lorentz observers~\cite{Muller:1967zza, Israel:1976tn, 1976PhLA...58..213I, Israel:1979wp, 1979RSPSA.365...43I}.

We will construct the model of MC-diffusion using an additional vector degree of freedom $\upsilon_\mu$ (satisfying $\upsilon_\mu u^\mu_0 = 0$). At this point, the new degree of freedom $\upsilon_\mu$ is unrelated to the flux $\cJ^\mu$ defined in eq.~\eqref{eq:param-J}.
We promote the equation of state to $\epsilon = \epsilon(s,n,\upsilon^2)$, so that the energy density can now also depend on the scalar $\upsilon^2 = \upsilon^\mu\upsilon_\mu$, and the first law of thermodynamics in \cref{eq:firstlaw} accordingly gets modified to
\begin{equation}
     \df\epsilon 
     = T_0 \df s 
     + \mu \df n 
     + \frac{\chi_\upsilon}{2}\df\upsilon^2,
     \label{eq:thermo-J}
\end{equation}
which defines $\chi_\upsilon$, a new ``thermodynamic'' coefficient. To get some intuition about the new term, let us look at the associated canonical free energy density ${\cal F}_{\text{can}} = \epsilon - T s$, whose variation at fixed temperature is given as $\df{\cal F}_{\text{can}} = \mu\df n + \chi_\upsilon\upsilon^\mu\df\upsilon_\mu$. The field $\upsilon_\mu$ is akin to a vector order parameter in Landau’s theory of phase transitions. Here we are studying the phase with $\upsilon_\mu=0$ in equilibrium, and neglecting the kinetic term for $\upsilon_\mu$, as it is irrelevant compared to the mass term.  The model essentially corresponds to model C of the Hohenberg-Halperin classification~\cite{Hohenberg:1977ym},\footnote{The free energy of model C of Hohenberg-Halperin also contains a term proportional to $\dow_\mu n \dow^\mu n$, which is second order in derivatives. Technically, we should count $\upsilon_\mu\sim{\cal O}(\dow)$ and thus this term is as important as the $\upsilon^2$ term in the free energy. However, given that our goal is not to rigorously describe the second order corrections to the diffusion model, but instead to probe the effects of a gapped mode, we will drop this and other similar second-derivative terms here for simplicity. \label{foot:halperin}} with a conserved scalar density $n$ and a non-conserved 
vector density $\upsilon_\mu$.\footnote{The additional contribution to the free energy is reminiscent of the kinetic term in fluid dynamics, with $\upsilon_i$ playing the role of momentum density that is relaxed (i.e. not conserved), possibly due to the presence of impurities, and $1/\chi_\upsilon$ that of mass density. The term might also be familiar from superfluid dynamics, with $\upsilon_i$ playing the role of superfluid velocity and $\chi_\upsilon$ that of the superfluid density. However, since the curl of $\upsilon_i$ is unconstrained, this model actually corresponds to the theory of superfluidity with vortices, the presence of which causes the superfluid velocity to relax.\label{foot:sf-relax}} In this way of thinking, $\upsilon_\mu$ is an arbitrary field we introduced in the effective theory without inherent microscopic definition and we are free to arbitrarily redefine it to suit our needs. We shall fix this redefinition freedom by choosing the flux $\cJ^\mu$ to be parametrised as
\begin{equation}
    \cJ^\mu = \alpha_\upsilon \upsilon^\mu,
    \label{eq:frame-choice-diff}
\end{equation}
for some arbitrary coefficient $\alpha_\upsilon$. We can, of course, fix $\alpha_\upsilon$ as well by rescaling the definition of $\upsilon_\mu$. For example, we can choose $\alpha_\upsilon=1$ to give $\upsilon^\mu$ the interpretation of flux or $\alpha_\upsilon = \chi_\upsilon$ to give it the interpretation of the thermodynamic conjugate of flux. With the former choice, the relation in \cref{eq:thermo-J} corresponds to extended irreversible thermodynamics~\cite{1988EJPh....9..329J}. For now, we will keep $\alpha_\upsilon$ arbitrary as it will prove to be useful later. We can obtain the dynamical equations for $\upsilon_\mu$ using the second law of thermodynamics as our guiding principle. The entropy production equation \eqref{eq:second-law} accordingly modifies to
\begin{equation}
    \dow_\mu S^\mu 
    = - \frac{\alpha_\upsilon}{T_0} \upsilon^\mu\!\lb
    \dow_\mu\mu - F_{\mu\nu} u^\nu_0
    + \frac{\chi_\upsilon}{\alpha_\upsilon} 
    u^\lambda_0 \dow_\lambda \upsilon_\mu 
    \rb \geq 0,
\end{equation}
again with $S^\mu=s u_0^\mu-\mu/T_0\,\cJ^\mu$.
Inspecting this equation, we can read out the constitutive equations for $\upsilon_\mu$ to be
\begin{align}
    \alpha_\upsilon\upsilon^\mu 
    &= - \sigma\, \Delta^{\mu\nu}\Big(
    \dow_\nu\mu - F_{\nu\rho}u^\rho_0
    + \frac{\chi_\upsilon}{\alpha_\upsilon} 
    u^\lambda_0 \dow_\lambda \upsilon_\nu
    \Big).
    \label{eq:calJ-consti}
\end{align}
The net rate of entropy production is still given by \cref{eq:ec-net-rate}, however $\sigma$ can now be a function of both $n$ and $\upsilon^2$. We note that the constitutive equations \eqref{eq:calJ-consti} are only meant to be representative and do not characterise the most general constitutive relations within a derivative-order scheme.\footnote{For example, noting that the flux is counted as $\upsilon^i \sim {\cal O}(\dow)$ and the time-derivatives are counted as $\dow_t \sim {\cal O}(\dow^2)$ in the diffusion model, we can also imagine additional terms like $\dow^i\dow_t n$, $\dow_k\dow^k\upsilon^i$, $\dow_k\dow^k\dow^i n$ in the constitutive equations, which we are going to omit here for simplicity. As we mentioned in \cref{foot:halperin}, we are only interested in probing the effects of a massive mode and not rigorously describe higher-derivative corrections to the diffusion model.} We can eliminate $\upsilon_\mu$ in favour of $\cJ^\mu$ to write down an equation for the relaxation of flux
\begin{equation}
    \cJ^\mu 
    + \tau\,\Delta^{\mu\nu} u^\lambda_0 \dow_\lambda\cJ_\mu 
    = - \sigma\, \Delta^{\mu\nu}\Big(
    \dow_\nu\mu - F_{\nu\rho}u^\rho_0
    \Big)
    + \tau
    u^\lambda_0 \dow_\lambda\ln\alpha_\upsilon\,\cJ^{\mu} ,
    \label{eq:consti-cJ-relax}
\end{equation}
where we have identified the relaxation time-scale $\tau = \sigma\chi_\upsilon/\alpha_\upsilon^2$. For larger time-scales $t\gg \tau$, the description effectively reduces to the one with just the conserved hydrodynamic mode.

We should emphasise that the theory described above is more general than the original Maxwell-Cattaneo theory of diffusion, because all coefficients $\mu$, $\tau$, $\sigma$, $\alpha_\upsilon$ can be functions of both $n$ and $\upsilon^2$ (or $\cJ^2$). We can construct a simpler model where all these are taken to be just functions of $n$. However, using the thermodynamic relation \eqref{eq:thermo-J}, we know that $\dow\chi_\upsilon/\dow n = 2\dow\mu/\dow\upsilon^2$, implying that $\chi_\upsilon = \alpha_\upsilon^2\tau/\sigma$ must be constant when $\mu$ just depends on $n$. Defining the charge susceptibility $\chi=(\dow\mu/\dow n)^{-1}$ and the diffusion coefficient $D = \sigma/\chi$, both generic functions of $n$, the constitutive relations reduce to
\begin{align}
    {\cal J}^\mu 
    + \tau(n)\,\Delta^{\mu\nu}
    u^\lambda_0\dow_\lambda {\cal J}_\nu
    &= - D(n)\, \Delta^{\mu\nu}\dow_\nu n 
    + \sigma(n)\, F^{\mu\rho}u^0_\rho
    + \frac{\sigma}{2} 
    \frac{\dow(\tau/\sigma)}{\dow n}
    \cJ^{\mu} \dow_\rho\cJ^\rho,
    \label{eq:MIS-diff-simple}
\end{align}
where we have eliminated $2\dow_\mu\ln\alpha_\upsilon =
\dow_\mu\ln(\sigma/\tau_\cJ)$ for constant $\chi_\upsilon$ together with the charge conservation equation \eqref{eq:conservation}. If we wish to get rid of the last term, as in the original Maxwell-Cattaneo theory, we must impose a condition on the derivatives of relaxation time $\tau'(n) = \tau(n)/\sigma(n)\,\sigma'(n)$. Since the bare $\sigma$, barring its dependence within $D=\sigma/\chi$, is only visible when coupling to background sources, the constraint between $\tau'(n)$ and $\sigma'(n)$ is not relevant for the classical dynamical equations of $n$ in the absence of sources.

We can use the variational formula in \cref{eq:retarded-var} to obtain the retarded correlation functions in the presence of the new relaxed vector degrees of freedom. For the theory \eqref{eq:MIS-diff-simple}, the expressions in \cref{eq:diff-response} for the retarded correlation functions of density and transverse-flux computed in the local rest frame modify to
\begin{align}
    G^R_{J^tJ^t}(\omega,k)
    &= \frac{- \sigma k^2}
    {i\omega (1 - i\omega\tau) - D k^2}, \nn\\
    G^R_{J^i_\perp J^j_\perp}(\omega,k)
    &= \frac{i\omega\sigma}{1-i\omega\tau} k^{ij}.
    \label{eq:retarded-J}
\end{align}
The form of the response functions is almost identical to the ones we saw before in \cref{eq:diff-response}, but with a modified pole structure. In a generically boosted frame of reference, obtained by implementing the Fourier-space boost transformation in \cref{eq:boost}, we find the boosted diffusion mode and a gapped mode in the longitudinal sector
\begin{subequations}
\begin{equation}
    \omega = k v_0 \cos\theta
    - iDk^2\sqrt{1-v_0^2}
    \lb 1 - v^2_0\cos^2\theta \rb + \ldots, \qquad 
    \omega = -i\frac{\sqrt{1-v_0^2}}{\tau - v_0^2 D}
    + \ldots,
    \label{eq:boosted-diff-modes-tau}
\end{equation}
and another gapped mode in the transverse sector
\begin{equation}
    \omega = -i\frac{\sqrt{1-v_0^2}}{\tau}
    + \ldots.
    \label{eq:boosted-diff-modes-tau-trans}
\end{equation}
\end{subequations}
For the gapped poles to be stable for all inertial frames of references $0 \leq |v_0| \leq 1$, we need to bound $\tau$ from below
\begin{equation}
    \tau > D > 0.
\end{equation}
Given the second law constraint $\sigma\geq 0$, this also implies an inequality for thermodynamic susceptibilities $0 < \alpha_\upsilon^2/\chi_\upsilon < \chi$. With the stability bounds in place, the theory \eqref{eq:MIS-diff-simple} is also causal and the modes at large wavevector propagating as
\begin{equation}
    \omega = \lb c_\infty v_0 \cos\theta 
    \pm \sqrt{(1-c_\infty)(1-c_\infty v_0^2\cos^2\theta)}
    + \cO(1/k) \rb k,
\end{equation}
with $c_\infty = (\tau - D)/(\tau -v^2_0 D)$. Note that $0 < c_\infty \leq 1$ due to the stability bounds, which implies that the speed of modes at large wavevector is bounded from above by 1 for all $\theta$.

\subsection{Stochastic fluctuations and effective action}
\label{sec:MSR-diff}

The Maxwell-Catanneo diffusion model is deterministic. Even in the strict long-distance late-time hydrodynamic regime, where the only relevant macroscopic degrees of freedom are the conserved charges, the ignored microscopic degrees of freedom, colloquially called ``stochastic noise'', can still leave qualitatively distinct signatures through random interactions. The standard method to include stochastic noise into classical deterministic equations was outlined by Martin-Siggia-Rose (MSR)~\cite{Martin:1973zz}. The starting point of this formalism is to modify the constitutive equations for $\upsilon_\mu$ with an arbitrary random noise term $\theta^\mu$, i.e.
\begin{align}
    \alpha_\upsilon\upsilon^\mu 
    &= - \sigma\, \Delta^{\mu\nu} \Big(
    \dow_\nu\mu - F_{r\nu\rho}u^\rho_0
    + \frac{\chi_\upsilon}{\alpha_\upsilon} 
    u^\lambda_0 \dow_\lambda \upsilon_\nu
    \Big)
    + \theta^\mu.
    \label{eq:J-stochastic}
\end{align}
In anticipation of our later Schwinger-Keldysh discussion, we will label the background electromagnetic and gauge fields with a subscript ``$r$'' from this section onward.
We can imagine solving the stochastic equations \eqref{eq:J-stochastic} to obtain the noisy expectation values $\langle \ldots \rangle_{\theta}$ in the presence of background fields $\phi_r = (A_{r\mu})$ and noise fields $\theta = (\theta^\mu)$. The physical expectation values can be obtained by performing a weighted average over arbitrary random noise configurations 
\begin{equation}
    \langle \ldots \rangle
    = \int {\cal D}\theta
    \,\langle \ldots \rangle_{\theta} 
    \exp(-\frac14 \int \df^{d+1}x
    \left\langle
    \frac{\Delta_{\mu\nu}}{T_0\tilde\sigma}
    \right\rangle_{\!\!\theta}
    \theta^\mu\theta^\nu).
    \label{eq:noise-integration-diff}
\end{equation}
Here $\tilde\sigma(n,\upsilon^2)$ is a new coefficient introduced to model the strength of stochastic fluctuations. Note the $\langle \ldots \rangle_{\theta}$ around the noise strength in the above expression, which means that the $\tilde\sigma(n,\upsilon^2)$ also needs to be evaluated on the solutions of the equations of motion \eqref{eq:J-stochastic}. This is a slight generalisation of the original MSR formulation where the noise strength is taken to be a constant and is needed for conformity with non-linear fluctuation-dissipation theorems; see the following. In practise, we can convert the onshell objects $\langle \ldots \rangle_{\theta}$ inside the path integral into the respective offshell versions by introducing Lagrange multipliers $\psi_a = (\varphi_a,V_{a\mu})$ (normalised as $u_0^\mu V_{a\mu} = 0$) for the equations of motion, leading to
\begin{align}
    \langle \ldots \rangle
    &= \int {\cal D}\theta\,\cD\psi_r\,
    \cD\psi_a
    \,(\ldots)
    \exp(-\frac14 \int \df^{d+1}x
    \frac{\Delta_{\mu\nu}}{T_0\tilde\sigma}
    \theta^\mu\theta^\nu) \nn\\
    &\qquad
    \times \exp(-i\int \df^{d+1}x\,
    \varphi_a \dow_\mu (nu^\mu_0 + \alpha_\upsilon\upsilon^\mu)) \nn\\
    &\qquad
    \times \exp(-i\int \df^{d+1}x\,
    V_{a\mu}\! \lb 
    \alpha_\upsilon\upsilon^\mu 
    + \sigma\, \Delta^{\mu\nu} \Big(
    \dow_\nu\mu - F_{r\nu\rho}u^\rho_0
    + \frac{\chi_\upsilon}{\alpha_\upsilon} 
    u^\lambda_0 \dow_\lambda \upsilon_\nu
    \Big)
    - \theta^\mu
    \rb
    )\,,
    \label{eq:lagrange-mult}
\end{align}
where $\psi_r = (\mu,\upsilon_\mu)$ collectively denote the dynamical hydrodynamic degrees of freedom. In the limit $\chi_\upsilon\to 0$, the field $\upsilon_\mu$ becomes a simple Lagrange multiplier to impose $V_{a\mu} = \partial_\mu \varphi_a$, and we recover the Schwinger-Keldysh effective action for the ordinary diffusion model reported in~\cite{Kovtun:2014hpa}. To compute stochastic correlation functions of operators $\cO = (J^\mu)$ using the path integral described above, it is convenient to introduce another set of background fields $\phi_a = (A_{a\mu})$ and consider the \emph{Schwinger-Keldysh generating functional}
\begin{equation}
    {\cal Z}[\phi_r,\phi_a]
    = \left\langle
    \exp(i\int \df^{d+1}x\,\phi_a {\cal O})
    \right\rangle.
    \label{eq:GF-1}
\end{equation}
In terms of this, the retarded 2-point Green's function given in \cref{eq:retarded-var} and the symmetric 2-point Green's function $G^S_{\cO\cO'} = \langle\cO\cO'\rangle
- \langle\cO\rangle\langle\cO'\rangle
\big|_{\phi_0}$ can be computed using the variational formulae
\begin{align}
    G^R_{\cO\cO'} 
    &= \bfrac{\delta}{\delta\phi'_r}
    \bfrac{-i\delta}{\delta\phi_a}
    \ln{\cal Z}[\phi_r,\phi_a]
    \bigg|_{\phi_0}, \nn\\
    G^S_{\cO\cO'} 
    &= \bfrac{-i\delta}{\delta\phi'_a}
    \bfrac{-i\delta}{\delta\phi_a}
    \ln{\cal Z}[\phi_r,\phi_a]
    \bigg|_{\phi_0},
    \label{eq:corr-from-gf}
\end{align}
and similarly for higher-point functions. To simplify the generating functional, we note that path integral over the noise fields $\theta^\mu$ in \eqref{eq:GF-1} is Gaussian and can be performed analytically to yield 
\begin{equation}
    {\cal Z}[\phi_r,\phi_a]
    = \int\cD\psi_r\,\cD\psi_a
    \exp(iS[\psi_r,\psi_a;\phi_r,\phi_a]),
    \label{eq:SK-path-integral}
\end{equation}
where the \emph{Schwinger-Keldysh effective action} is given as
\begin{align}
    S
    &= 
    \int \df^{d+1}x
    \bigg[ B_{a\mu} n u^\mu_0
    + \alpha_\upsilon (B_{a\mu} - V_{a\mu}) \upsilon^\mu \nn\\
    &\hspace{8em}
    - \sigma\, V_{a}^\mu \lb 
    \dow_\mu \mu - F_{r\mu\rho} u^\rho_0
    + \frac{\chi_\upsilon}{\alpha_\upsilon}
    u_0^\lambda\dow_\lambda \upsilon_\mu
    \rb
    + iT_0\tilde\sigma\,V_{a}^\mu V_{a\mu}
    \bigg],
    \label{eq:action-MIS}
\end{align}
and we have defined $B_{a\mu} = \dow_\mu\varphi_a + A_{a\mu}$. In the limit $\chi_\upsilon\to 0$, the field $\upsilon_\mu$ becomes a simple Lagrange multiplier to impose $V_{a\mu} = B_{a\mu}$ and we recover the Schwinger-Keldysh effective action for the ordinary diffusion model reported in~\cite{Harder:2015nxa}.

\paragraph*{FDT and dynamical KMS symmetry:}

Using this action, it is straight-forward to verify that we recover the 2-point retarded correlation functions as given in \cref{eq:retarded-J}. We can also obtain the  symmetric 2-point correlation functions
\begin{align}
    G^S_{J^tJ^t}(\omega,k)
    &= \frac{2k^2T_0\tilde\sigma}
    {|i\omega (1 - i\omega\tau) - D k^2|^2}, \nn\\
    G^S_{J^i_\perp J^i_\perp}(\omega,k)
    &= \frac{2T_0\tilde\sigma}{|1-i\omega\tau|^2},
    \label{eq:symmetric-J}
\end{align}
which are proportional to the strength of stochastic noise in the model.
A characteristic feature of finite-temperature field theories are fluctuation-dissipation theorems. They posit that retarded and symmetric 2-point Green's functions of any operator, when computed in a thermal state with global temperature $T_0$ and velocity $u_0^\mu = \delta^\mu_t$, must satisfy the fluctuation-dissipation theorem (FDT) in \cref{eq:fdt}.
The boosted version of \cref{eq:fdt} can be obtained by replacing $\omega$ with $-u_0^\mu p_\mu$.
Applying this to the symmetric functions in \cref{eq:symmetric-J} fixes the strength of noise to be given by the dissipative conductivity transport coefficient
\begin{equation}
    \tilde\sigma = \sigma.
    \label{eq:sigma-KMS}
\end{equation}
Technically, the 2-point FDT only imposes this equality in thermodynamic equilibrium at $\mu=\mu_0$ and $\upsilon_\mu = 0$. However, similar non-linear FDTs exist for higher-point functions, which impose the relation \eqref{eq:sigma-KMS} for arbitrary thermodynamic arguments. Of course, checking this order-by-order for all higher-point correlation functions is a tedious exercise and it would be great if we could realise it as a symmetry directly at the level of the effective action. This is precisely the dynamical Kubo-Martin-Schwinger (KMS) symmetry as we discuss in the next subsection. 

To avoid getting lost into technicalities, let us already summarise the dynamical KMS symmetry of the effective action \eqref{eq:action-MIS}. First, we need to identify certain discrete symmetry $\Theta$ of the underlying microscopic theory including the time-reversal transformation T. Depending on the physical system under consideration, this could be just the time-reversal $\Theta=\rmT$, spacetime-parity $\Theta=\rmP\rmT$, time-reversal with charge conjugation $\Theta = {\rm CT}$, or spacetime-parity together with charge conjugation $\Theta={\rm CPT}$. The eigenvalues of various fields under different $\Theta$ symmetries is given in \cref{tab:CPT-rel}. The KMS symmetry is defined as the invariance of the Schwinger-Keldysh generating functional $\cZ[\phi_r,\phi_a]$ under a transformation of the background fields $\phi_{r,a}=(A_{r,a\mu})$ given by%
\footnote{Given a field $f(x)$ with $\Theta$-eigenvalue $\eta_f^\Theta = \pm 1$, defined for complex spacetime coordinates, its $\Theta$-transformation is given as $\Theta f(x) = \eta_f^\Theta f(\Theta x)$, where $\Theta t = -t$ and $\Theta \vec x = \pm\vec x$ depends on whether $\Theta$ includes the spatial parity operator P. For example, taking $\Theta={\rm PT}$ and $f$ to be PT-even, we have $\Theta f(x) = f(-x)$. Another field $g(x) = f(x+iV)$ is an eigenfunction of $\Theta$ if and only if the vector $V^\mu$ is real, with the $\Theta$-eigenvalue $\eta_g^\Theta = \eta_f^\Theta$. The $\Theta$-transformation of this equation is $\Theta g(x^*) = \Theta f(x^*{-\,}iV)$. In particular, the $\Theta$-transformation of an equation also implements a complex conjugation $\Theta:i\to -i$. The C,P,T eigenvalues of the derivatives $F_t = \dow_t f$ and $F_i = \dow_i f$ are defined as $(\eta^\rmC_f,\eta^\rmP_f,-\eta^\rmT_f)$ and $(\eta^\rmC_f,-\eta^\rmP_f,\eta^\rmT_f)$ respectively. We will define $\Theta\dow_\mu f(x) = \eta^\Theta_{\dow_\mu\!f} (\dow_\mu f)(\Theta x)$; this convention is different from~\cite{Glorioso:2018wxw} which uses $\eta^\Theta_f$ instead of $\eta^\Theta_{\dow_\mu\!f}$.
The benefit of this convention is that we simply have $\Theta F_\mu = \Theta\dow_\mu f$, unlike the conventions of~\cite{Glorioso:2018wxw} that require $\Theta F_\mu = \eta^\Theta_{\dow_\mu}\Theta\dow_\mu f$. } 
\begin{align}
    \phi_{r}
    &\to \Theta\phi_{r}, \qquad 
    \phi_{a}
    \to \Theta\lb\phi_{a}
    + i\beta_0^\mu\dow_\mu\phi_{r}\rb.
    \label{eq:KMS-back}
\end{align}
Note that the auxiliary field $\varphi_a$ appears in the effective action via the combination $B_{a\mu} = \dow_\mu\varphi_a + A_{a\mu}$, so its $\Theta$-eigenvalue is forced upon us by the $\Theta$-eigenvalue of the background fields, which turns out to be opposite the $\Theta$-eigenvalue of $\mu$. However, there is no such constraint for the auxiliary field $V_{a\mu}$, resulting in two independent KMS prescriptions for the dynamical fields. If we take the $\Theta$-eigenvalue of $V_{a\mu}$ to be the same as $\upsilon_\mu$, we find the ``standard'' prescription of KMS symmetry 
\begin{gather}
    \mu \to \Theta\mu, \qquad 
    \varphi_a \to \Theta\!\lb \varphi_a 
    + i\lb \frac{\mu-\mu_0}{T_0} - \beta_0^\mu A_{r\mu} \rb \rb, \nn\\
    \upsilon_\mu \to \Theta\upsilon_{\mu}, \qquad
    V_{a\mu} \to  
    \Theta\!\lb
    V_{a\mu} 
    + \frac{i}{T_0}\Delta_\mu^\nu \lb \dow_\nu\mu 
    + u_0^\lambda F_{r\lambda\nu}
    + \frac{\chi_\upsilon}{\alpha_\upsilon}
    u_0^\lambda\dow_\lambda\upsilon_{\nu}\rb \rb.
    \label{eq:KMS-diff-standard}
\end{gather}
On the other hand, if we take the $\Theta$-eigenvalue of $V_{a\mu}$ to be the opposite that of $\upsilon_\mu$, its KMS transformation has to be replaced with the ``alternate'' prescription
\begin{gather}
    V_{a\mu} \to  
    \Theta\!\lb
    V_{a\mu} 
    + \frac{i}{T_0}\frac{\alpha_\upsilon}{\sigma} 
    \upsilon_{\mu} \rb.
    \label{eq:KMS-diff-alternate}
\end{gather}
It can be explicitly checked that both these prescriptions leave the effective action \eqref{eq:action-MIS} invariant. We will discuss the physical principles underlying the standard prescription in detail in \cref{app:SK-MIS}, while the details of the alternate prescription can be found in~\cref{app:alt}.\footnote{The authors in~\cite{Mullins:new} arrived at the alternate prescription of KMS transformations by invoking the principle of detailed balance in the MSR effective action. To make contact with their language, let us define $\delta n = n(\mu) - n(\mu_0)$, $\delta\cJ_\mu = \alpha_\upsilon\upsilon_\mu$, $\delta \bar n = T_0\chi\varphi_a$, and $\delta\bar\cJ_\mu = T_0\sigma V_{a\mu}$. Choosing $\Theta={\rm PT}$ and turning off background fields, the alternate KMS transformations act as $\delta n \to \delta n$, $\delta\cJ_\mu \to \delta\cJ_\mu$, $\delta\bar n \to -\delta\bar n - i\delta n$, and $\delta\bar\cJ_\mu \to -\delta\bar\cJ_\mu - i\delta\cJ_\mu$, with the respective right-hand sides evaluated at $-x^\mu$, where we have ignored higher-order terms in fluctuations. These are precisely the KMS transformations obtained in~\cite{Mullins:new}, although we note that this simple form only applies at linear order in fluctuations and in the absence of background fields. On the other hand, under the standard KMS prescription, the transformation of $\delta\bar\cJ_\mu$ is modified to $\delta\bar\cJ_\mu \to \delta\bar\cJ_\mu + iD\Delta_\mu^\nu\dow_\nu n + i \tau u_0^\lambda\dow_\lambda \delta\cJ_\mu$, which is also a symmetry of the effective action derived in~\cite{Mullins:new}.\label{foot:pseudo-kms}}

\subsection{Schwinger-Keldysh formalism}
\label{app:SK-MIS}

In the previous subsection, we used the MSR formalism to phenomenologically derive the effective action for MC-diffusion, by integrating over random noise configurations dictated by FDT. In recent years, a more systematic SK-EFT framework has been formalised, which can be used to derive such effective actions starting from symmetry principles~\cite{Crossley:2015evo, Haehl:2018lcu, Jensen:2017kzi}. Furthermore, while the MSR formalism is only apt for describing Gaussian stochastic noise, the symmetry principles of the SK framework can also be used to systematically include higher non-Gaussian stochastic interactions into the hydrodynamic framework; see e.g.~\cite{Jain:2020zhu}. In this subsection, we review the SK formalism for relativistic diffusion, with appropriate modifications to account for the gapped MC degrees of freedom.

\paragraph*{Dynamical fields and global symmetries:} 

SK field theories are set-up on a closed-time contour, with leg ``1'' going forward in time and leg ``2'' returning backward in time to the initial state. While the conventional single-time contour from textbook quantum field theory can only access correlation functions at zero temperature, this exotic closed-time contour can be used to compute symmetric, retarded, and advanced correlators at finite temperature.
Each leg of the contour is equipped with its own set of degrees of freedom and background fields. For the present case of interest, the dynamical field content is a pair of phase fields $\varphi_{1,2}$, with the subscripts labelling the respective legs of the contour.
The theory is required to obey independent global U(1) symmetries on the two legs, i.e.
\begin{subequations}
\begin{equation}
    \varphi_{1,2}\to \varphi_{1,2} - \Lambda_{1,2}.
\end{equation}
We can gauge these symmetries by introducing a set of background gauge fields $A_{1,2\mu}$, with transformation rules
\begin{equation}
    A_{1,2\mu} \to A_{1,2\mu} + \dow_\mu \Lambda_{1,2}.
\end{equation}
\label{eq:global-U1-SK}%
\end{subequations}
Using these, we can define the global symmetry invariants
\begin{equation}
    B_{1,2\mu} = \dow_\mu\varphi_{1,2} + A_{1,2\mu},
\end{equation}
which are manifestly invariant under the global U(1) symmetries. The theory also carries a fixed thermal vector $\beta^\mu_0 = u^\mu_0/T_0$, characterising the four-velocity and temperature of the inertial observer with respect to which the global thermal equilibrium state is defined.

In addition, to implement the gapped degrees of freedom present in MC-diffusion, we also need to add a pair of dynamical spatial vector fields $\upsilon_{1,2\mu}$ satisfying
\begin{equation}
    \beta^\mu_0 \upsilon_{1,2\mu} = 0.
\end{equation}
The new degrees of freedom are taken to be invariant under the pair of global U(1) symmetries.

It is useful to introduce an average-difference basis $f_r = (f_1+f_2)/2$, $f_a = (f_1-f_2)/\hbar$ for various dynamical and background fields. The average ``$r$'' combinations are understood as the ``physical'' macroscopic fields, while the difference ``$a$'' combinations as the stochastic noise associated with them. The Schwinger-Keldysh effective action $S[\Phi_{r},\Phi_{a};\beta^\mu_0]$ of the theory can be expressed in terms of the global symmetry invariants $\Phi_{r,a}=(B_{r,a\mu},\upsilon_{r,a\mu})$ and the thermal vector $\beta^\mu_0$.

\paragraph*{Diagonal shift symmetry:}

In addition to the global symmetries above, we impose a local diagonal shift symmetry between the two phase fields
\begin{equation}
    \varphi_{1,2} \to \varphi_{1,2} 
    + \lambda,
    \label{eq:shift-SK}
\end{equation}
that acts uniformly on the two legs of the Schwinger-Keldysh contour. This symmetry is required to be time-independent, i.e.
\begin{equation}
    \beta^\mu_0\dow_\mu\lambda = 0,    
\end{equation}
Among the global symmetry invariants defined above, it is easy to see that the difference combination $B_{a\mu}$ is invariant under the diagonal shift symmetry, while average combination $B_{r\mu}$ transforms as a gauge field
\begin{equation}
    B_{r\mu} \to B_{r\mu} + \dow_\mu\lambda.
\end{equation}
Due to the time-independent nature of the symmetry, the time-component $u^\mu B_{r\mu}$ is invariant, which is identified as the chemical potential $\mu$ in the theory of diffusion.

\paragraph*{Schwinger-Keldysh generating functional:}

The fundamental object of interest in non-equilibrium field theory is the SK generating functional $\cZ[\phi_r,\phi_a]$, prescribed as a functional of the two sets of background fields $\phi_{r,a}=(A_{r,a\mu})$, which can be used to compute retarded and symmetric correlation functions of operators in the theory using the variational formulae \eqref{eq:corr-from-gf}. In the SK-EFT formalism, the generating functional is obtained by performing a path integral of the effective action $S[\Phi_{r},\Phi_{a};\beta^\mu]$ over the two sets of dynamical fields $\psi_{r,a}=(\varphi_{r,a},\upsilon_{r,a\mu})$, as in 
\begin{equation}
    {\cal Z}[\phi_r,\phi_a]
    = \int\cD\psi_r\,\cD\psi_a
    \exp(iS[\Phi_{r},\Phi_{a};\beta^\mu_0])
\end{equation}
The generating functional is required to satisfy the following three conditions
\begin{equation}
    \cZ[\phi_r,\phi_a = 0] = 1, \qquad 
    \cZ[\phi_r,-\phi_a] = \cZ^*[\phi_r,\phi_a], \qquad 
    \Re\cZ[\phi_r,\phi_a] \leq 0,
    \label{eq:SK-conditions-pf}
\end{equation}
arising from generic properties of quantum field theories defined on a closed-time contour.
More details regarding the underlying physics can be found in the review of~\cite{Glorioso:2018wxw}. These conditions can naturally be extended to the effective action as
\begin{equation}
    S[\Phi_r,\Phi_a = 0;\beta^\mu_0] = 0, \qquad 
    S[\Phi_r,-\Phi_a;\beta^\mu_0] 
    = -S^*[\Phi_r,\Phi_a;\beta^\mu_0], \qquad 
    \Im S[\Phi_r,\Phi_a;\beta^\mu_0] \geq 0.
    \label{eq:SK-conditions}
\end{equation}
We can arrange $S$ as a series in powers of $\Phi_a$, in which case the three conditions mean that: $S$ must at least be linear in $\Phi_a$, the terms with even-powers of $\Phi_a$ must be imaginary, and these imaginary terms must be arranged into a quadratic form with non-negative coefficients.

\paragraph*{Dynamical KMS symmetry:}

While the SK conditions \eqref{eq:SK-conditions-pf} are satisfied for arbitrary non-equilibrium field theories defined on a closed-time contour, to describe correlation functions in a thermal state, the SK generating functional $\cZ[\phi_r,\phi_a]$ is also required to satisfy a discrete dynamical KMS (Kubo-Martin-Schwinger) symmetry that is responsible for implementing the FDT \eqref{eq:fdt} and its higher-point generalisations.  Using the discrete microscopic $\Theta$ symmetry identified at the end of the previous subsection, the KMS symmetry is defined as a transformation of the background fields
\begin{gather}
    \phi_{1}(x)
    \to \Theta\phi_{1}(x), \qquad
    \phi_{2}(x)
    \to \Theta\phi_{2}(x + i\hbar\,\Theta\beta_0),
    \label{eq:KMS-full}
\end{gather}
where the $\Theta\phi_2$ is evaluated on the complex spacetime arguments $x^\mu + i\hbar\,\Theta\beta^\mu_0$, with $\beta^\mu_0 = u_0^\mu/T_0$ being the thermal vector associated with the inertial equilibrium observer. It can be check that KMS is a $\bbZ_2$ transformation: repeating it returns to the original background field configuration. More details can be found in textbook treatments of thermal field theory, e.g. in~\cite{Bellac:2011kqa}, or in the more recent review of~\cite{Glorioso:2018wxw}. Since the KMS condition acts differently on the two copies of background fields $\phi_{1,2}$, its action on the average-different fields $\phi_{r,a}$ is highly non-local. However, given that the background fields are sufficiently smooth functions, we can derive a simple expression in the classical ($\hbar\to 0$) limit, given in \cref{eq:KMS-back}.
This classical version of the dynamical KMS symmetry is more relevant for classical stochastic field theories valid at small frequencies $\omega \ll T_0/\hbar$, where quantum effects are suppressed. At quadratic order in background fields, this condition precisely yields the 2-point fluctuation-dissipation theorem in \cref{eq:fdt}. Similar higher-point relations can be derived by using this symmetry at higher non-linear orders in background fields.

\begin{table}[t]
  \centering
  \begin{tabular}{c|ccc|c|c|c}
    \toprule
    & C & P & T & PT & CT & CPT \\
    \midrule
    $X^0$, $t = x^0$, $\sigma^0$  & $+$ & $+$ & $-$ & $-$ & $-$ & $-$ \\
    $X^i$, $x^i$, $\sigma^i$ & $+$ & $-$ & $+$ & $-$ 
    & $+$ & $-$ \\
    $\varphi$ & $-$ & $+$ & $-$ & $-$ & $+$ & $+$ \\
    \midrule

    $u^i$ & $+$ & $-$ & $-$ & $+$ & $-$ & $+$ \\
    $T$ & $+$ & $+$ & $+$ & $+$ & $+$ & $+$ \\
    $\mu$ & $-$ & $+$ & $+$ & $+$ & $-$ & $-$ \\

    \midrule

    $\kappa_{ij}$ & $+$ & $+$ & $+$ & $+$ & $+$ & $+$ \\
    $\upsilon_i$ & $-$ & $-$ & $-$ & $+$ & $+$ & $-$ \\
    
    \midrule
    $T^{00}$, $g_{00}$ & $+$ & $+$ & $+$ & $+$ & $+$ & $+$ \\
    $T^{0i}$, $g_{0i}$ & $+$ & $-$ & $-$ & $+$ & $-$ & $+$ \\
    $T^{ij}$, $g_{ij}$ & $+$ & $+$ & $+$ & $+$ & $+$ & $+$ \\
    $J^0$, $A_0$ & $-$ & $+$ & $+$ & $+$ & $-$ & $-$ \\
    $J^i$, $A_i$ & $-$ & $-$ & $-$ & $+$ & $+$ & $-$ \\
    \bottomrule
  \end{tabular}
  \caption{Action of parity (P), time-reversal (T), and charge conjugation (C)
    of various quantities in classical relativistic hydrodynamics and effective
    field theory. Schwinger-Keldysh double copies of various quantities in the effective
    theory, ``$1/2$'' or ``$r/a$'', behave the same as their unlabelled
    versions.\label{tab:CPT-rel}}
\end{table}

In our SK-EFT for diffusion, the dynamical KMS symmetry is naturally realised on the dynamical fields as
\begin{gather}
    \varphi_1(x) 
    \to \Theta\varphi_1(x), \qquad 
    \varphi_2(t,\vec x) 
    \to \Theta\varphi_2(x+i\hbar\,\Theta\beta_0), \nn\\
    \upsilon_{1\mu}(x) 
    \to \Theta\upsilon_{1\mu}(x), \qquad 
    \upsilon_{2\mu}(x) 
    \to \Theta\upsilon_{2\mu}(x+i\hbar\,\Theta\beta_0).
    \label{eq:KMS-diff-def}
\end{gather}
In the classical limit, these transformations are realised on the average-difference basis as
\begin{gather}
    \varphi_r \to \Theta\varphi_r, \qquad 
    \varphi_a \to \Theta\lb \varphi_a 
    + i\beta^\lambda_0\dow_\lambda\varphi_r\rb, \nn\\
    \upsilon_{r\mu} \to \Theta\upsilon_{r\mu}, \qquad 
    \upsilon_{a\mu} \to \Theta\lb \upsilon_{a\mu}
    + i\beta^\lambda_0\dow_\lambda\upsilon_{r\mu} \rb.
    \label{eq:KMS-diff-classical}
\end{gather}
The global symmetry invariants $\Phi_{r,a}$ and the thermal vector $\beta^\mu$ comprising the Schwinger-Keldysh effective action $S[\Phi_r,\Phi_a;\beta^\mu_0]$ transform under the classical KMS symmetry as
\begin{equation}
    \Phi_r \to \Theta\Phi_r, \qquad 
    \Phi_a \to \Theta\lb \Phi_a
    + i\beta^\lambda_0\dow_\lambda\Phi_r \rb, \qquad 
    \beta^\mu_0 \to \Theta\beta^\mu_0.
    \label{eq:compound-diff}
\end{equation}
Given that KMS is a $\bbZ_2$ transformation, is also follows that $\Phi_a + i\beta^\lambda_0\dow_\lambda\Phi_r \to \Theta\Phi_a$.

\paragraph*{Effective action and hydrodynamic frames:}

As an example, truncating the theory to at most quadratic order in $\Phi_{a}$ noise fields and dropping any explicit spatial derivatives, the most general effective action for MC-diffusion is given as
\begin{align}
    S
    &= \int \df^{d+1}x
    \bigg[
    B_{a\mu} n u_0^\mu
    - \chi_\upsilon \upsilon_{a\mu}\upsilon_{r}^\mu \nn\\
    &\hspace{8em} 
    + iT_0\lb -\lambda\, u^\mu_0 u^\nu_0 
    + \sigma\,\Delta^{\mu\nu}
    \rb B_{a\mu}
    \lb B_{a\nu} 
    + i\beta_0^\lambda \dow_\lambda B_{r\nu} \rb\nn\\
    &\hspace{8em} 
    + iT_0\sigma_{\upsilon}\, \Delta^{\mu\nu}\upsilon_{a\mu}
    \lb \upsilon_{a\nu} 
    + i\beta_0^\lambda \dow_\lambda\upsilon_{r\nu} \rb \nn\\
    &\hspace{8em} 
    + iT_0\gamma_\times\,\Delta^{\mu\nu} 
    \Big( B_{a\mu} \lb \upsilon_{a\nu} 
    + i\beta_0^\lambda \dow_\lambda\upsilon_{r\nu} \rb
    + \upsilon_{a\mu}
    \lb B_{a\nu} + i\beta_0^\lambda \dow_\lambda B_{r\nu} \rb
    \Big) \nn\\
    &\hspace{8em} 
    + iT_0\bar\gamma_\times\,\Delta^{\mu\nu} 
    \Big( B_{a\mu} \lb \upsilon_{a\nu} 
    + i\beta_0^\lambda \dow_\lambda \upsilon_{r\nu} \rb
    - \upsilon_{a\mu}
    \lb B_{a\nu} + i\beta_0^\lambda \dow_\lambda B_{r\nu} \rb
    \Big)
    \bigg],
    \label{eq:action-SK-raw}   
\end{align}
where we have identified $\mu \equiv B_{rt}$ and $\upsilon_\mu \equiv \upsilon_{r\mu}$, and various coefficients $n$, $\chi_\upsilon$, $\lambda$, $\sigma$, $\sigma_\upsilon$, $\gamma_\times$, and $\gamma'_\times$ are functions of $\mu$ and $\upsilon^2$.
The action is manifestly invariant under the global U(1) and spatially-local diagonal shift symmetries. It trivially obeys the first two SK conditions; for the third condition, we can arrange the imaginary part of the action into a sum of squares 
\begin{align}
    &
    -T_0\lambda\, \lb u^\mu_0B_{a\mu}\rb^2
    + iT_0\sigma\,
    \lb B_{a\mu} + \frac{\gamma_\times}{\sigma}\upsilon_{a\mu} \rb 
    \lb B_{a}^\mu + \frac{\gamma_\times}{\sigma}\upsilon_{a}^\mu \rb 
    + T_0
    \lb \sigma_\upsilon
    - \frac{\gamma_\times^2}{\sigma} \rb 
    \upsilon_{a\mu}\upsilon_a^\mu
    \geq 0,
\end{align}
leading to the inequality constraints
\begin{equation}
    \lambda \leq 0, \qquad 
    \sigma \geq 0, \qquad 
    \sigma_\upsilon \geq 
    \frac{\gamma_\times^2}{\sigma}.
    \label{eq:SK-ineq}
\end{equation}
Finally, let us look at the dynamical KMS symmetry. Using $\Theta = {\rm T,PT}$, the terms in the first line map to themselves (up to a coordinate flip in the integral $x^\mu \to \Theta x^\mu$), with the residual terms that sum to a total derivative and drop out from the effective action, i.e.
\begin{equation}
    i\beta_0^\mu \lb n\, \dow_\mu \mu
    - \chi_\upsilon \upsilon^\nu
    \dow_\mu \upsilon_{\nu} \rb
    = \dow_\mu\!\lb ip\beta_0^\mu \rb,
    \label{eq:totalder-diff}
\end{equation}
Here we have used the thermodynamic relations in \cref{eq:thermo-J} and identified the thermodynamic pressure $p = T_0 s + n\mu - \epsilon$. Since $\Phi_a$ and $\Phi_a + i\beta^\lambda_0\dow_\lambda\Phi_r$ map to each other under KMS transformations (up to $x^\mu \to \Theta x^\mu$), the terms in the next three lines are manifestly invariant. Finally, the final $\bar\gamma_\times$ term is disallowed by KMS for $\Theta = {\rm T,PT}$. However, for $\Theta = {\rm CT,CPT}$, this term is allowed as long as $\bar\gamma_\times$ is an odd function of $\mu$. Imposing the KMS symmetry with $\Theta = {\rm CT,CPT}$ also requires $n$ to be an odd function of $\mu$, while the remaining coefficients $\chi_\upsilon$, $\lambda$, $\sigma$, $\sigma_\upsilon$, and $\gamma_\times$ must be even functions of $\mu$. 

The SK effective action \eqref{eq:action-SK-raw} has more parameters compared to the action \eqref{eq:action-MIS} derived using the MSR formalism, so is clearly more general. However, this is merely a manifestation of choice of hydrodynamic frame. To see this, let us look at the vector equation of motion obtained by varying with respect to $\upsilon_{ai}$; in the absence of noise fields we get
\begin{equation}
    \chi_\upsilon \upsilon^\mu
    = - (\gamma_\times-\bar\gamma_\times) 
    \Delta^{\mu\nu}\!\lb \dow_\nu \mu - F_{r\nu\rho}u^\rho_0 \rb
    - \sigma_\upsilon
    \Delta^{\mu\nu}
    u_0^\lambda \dow_\lambda \upsilon_{\nu},
    \label{eq:non-frame-upsilon-eq}
\end{equation}
where we have used $u^\lambda_0\dow_\lambda B_{r\mu} = \dow_\mu \mu - F_{r\mu\nu}u^\nu_0$.
On the other hand, we can find the physical current $J^\mu$ by varying the action with respect to $A_{a\mu}$; using the above equations and in the absence of noise fields, we find
\begin{align}
    J^\mu
    &= \lb n + \lambda\, u_0^\lambda\dow_\lambda\mu \rb u_0^\mu
    - \sigma\, \Delta^{\mu\nu} \lb \dow_\nu \mu - F_{r\nu\rho}u^\rho_0 \rb 
    - (\gamma_\times+\bar\gamma_\times) \Delta^{\mu\nu}
    u_0^\lambda \dow_\lambda \upsilon_{\nu} \nn\\
    &= \lb n + \lambda\, u_0^\lambda\dow_\lambda\mu \rb u_0^\mu
    + \frac{\sigma\chi_\upsilon}{\gamma_\times - \bar\gamma_\times} \upsilon^\mu
    + \frac{\sigma\sigma_\upsilon - \gamma_\times^2 + \bar\gamma_\times^2
    }{\gamma_\times-\bar\gamma_\times} \Delta^{\mu\nu}
    u_0^\lambda \dow_\lambda\upsilon_{\nu}.
    \label{eq:general-frame}
\end{align}
Reconciling this with the frame choice we made in \cref{eq:param-J,eq:frame-choice-diff}, and noting that the $\bar\gamma_\times$ term is not KMS-invariant for $\Theta={\rm T,PT}$, we are required to set
\begin{equation}
    \lambda = 0, \qquad 
    \sigma_\upsilon = \frac{\gamma_\times^2}{\sigma}, \qquad 
    \gamma_\times 
    = \frac{\sigma\chi_\upsilon}{\alpha_\upsilon}, \qquad 
    \bar\gamma_\times = 0.
    \label{eq:extra-coeff-frame}
\end{equation}
On the other hand, for $\Theta = {\rm CPT,CT}$, since $\bar\gamma_\times$ is an odd function of $\mu$, requring the coefficient of $\upsilon^\mu_r$ in \cref{eq:general-frame} to be $\alpha_\upsilon$ for all $\mu$ ultimately yields the same constraints as above.
The final effective action is given as
\begin{align}
    S
    &= 
    \int \df^{d+1}x
    \bigg[ B_{a\mu} n u_0^\mu
    - \chi_\upsilon\upsilon_{a\mu}\upsilon_{r}^\mu  \nn\\
    &\hspace{5em}
    + iT_0\sigma \,\Delta^{\mu\nu}
    \lb B_{a\mu} 
    + \frac{\chi_\upsilon}{\alpha_\upsilon}\upsilon_{a\mu}\rb\!\!
    \lb B_{a\nu}
    + i\beta^\lambda_0\dow_\lambda B_{r\nu}
    + \frac{\chi_\upsilon}{\alpha_\upsilon}\!\lb
    \upsilon_{a\nu}
    + i\beta^\lambda_0\dow_\lambda\upsilon_{r\nu} \rb  \rb 
    \bigg].
    \label{eq:final-action-diff}
\end{align}
This is precisely the effective action \eqref{eq:action-MIS} derived using the MSR formalism, with the KMS condition \eqref{eq:sigma-KMS} already imposed, with the field identifications
\begin{equation}
    \upsilon_\mu = \upsilon_{r\mu}, \qquad
    V_{a\mu} = \Delta_\mu^\nu B_{a\nu} 
    + \frac{\chi_\upsilon}{\alpha_\upsilon}\upsilon_{a\mu}.
    \label{eq:field-redef-msr}
\end{equation}
One can check that this results in the standard prescription of KMS symmetry given in \cref{eq:KMS-diff-standard}. To realise the alternate prescription in \cref{eq:KMS-diff-alternate}, we need to modify the SK-EFT by imposing a diagonal shift symmetry between the vector degrees of freedom; see \cref{app:alt}.

\section{M\"uller-Israel-Stewart model of hydrodynamics}
\label{sec:full-hydro}

In this section, we will implement the lessons learnt from the diffusion model and setup an EFT framework for the M\"uller-Israel-Stewart (MIS) theory of relativistic hydrodynamics. Unlike the internal U(1) symmetry in the diffusion model, we will now need to account for the spacetime Poincar\'e transformations in relativistic hydrodynamics. Spacetime symmetries are highly non-linear and give rise to certain subtleties in the MIS framework, causing the subsequent effective field theory framework to be considerably more involved. Nonetheless, on a conceptual front, the construction here will be a straightforward generalisation of our discussion in \cref{sec:MIS-diffusion}.

\subsection{Relativistic hydrodynamics}

The starting point of relativistic hydrodynamics are a set of conservation equations for the energy-momentum tensor $T^{\mu\nu}$ and the charge current $J^\mu$, associated with spacetime Poincar\'e and internal U(1) symmetries, i.e.
\begin{align}
    \nabla_\mu T^{\mu\nu} &= F^{\nu\rho}J_\rho, \nn\\
    \nabla_\mu J^\mu &= 0.
    \label{eq:hydro-conservation}
\end{align}
We have implicitly introduced a background spacetime metric $g_{\mu\nu}$ coupled to the energy-momentum tensor $T^{\mu\nu}$ via the associated covariant derivative operator $\nabla_\mu$. Note that energy-momentum tensor is sourced by the Lorentz force term in the presence of background electromagnetic field strength $F_{\mu\nu} = 2\dow_{[\mu}A_{\nu]}$. The energy-momentum conservation equation can be used to determine the dynamics of the energy density $\epsilon$ and fluid velocity $u^\mu$, defined as the time-like eigenvalue and eigenvector of the energy-momentum tensor $T^{\mu\nu}u_\nu = - \epsilon\, u^\mu$ (with $u^\mu u_\mu = -1$). This definition of $\epsilon$ and $u^\mu$ is usually referred to as the ``Landau frame'' in the literature, following \cite{landau1959fluid}. Similarly, the charge conservation equation can be used to determine the dynamics of the charge density $n$, defined via $J^\mu u_\mu = -n$. However, to complete these equations, we need to specify a set of constitutive relations for the remaining components of the $T^{\mu\nu}$ and $J^\mu$. 

This is where the second law of thermodynamics comes in: we require that there exists an entropy current $S^\mu$, whose divergence is locally non-negative for all solutions of the conservation equations. We start, as with the diffusion model, with the equation of state $\epsilon(s,n)$. However, this time around, energy conservation is included within our set of conservation equations, so we do not need to make a separate ansatz for $\dow_t\epsilon$ like we did in the diffusion model. We can parametrise the first law of thermodynamics as
\begin{equation}
    \df\epsilon = T\df s + \mu\df n,
    \label{eq:thermo-hydro}
\end{equation}
where $T$ and $\mu$ are the local thermodynamic temperature and chemical potential. Note that the temperature is not a constant and we have allowed it to have a nontrivial spacetime profile. Let us also parametrise the constitutive relations as
\begin{align}
    T^{\mu\nu} 
    &= (\epsilon+p) u^\mu u^\nu + p\,g^{\mu\nu}
    + \Pi^{\mu\nu}, \nn\\
    J^\mu 
    &= n\, u^\mu + {\cal J}^\mu,
    \label{eq:hydro-decomposition}
\end{align}
where $p(\epsilon,n) = Ts+\mu n - \epsilon$ denotes the thermodynamic pressure of the fluid and $\Delta^{\mu\nu} = g^{\mu\nu} + u^\mu u^\nu$ is the spatial projection operator. The transverse tensor $\Pi^{\mu\nu}$ and vector ${\cal J^\mu}$ (satisfying $\Pi^{\mu\nu}u_\nu = {\cal J}^\mu u_\mu = 0$ and $\Pi^{\mu\nu}=\Pi^{\nu\mu}$) parametrise the to-be-determined information in the stress tensor (spatial components of the energy-mometum tensor) and charge flux. The utility of this parametrisation is that the second law statement reduces to a simple form
\begin{equation}
    \nabla_\mu S^\mu 
    = - \Pi^{\mu\nu} \nabla_{(\mu} \beta_{\nu)}
    - {\cal J}^\mu \lb \dow_\mu\frac{\mu}{T}
    + \beta^\lambda F_{\lambda\mu} \rb
    \geq 0,
\end{equation}
where $\beta^\mu = u^\mu/T$ and the entropy current is given as $S^\mu = s\,u^\mu - \mu/T\,\cJ^\mu$. A simple entropy-conserving solution of this equation is $\Pi^{\mu\nu} = {\cal J}^\mu = 0$; these are nothing but the ideal non-dissipative relativistic fluids. The leading order dissipative corrections are given as
\begin{align}
    \Pi^{\mu\nu} &= 
    - \zeta\,\Delta^{\mu\nu}
    \nabla_\lambda u^\lambda
    - 2\eta\,\Delta^{\rho\langle\mu}
    \Delta^{\nu\rangle\sigma}
    \nabla_\rho u_\sigma, \nn\\
    {\cal J}^\mu &= -\sigma\,\Delta^{\mu\nu}
    \lb T \nabla_\nu\frac{\mu}{T}
    + u^\lambda F_{\lambda\nu} \rb,
\end{align}
where $\zeta$, $\eta$, $\sigma$ are non-negative transport coefficients identified as fluid bulk viscosity, shear viscosity, and conductivity respectively. The angular brackets denote the symmetric transverse-traceless combination of indices: $X_{\langle\mu\nu\rangle} = 1/2\,(X_{\mu\nu}+X_{\nu\mu}) - 1/d\,\Delta_{\mu\nu} \Delta^{\rho\sigma}X_{\rho\sigma}$. Above, we have used the fact that $\Delta^{\mu\nu}\nabla_\mu\beta_\nu = 1/T\,\nabla_\mu u^\mu$ and $\Delta_{\langle\mu}^\rho
    \Delta_{\nu\rangle}^\sigma 
    \nabla_\rho\beta_\sigma = 1/T\,\Delta_{\langle\mu}^\rho
    \Delta_{\nu\rangle}^\sigma 
    \nabla_\rho u_\sigma$.

Consider the equilibrium configuration $\beta^\mu = \delta^\mu_t/T_0$ and $\mu/T = (\mu_0 + A_t)/T_0$ in the presence of time-independent background sources $g_{\mu\nu}(\vec x)$, $A_\mu(\vec x)$. It is easy to see that in this configuration 
\begin{equation}
    2 \nabla_{(\mu} \beta_{\nu)} 
    = \frac1T_0 \dow_t g_{\mu\nu} = 0, \qquad 
    \nabla_\mu\frac{\mu}{T}
    + \beta^\lambda F_{\lambda\mu}
    = \frac{1}{T_0}\dow_t A_\mu = 0.
\end{equation}
As a consequence, entropy is identically conserved in an equilibrium configuration. It is also easy to see that the dissipative corrections $\Pi^{\mu\nu}$, ${\cal J}^\mu$ identically vanish in equilibrium and the ideal fluid constitutive relations trivially satisfy the conservation equations. Correspondingly, the equilibrium configuration is a trivial solution of the dissipative hydrodynamic equations.

Having set up the hydrodynamic model, we can use the 
variational formulae in \cref{eq:retarded-var} to compute the retarded correlation functions of the operators ${\cal O} = (\sqrt{-g}\,T^{\mu\nu},\sqrt{-g}\,J^\mu)$ by varying with respect to the associated background sources $\psi = (\half g_{\mu\nu}, A_\mu)$, with the flat values $\psi_0 = (\half g_{\mu\nu} = \half\eta_{\mu\nu},A_\mu = 0)$. As an example, let us re-evaluate the 2-point retarded Green's functions for the density $J^t$ and transverse flux $J^i_\perp$ using the hydrodynamic model in a state with equilibrium velocity $u^\mu_0=\delta^\mu_t$, i.e.
\begin{align}
    G^R_{J^tJ^t}(\omega,k)
    &= - \frac{\sigma k^2 
    }{i\omega - D_n k^2}
    - \frac{
    \frac{n^2}{\epsilon+p} k^2}{
    \omega^2 - v_s^2k^2 + i\omega D_\pi^\| k^2}, \nn\\
    G^R_{J^i_\perp J_\perp^j}(\omega,k)
    &= \lb i\omega\sigma 
    - \frac{\frac{n^2}{\epsilon+p}i\omega}
    {i\omega - D_\pi^\perp k^2} \rb k^{ij},
    \label{eq:response-hydro}
\end{align}
where we have defined 
\begin{gather}
    v_s^2 = \frac{\dow p}{\dow\epsilon}\bigg|_{s/n}, \qquad 
    \frac1\chi 
    = \frac{T\dow(\mu/T)}{\dow n}\bigg|_\epsilon, \qquad 
    D_n = \frac{\sigma}{\chi}, \qquad 
    D_\pi^\perp = \frac{\eta}{\epsilon+p}, \qquad 
    D_\pi^\| = \frac{\zeta+2\frac{d-1}{d}\eta}{\epsilon+p}.
\end{gather}
For clarity of presentation, we have taken $\dow p/\dow n|_\epsilon = 0$, which causes the mode spectrum in the charge and energy-momentum sectors to neatly decouple: in the longitudinal sector we find the familiar charge diffusion mode and a new fluid sound mode
\begin{subequations}
\begin{equation}
    \omega = -iD_nk^2 + \ldots, \qquad 
    \omega = \pm v_s k 
    - \frac{i}{2} D_\pi^\| k^2
    + \ldots.
\end{equation}
In the transverse sector, we find the shear diffusion mode
\begin{equation}
    \omega  = -iD_\pi^\perp k^2 + \ldots.
\end{equation}
\label{eq:hydromodes}%
\end{subequations}
More details on the retarded correlation functions and mode spectrum of relativistic hydrodynamics can be found in the lecture notes of~\cite{Kovtun:2012rj}. The instabilities we encountered in a boosted frame of reference in the relativistic diffusion model near \cref{eq:boosted-diff-modes} also plague dissipative relativistic hydrodynamics. In a boosted frame of reference, obtained using the transformation \eqref{eq:boost}, the sound and diffusion modes above get appropriately boosted, but we also get a few unstable gapped modes
\begin{gather}
    \omega 
    = i\frac{\sqrt{1-v^2_0}}{v^2_0 D_n} + \ldots, \qquad 
    \omega 
    = i\frac{\sqrt{1-v^2_0}(1 - v^2_0v_s^2)}{v_0^2 D_\pi^\|}
    + \ldots, \qquad 
    \omega 
    = i\frac{\sqrt{1-v^2_0}}{v_0^2 D_\pi^\perp} + \ldots.
\end{gather}
To cure these instabilities, we will need to introduce new gapped modes near the frequency scales of instabilities, which will be the goal of MIS-hydrodynamics in the next subsection.

\subsection{M\"uller-Israel-Stewart model}
\label{sec:MIS-hydro}

To construct the M\"uller-Israel-Stewart (MIS) model of hydrodynamics, let us introduce new tensor $\kappa_{\mu\nu}$ and vector $\upsilon_\mu$ degrees of freedom (satisfying $\kappa_{\mu\nu}u^\nu = \upsilon_\mu u^\mu = 0$ and $\kappa_{\mu\nu} = \kappa_{\nu\mu}$) to allow for all the components in the energy-momentum tensor $T^{\mu\nu}$ and $J^\mu$ to fluctuate independently. The first law of thermodynamics from \cref{eq:thermo-hydro} needs to be appropriately modified to include the additional dependence on the scalars $\tr\kappa = \kappa^{\mu}_{~\mu}$, $\kappa^2 = \kappa^{\mu\nu}\kappa_{\mu\nu}$, and $\upsilon^2 = \upsilon^\mu \upsilon_\mu$, i.e.
\begin{align}
    \df\epsilon
    &= T \df s + \mu\df n
    + \frac{\chi_\kappa^\ssfS}{8}
    \df (\tr\kappa)^2
    + \frac{\chi_\kappa^\ssfT}{4}
    \df\!\lb \kappa^2 - \frac1d (\tr\kappa)^2 \rb
    + \frac{\chi_\upsilon}{2}\df\upsilon^2,
    \label{eq:thermo-hydro-Pi}
\end{align}
where $\chi_\kappa^\ssfS$, $\chi_\kappa^\ssfT$, and $\chi_\upsilon$ are new thermodynamic coefficients and $d$ is the number of spatial dimensions.\footnote{This setup is reminiscent of the theory of plasticity, with $\half\kappa_{\mu\nu}$ serving as the strain tensor, and the coefficients $\chi_\kappa^\ssfS$ and $\chi_\kappa^\ssfT$ as the bulk and shear moduli respectively. However, note that here $\kappa_{\mu\nu}$ is not required to be given by the symmetric derivative of a displacement field, hence this setup applies to a plastic material, not elastic, leading to a relaxed plastic stress tensor; see e.g.~\cite{Fukuma:2011pr, Armas:2022vpf}. Correspondingly, the dynamical equation for $\Pi_{\mu\nu}$ in \eqref{eq:mis-hydro-consti-Pi} that we will soon derive, correspond to the rheology equations of a Maxwell material.\label{foot:plastic}} A priori, $\kappa_{\mu\nu}$ and $\upsilon_\mu$ are introduced as arbitrary degrees of freedom unrelated to $T^{\mu\nu}$ and $J^\mu$. To fix this, let us decompose the grand-canonical free-energy density $\cF(\epsilon,n,\tr\kappa,\kappa^2,\upsilon^2) = \epsilon - Ts-\mu n$ into a fluid part $-p(\epsilon,n)$ and the remainder $-\cP(\epsilon,n,\tr\kappa,\kappa^2,\upsilon^2)$ that vanishes when both $\kappa_{\mu\nu}$ and $\upsilon_\mu$ are taken to zero. Having done this, we choose the parametrisation for the constitutive relations in \cref{eq:hydro-decomposition} and choose $\Pi^{\mu\nu}$ and $\cJ^\mu$ to be related to $\kappa_{\mu\nu}$ and $\upsilon_\mu$ as
\begin{align}
    \Pi^{\mu\nu} 
    &= \frac{\alpha_\kappa^\ssfS}{2}\Delta^{\mu\nu}\tr\kappa
    + \alpha_\kappa^\ssfT \kappa^{\langle\mu\nu\rangle}
    + \lb 
    \cP \Delta^{\mu\nu}
    + \frac{\chi^\ssfS_\kappa}{2}\kappa^{\mu\nu} \tr\kappa
    + \chi_\kappa^\ssfT
    \lb \kappa^{\mu\rho}\kappa^{\nu}_{~\rho}
    - \frac1d \kappa^{\mu\nu}\tr\kappa
    \rb
    + \chi_\upsilon \upsilon^\mu\upsilon^\nu\rb
    , \nn\\
    \cJ^\mu 
    &= \alpha_\upsilon \upsilon^\mu,
    \label{eq:hydro-reparametrisation}
\end{align}
where we have introduced arbitrary coefficients $\alpha_\kappa^\ssfS$, $\alpha_\kappa^\ssfT$, and $\alpha_\upsilon$ akin to our choice in \cref{eq:frame-choice-diff}. Similar to our discussion below \cref{eq:frame-choice-diff}, we can always arbitrarily redefine the fields $\kappa_{\mu\nu}$ and $\upsilon_\mu$ to set these coefficients to 1, which, linearly, would give them the physical interpretation of spatial stress tensor and flux directly, or to $\chi_\kappa^\ssfS$, $\chi_\kappa^\ssfT$, and $\chi_\upsilon$ respectively, giving them the physical interpretation that of the respective thermodynamic conjugates. The additional terms in the parenthesis in \cref{eq:hydro-reparametrisation} have been included for later compatibility with the SK-EFT.\footnote{This is the so-called ``thermodynamic frame'', where the conserved currents are decomposed into a ``hydrostatic part'' obtained from the grand-canonical free energy density $\cF = -p-\cP$ and an arbitrary ``non-hydrostatic part'' transverse to the fluid velocity, i.e.
\begin{equation}
    T^{\mu\nu} = - \frac{2}{\sqrt{-g}} \frac{\delta(\sqrt{-g}\,\cF) }{\delta g_{\mu\nu}}
    + T^{\mu\nu}_{\text{nhs}}, \qquad 
    J^\mu = - \frac{1}{\sqrt{-g}} \frac{\delta(\sqrt{-g}\,\cF) }{\delta A_{\mu}}
    + J^\mu_{\text{nhs}}.
\end{equation}
The variations above are evaluated at fixed $\beta^\mu = u^\mu/T$, $\Lambda_\beta = \mu/T - \beta^\mu A_\mu$, $\kappa_{\mu\nu}$, and $\upsilon_\mu$. The fields $\kappa_{\mu\nu}$ and $\upsilon_\mu$ are related to $T^{\mu\nu}_{\text{nhs}}$ and $J^\mu_{\text{nhs}}$ as
\begin{equation}
    T^{\mu\nu}_{\text{nhs}} = \frac{\alpha_\kappa^\ssfS}{2}\Delta^{\mu\nu}\tr\kappa
    + \alpha_\kappa^\ssfT \kappa^{\langle\mu\nu\rangle}, \qquad 
    J^\mu_{\text{nhs}}
    = \alpha_\upsilon\,\upsilon^\mu.
\end{equation}
The SK-EFT for hydrodynamics, realising the non-linear discrete KMS symmetry, naturally give rise to the hydrodynamic equations in the thermodynamic Landau frame, as we shall review in the following.

Note that with our particular choice of thermodynamics in \cref{eq:thermo-hydro-Pi}, the thermodynamic frame agrees with the definition of the Landau frame $T^{\mu\nu}u_\nu = -\epsilon\, u^\mu$ and $J^\mu u_\mu = -n$. However, this does not hold more generally. In particular, as explained in~\cite{Kovtun:2022vas}, if we allow the thermodynamic equation of state in \cref{eq:thermo-hydro-Pi} to also depend on the spacetime curvature (as is the case even in non-interacting field theories), the thermodynamic frame would no longer satisfy the Landau frame condition. This means that the definition of thermodynamic variables, such as temperature $T$, in the thermodynamic frame are generically different from their Landau frame definitions.}  For the time being, the utility of this rather non-trivial choice is that the expression for entropy production only modifies in terms of Lie derivatives of $\kappa_{\mu\nu}$ and $\upsilon_\mu$ along $\beta^\mu = u^\mu/T$, denoted by $\lie_\beta$, i.e.
\begin{align}
    \nabla_\mu S^\mu
    &= 
    - \frac{\alpha_\kappa^\ssfS}{4} \tr\kappa
    \lb 2\nabla_\lambda\beta^\lambda
    + \frac{\chi^\ssfS_\kappa}{\alpha_\kappa^\ssfS}
    \Delta^{\rho\sigma}
    \lie_\beta \kappa_{\rho\sigma} \rb
    - \!
    \half \alpha_\kappa^\ssfT \kappa^{\langle\mu\nu\rangle} 
    \lb 2\nabla_{\mu}\beta_{\nu}
    + \frac{\chi^\ssfT_\kappa}{\alpha_\kappa^\ssfT}
    \lie_\beta \kappa_{\mu\nu} \rb \nn\\
    &\qquad 
    - \alpha_\upsilon \upsilon^\mu\!
    \lb  \dow_\mu\frac{\mu}{T}
    + \beta^\lambda F_{\lambda\mu} 
    + \frac{\chi_\upsilon}{\alpha_\upsilon} 
    \lie_\beta\upsilon_\mu
    \rb \geq0.
    \label{eq:dS-mis-diff}
\end{align}
The entropy current is again given as $S^\mu = s\,u^\mu - \mu/T\,\cJ^\mu$.
Physically, the Lie derivatives in this expression mean that the new degrees of freedom only modify the entropy production of relativistic hydrodynamics if they evolve in the local rest frame of the fluid. 
Requiring that entropy is strictly produced for all hydrodynamic configurations immediately allows us identify the dynamical equations for $\kappa_{\mu\nu}$ and $\upsilon_\mu$, i.e.
\begin{align}
    \frac{\alpha_\kappa^\ssfS}{2}\Delta^{\mu\nu}\tr\kappa
    + \alpha_\kappa^\ssfT \kappa^{\langle\mu\nu\rangle}
    &= 
    - \zeta\,\Delta^{\mu\nu}
    \lb \nabla_\lambda u^\lambda 
    + \frac{T\chi^\ssfS_\kappa}{2\alpha_\kappa^\ssfS}
    \Delta^{\rho\sigma}
    \lie_\beta \kappa_{\rho\sigma}
    \rb 
    - 2\eta\,\Delta^{\rho\langle\mu}
    \Delta^{\nu\rangle\sigma}
    \lb \nabla_\rho u_\sigma
    + \frac{T\chi^\ssfT_\kappa}{2\alpha_\kappa^\ssfT}
    \lie_\beta \kappa_{\rho\sigma}
    \rb, \nn\\
    \alpha_\upsilon\upsilon^\mu
    &= -\sigma\,\Delta^{\mu\nu}
    \lb T \nabla_\nu\frac{\mu}{T}
    + u^\lambda F_{\lambda\nu} 
    + \frac{T\chi_\upsilon}{\alpha_\upsilon} 
    \lie_\beta\upsilon_\nu
    \rb.
    \label{eq:mis-hydro-consti}
\end{align}
Due to the non-linear relation in \cref{eq:hydro-reparametrisation}, the relaxation equations for $\Pi^{\mu\nu}$ and $\cJ^\mu$ are quite involved. At the linearised level, we find
\begin{align}
    \Pi^{\mu\nu}
    + \lb \frac1d \tau_\Pi^\ssfS
    \Delta^{\mu\nu}\Delta^{\rho\sigma}
    + \tau_\Pi^\ssfT
    \Delta^{\rho\langle\mu}
    \Delta^{\nu\rangle\sigma}
    \rb T\lie_\beta \Pi_{\rho\sigma}
    &= 
    - \zeta\,\Delta^{\mu\nu}\nabla_\lambda u^\lambda 
    - 2\eta\,\Delta^{\rho\langle\mu}
    \Delta^{\nu\rangle\sigma}
    \nabla_\rho u_\sigma + \ldots, \nn\\
    \cJ^\mu + \tau_\cJ\,
    \Delta^{\mu\nu}T\lie_\beta \cJ_\nu
    &= -\sigma\,\Delta^{\mu\nu}
    \lb T \nabla_\nu\frac{\mu}{T}
    + u^\lambda F_{\lambda\nu} 
    \rb
    + \ldots,
    \label{eq:mis-hydro-consti-Pi}
\end{align}
where we have identified the three distinct relaxation time-scales in the scalar, tensor, and vector sectors of the theory, i.e. $\tau_\Pi^\ssfS = \zeta\chi^\ssfS_\kappa/(\alpha^\ssfS_\kappa)^2$, $\tau_\Pi^\ssfT = \eta\chi^\ssfT_\kappa/(\alpha^\ssfT_\kappa)^2$, and $\tau_{\cal J} = \sigma\chi_\upsilon/\alpha_\upsilon^2$. More complete expressions for conformal fluids are provided later in \cref{sec:conformal}.

We can use the modified theory MIS-hydrodynamics to obtain the retarded Green's functions of hydrodynamic observables. In practice, this amounts to replacing $\zeta \to \zeta/(1-i\omega\tau_\Pi^\ssfS)$, $\eta \to \eta/(1-i\omega\tau_\Pi^\ssfT)$, and $\sigma \to \sigma/(1-i\omega\tau_\cJ)$ in the response functions we found in \cref{eq:response-hydro}. Choosing $\tau_\Pi^\ssfS =\tau_\Pi^\ssfT = \tau_\Pi$ for simplicity, we are led to
\begin{align}
    G^R_{J^tJ^t}(\omega,k)
    &= - \frac{\sigma k^2 
    }{i\omega(1-i\omega\tau_\cJ) - D_n k^2} 
    - \frac{
    \frac{n^2}{\epsilon+p} k^2
    (1-i\omega\tau_\Pi)
    }{
    (\omega^2 - v_s^2k^2)(1-i\omega\tau_\Pi)
    + i\omega D_\pi^\| k^2
    }, \nn\\
    G^R_{J^i_\perp J_\perp^j}(\omega,k)
    &= \lb \frac{i\omega\sigma}{1-i\omega\tau_\cJ}
    - \frac{\frac{n^2}{\epsilon+p}i\omega
    (1-i\omega\tau_\Pi^\ssfT)}
    {i\omega(1-i\omega\tau_\Pi) - D_\pi^\perp k^2} \rb k^{ij}.
\end{align}
The pole structure of these correlators still admits the hydrodynamic modes presented in \cref{eq:hydromodes}. We also find new gapped modes, which in a generic boosted frame of reference take the form
\begin{equation}
    \omega = -i\frac{\sqrt{1-v^2_0}}{\tau_\cJ - v^2_0 D_n}
    + \ldots, \qquad 
    \omega = -i\frac{\sqrt{1-v^2_0}}{
    \tau_\Pi - v^2_0/(1-v^2_0v_s^2)\, D_\pi^\|} + \ldots,
\end{equation}
in the longitudinal sector and
\begin{equation}
    \omega = -i\frac{\sqrt{1-v^2_0}}{\tau_\cJ}
    + \ldots, \qquad 
    \omega = -i\frac{\sqrt{1-v^2_0}}{\tau_\Pi - v^2_0 D_\pi^\perp}
    + \ldots,
\end{equation}
in the transverse sector. For stability of these modes, we need to take the relaxation times to be bounded from below as
\begin{equation}
    \tau_\cJ \geq D_n \geq 0, \qquad 
    \tau_\Pi \geq \frac{1}{1-v_s^2} D_\pi^\| 
    \geq D_\pi^\perp \geq 0.
\end{equation}
More intricate bounds exist when we allow for $\tau_\Pi^\ssfS \neq \tau_\Pi^\ssfT$ and $\dow p/\dow n|_\epsilon \neq 0$, which we will not explore here. Generically, given a theory of dissipative relativistic hydrodynamics, we can always setup a theory of MIS-hydrodynamics with sufficiently large relaxation time-scales $\tau_\Pi^\ssfS$, $\tau_\Pi^\ssfT$, and $\tau_\cJ$, so that the theory remains linearly stable in all boosted frames of references.

Had we used a more traditional parametrisation of the constitutive relations in terms of $\kappa_{\mu\nu}$ and $\upsilon_\mu$, i.e. absorbed $\cP$ in \cref{eq:hydro-reparametrisation} into the thermodynamic pressure $p$ in \cref{eq:hydro-decomposition} and dropped the remaining terms in the parenthesis in \cref{eq:hydro-reparametrisation}, the Lie derivatives $\lie_\beta$ in \cref{eq:dS-mis-diff,eq:mis-hydro-consti,eq:mis-hydro-consti-Pi} above would have been replaced with $\beta^\mu\nabla_\mu$. This would result in a more familiar version of MIS-hydrodynamics as found in~\cite{Muller:1967zza, Israel:1976tn, 1976PhLA...58..213I, Israel:1979wp, 1979RSPSA.365...43I, Hiscock:1983zz, Baier:2007ix, Denicol:2012cn}. However, as we will see later in \cref{app:SK-hydro}, the associated EFT derived using the MSR formalism would not agree with the dynamical KMS symmetry non-linearly and thus would violate higher-point FDTs. Note that the two choices of parametrisations only differ from each-other non-linearly and yield the same results for mode spectrum within the regime of linearised hydrodynamics.

\subsection{Stochastic fluctuations and effective action}
\label{sec:MSR-hydro}

We can follow the MSR procedure in \cref{sec:MSR-diff} to promote the MIS theory of hydrodynamics presented above into a effective field theory. We start by introducing random noise terms $\theta^{\mu\nu}_\cT$ and $\theta^\mu_\cJ$ in the constitutive relations \eqref{eq:mis-hydro-consti} respectively, i.e.
\begin{align}
    \frac{\alpha_\kappa^\ssfS}{2}\Delta^{\mu\nu}\tr\kappa
    + \alpha_\kappa^\ssfT \kappa^{\langle\mu\nu\rangle}
    &= 
    - \zeta\,\Delta^{\mu\nu}
    \lb \nabla_\lambda u^\lambda 
    + \frac{T\chi^\ssfS_\kappa}{2\alpha_\kappa^\ssfS}
    \Delta^{\rho\sigma}
    \lie_\beta \kappa_{\rho\sigma}
    \rb 
    - 2\eta\,\Delta^{\rho\langle\mu}
    \Delta^{\nu\rangle\sigma}
    \lb \nabla_\rho u_\sigma
    + \frac{T\chi^\ssfT_\kappa}{2\alpha_\kappa^\ssfT}
    \lie_\beta \kappa_{\rho\sigma}
    \rb \nn\\
    &\qquad
    + \theta^{\mu\nu}_\Pi, \nn\\
    \alpha_\upsilon\upsilon^\mu
    &= -\sigma\,\Delta^{\mu\nu}
    \lb T \nabla_\nu\frac{\mu}{T}
    + u^\lambda F_{\lambda\nu} 
    + \frac{T\chi_\upsilon}{\alpha_\upsilon} 
    \lie_\beta\upsilon_\nu
    \rb
    + \theta^\mu_\cJ,
    \label{eq:mis-hydro-consti-st}
\end{align}
Physical correlations are computed similar to \cref{eq:noise-integration-diff} by integrating the noisy expectation values of operators $\langle\ldots\rangle_{\theta}$, evaluated on the solutions of the stochastic equations in the presence of the background fields $\phi_r=(\half g_{r\mu\nu},A_{r\mu})$ and noise fields $\theta = (\theta^{\mu\nu}_\Pi, \theta^\mu_\cJ)$. The noise configurations are sampled from the weight distribution
\begin{equation}
    \exp(-    \frac{1}{4}
    \int\df^{d+1}x\sqrt{-g_r}
    \lB
    \left\langle
    \frac{\Delta_{\mu\nu}\Delta_{\rho\sigma}}{d^2 T\tilde\zeta}
    + \frac{\Delta_{\rho\langle\mu}\Delta_{\nu\rangle\sigma}}{2T\tilde\eta}
    \right\rangle_{\!\!\theta}
    \theta_\Pi^{\mu\nu}\theta^{\rho\sigma}_\Pi
    + \left\langle\frac{\Delta_{\mu\nu}}{T\tilde\sigma}
    \right\rangle_{\!\!\theta}
    \theta^\mu_\cJ \theta^\nu_\cJ \rB),
    \label{eq:gaussian-noise-hydro}
\end{equation}
where the new coefficients $\tilde\zeta$, $\tilde\eta$, and $\tilde\sigma$ control the strength of stochastic fluctuations. Next, similar to \cref{eq:lagrange-mult}, we can convert the noisy expectation values $\langle\ldots\rangle_{\theta}$ into offshell fields by integrating over the hydrodynamic fields $\psi_r = (\beta^\mu,\mu,\kappa_{\mu\nu},\upsilon_\mu)$ and a set of Lagrange multipliers $\psi_a=(X^\mu_a,\varphi_a+X_a^\mu A_{r\mu},\half W_{a\mu\nu},V_{a\mu})$ for the two conservation equations in \cref{eq:hydro-conservation} and the two sets of constitutive relations in \cref{eq:mis-hydro-consti-st} respectively. To conveniently compute correlation functions, we can define the Schwinger-Keldysh path integral as \cref{eq:GF-1}, by introducing ``$a$''-type background sources $\phi_a=(\half g_{a\mu\nu},A_{a\mu})$ to probe the operators $\cO = (\sqrt{-g}\,T^{\mu\nu},\sqrt{-g}\,J^\mu)$. Finally, we can analytically perform the Gaussian path integral over the noise fields $\theta = (\theta^{\mu\nu}_\Pi, \theta^\mu_\cJ)$ and arrive at the generating functional in \cref{eq:SK-path-integral}, with the associated effective action
\begin{align}
    S &= \int\df^{d+1}x\sqrt{-g_r}
    \bigg[
    \half G_{a\mu\nu} 
    \lb \epsilon\, u^\mu u^\nu 
    - \cF\Delta^{\mu\nu} \rb
    + B_{a\mu} n u^\mu \nn\\
    &\hspace{10em}
    + \half (G_{a\mu\nu} - W_{a\mu\nu}) 
    \lb \frac{\alpha_\kappa^\ssfS}{2}\Delta^{\mu\nu}\tr\kappa
    + \alpha_\kappa^\ssfT \kappa^{\langle\mu\nu\rangle} \rb
    + (B_{a\mu} - V_{a\mu}) \alpha_\upsilon\upsilon^{\mu}
    \nn\\
    &\hspace{10em}
    + \half G_{a\mu\nu} 
    \lb \frac{\chi^\ssfS_\kappa}{2}  
    \kappa^{\mu\nu}\tr\kappa
    + \chi^\ssfT_\kappa
    \lb \kappa^{\mu\rho}\kappa^{\nu}_\rho
    - \frac1d \kappa^{\mu\nu}\tr\kappa \rb
    + \chi_\upsilon\upsilon^\mu\upsilon^\nu \rb
    \nn\\
    &\hspace{10em}
    - \half\zeta\, \Delta^{\mu\nu} \Delta^{\rho\sigma} 
    W_{a\mu\nu}\!\lb 
    \nabla_\rho u_\sigma 
    + \frac{T\chi^\ssfS_\kappa}{2\alpha^\ssfS_\kappa} \lie_\beta\kappa_{\rho\sigma}
    \rb 
    + \frac{iT}{4} \tilde\zeta\,\Delta^{\mu\nu}\Delta^{\rho\sigma}
    W_{a\mu\nu}W_{a\rho\sigma}
    \nn\\
    &\hspace{10em}
    - \eta\, \Delta^{\rho\langle\mu}
    \Delta^{\nu\rangle\sigma} W_{a\mu\nu}\!
    \lb 
    \nabla_\rho u_\sigma 
    + \frac{T\chi_\kappa^\ssfT}{2\alpha_\kappa^\ssfT}
     \lie_\beta\kappa_{\rho\sigma}
    \rb 
    + \frac{iT}{2}
    \tilde\eta\,\Delta^{\rho\langle\mu}
    \Delta^{\nu\rangle\sigma}
    W_{a\mu\nu}W_{a\rho\sigma} \nn\\
    &\hspace{10em}
    - \sigma\,\Delta^{\mu\nu}V_{a\mu}\!\lb 
    T \nabla_\nu\frac{\mu}{T}
    + u^\lambda F_{r\lambda\nu}
    + \frac{T\chi_\upsilon}{\alpha_\upsilon} 
    \lie_\beta\upsilon_\nu
    \rb
    + iT\tilde\sigma\, \Delta^{\mu\nu}V_{a\mu}V_{a\nu}
    \bigg],
    \label{eq:action-hydro-raw}
\end{align}
where $G_{a\mu\nu} = g_{a\mu\nu} + \lie_{X_a}g_{r\mu\nu}$ and $B_{a\mu} = A_{a\mu} + \dow_\mu \varphi_a + \lie_{X_a} A_{r\mu}$. In the limit $\chi_\kappa^\ssfS,\chi_\kappa^\ssfT,\chi_\upsilon\to 0$, the fields $\kappa_{\mu\nu}$ and $\upsilon_\mu$ become Lagrange multipliers to impose the conditions $W_{a\mu\nu} = G_{a\mu\nu}$ and $V_{a\mu} = B_{a\mu}$ respectively, and we recover the effective field theory for ordinary relativistic hydrodynamics developed in~\cite{Crossley:2015evo, Haehl:2018lcu, Jensen:2017kzi}.

\paragraph*{FDT and KMS symmetry:}

We can use the effective action derived above to compute the retarded and symmetric correlation functions of relativistic MIS-hydrodynamics. Requiring the 2-point correlation functions to satisfy FDT imposes
\begin{equation}
    \tilde\zeta = \zeta, \qquad 
    \tilde\eta = \eta, \qquad 
    \tilde\sigma = \sigma.
    \label{eq:KMS-conditions}
\end{equation}
As we will see in the next subsection in detail, this choice also guarantees that the effective action realises the non-linear dynamical KMS symmetry and thus is compatible with all higher-point FDTs as well. The KMS symmetry is realised on the background fields $\phi_{r,a}=(\half g_{r,a\mu\nu},A_{r,a\mu})$ the same as \cref{eq:KMS-back}. However, similar to our discussion around \cref{eq:KMS-diff-standard}, the dynamical fields can realise four different prescriptions of KMS transformations depending on the choice of $\Theta$-eigenvalues of the auxiliary fields $W_{a\mu\nu}$ and $V_{a\mu}$. If we take these to transform the same way as $\kappa_{\mu\nu}$ and $\upsilon_\mu$ respectively, the effective action realises the ``standard'' prescription of the dynamical KMS symmetry
\begin{gather}
    \beta^\mu \to \Theta\beta^\mu, \qquad 
    X^\mu_a \to  \Theta\Big( X^\mu_a
    + i(\beta^\mu - \beta^\mu_0) \Big), \nn\\
    \mu \to \Theta\mu, \qquad 
    \varphi_a \to \Theta\!\lb \varphi_a 
    + i\lb \frac{\mu}{T} - \frac{\mu_0}{T_0} - \beta^\mu A_{r\mu} \rb \rb, \nn\\
    \kappa_{\mu\nu} \to \Theta\kappa_{\mu\nu}, \quad 
    W_{a\mu\nu}
    \to \Theta\!\lb
    W_{a\mu\nu}
    + \frac{i}{T}\!\lb
    2\Delta_{(\mu}^\rho\Delta_{\nu)}^\sigma \nabla_\rho u_\sigma
    + \lb \frac1d \frac{\chi_\kappa^\ssfS}{\alpha_\kappa^\ssfS} \Delta_{\mu\nu}
    \Delta^{\rho\sigma}
    + \frac{\chi_\kappa^\ssfT}{\alpha_\kappa^\ssfT}
    \Delta_{\langle\mu}^\rho\Delta_{\nu\rangle}^\sigma
    \rb T\lie_\beta\kappa_{\rho\sigma}
    \rb
    \rb, \nn\\
    \upsilon_\mu \to \Theta\upsilon_{\mu}, \qquad 
    V_{a\mu} \to  
    \Theta\!\lb
    V_{a\mu} 
    + \frac{i}{T}\Delta_\mu^\nu \lb 
    T\dow_\nu\frac{\mu}{T} 
    + u^\lambda F_{r\lambda\nu}
    + \frac{T\chi_\upsilon}{\alpha_\upsilon}
    \lie_\beta\upsilon_{\nu}\rb \rb.
    \label{eq:KMS-hydro-standard}
\end{gather}
On the other hand, if $W_{a\mu\nu}$ and/or $V_{a\mu}$ transform oppositely to $\kappa_{\mu\nu}$ and/or $\upsilon_\mu$, their KMS transformations are given by the alternate prescriptions
\begin{subequations}
\begin{equation}
    W_{a\mu\nu}
    \to \Theta \!\lb W_{a\mu\nu}
    + \frac{i}{T}\lb
    \frac1d\frac{\alpha_\kappa^\ssfS}{\zeta}
    \Delta_{\mu\nu}\Delta^{\rho\sigma}
    + \frac{\alpha_\kappa^\ssfT}{\eta} 
    \Delta_{\langle\mu}^\rho\Delta_{\nu\rangle}^\rho
    \rb
    \kappa_{\rho\sigma} \rb,
\end{equation}
and/or
\begin{gather}
    V_{a\mu} \to  
    \Theta\!\lb
    V_{a\mu} 
    + \frac{i}{T}\frac{\alpha_\upsilon}{\sigma} 
    \upsilon_{\mu} \rb.
\end{gather}
\label{eq:KMS-hydro-alt}%
\end{subequations}
respectively. The derivation of these transformation rules in presented in \cref{app:SK-hydro,app:alt}.

Once again, had we used a more traditional parametrisation of the constitutive relations in terms of $\kappa_{\mu\nu}$ and $\upsilon_\mu$, as described towards the end of \cref{sec:MIS-hydro}, the occurrences of $\lie_\beta$ in the effective action \eqref{eq:action-hydro-raw} would be replaced with $\beta^\mu\nabla_\mu$. However, as we discuss in the next subsection, the resulting effective action cannot be made non-linearly compatible with the dynamical KMS symmetry. In other words, the simple choice of Gaussian weight distribution for noise configurations in \cref{eq:gaussian-noise-hydro} would not be sufficient to guarantee agreement with all higher-point FDTs.

\subsection{Schwinger-Keldysh formalism}
\label{app:SK-hydro}

We will now derive the effective action \eqref{eq:action-hydro-raw} for MIS-hydrodynamics using the Schwinger-Keldysh formalism and illustrate how it non-linearly realises the dynamical KMS symmetry. The following discussion is a straightforward generalisation of the original construction of SK-EFTs for relativistic hydrodynamics~\cite{Crossley:2015evo, Haehl:2018lcu, Jensen:2017kzi}, extended to include the gapped MIS fields.

\paragraph*{Fluid worldvolume, dynamical fields, and global symmetries:}

The presence of global spacetime symmetries makes the effective field theory for hydrodynamics a little more subtle than the linear diffusion model. Just like the doubled phase fields in the diffusion model on which the internal U(1) symmetries act, the spacetime on which the spacetime symmetries act also needs to be doubled. The elegant insight of~\cite{Crossley:2015evo, Haehl:2018lcu, Jensen:2017kzi} is to setup the effective field theory for hydrodynamics as a sigma-model on an auxiliary ``fluid worldvolume'' with coordinates $\sigma^\alpha$. Dynamical fields live on the fluid worldvolume: a pair of coordinate fields $X^\mu_{1,2}(\sigma)$ defining the two copies of spacetime and a pair of U(1) phase fields $\varphi_{1,2}(\sigma)$ akin to the diffusion model. 

The hydrodynamic effective field theory realises the global spacetime and internal symmetries independently on the two spacetimes. We will directly start with the gauged versions of these symmetries: spacetime diffeomorphisms and U(1) gauge transformations
\begin{align}
    X^\mu_{1,2}
    &\to X'^{\mu}_{1,2}(X_{1,2}), \nn\\
    \varphi_{1,2}
    &\to \varphi_{1,2}
    - \Lambda_{1,2}(X_{1,2}).
\end{align}
with the associated background fields: a pair of background spacetime metrics $g_{1,2\mu\nu}(X_{1,2})$ and U(1) gauge fields $A_{1,2\mu}(X_{1,2})$. Background U(1) gauge transformations act on $A_{1,2\mu}$ as
\begin{equation}
    A_{1,2\mu}(X_{1,2})
    \to 
    A_{1,2\mu}(X_{1,2})
    + \frac{\dow}{\dow X^\mu_{1,2}} \Lambda_{1,2}(X_{1,2}),
\end{equation}
whereas background diffeomorphisms act on $g_{1,2\mu\nu}$ and $A_{1,2\mu}$ as usual. When coupled to a flat background, $g_{1,2\mu\nu} = \eta_{\mu\nu}$, $A_{1,2\mu} = 0$, these symmetries reduce the global symmetries: spacetime Poincar\'e transformations and constant U(1) transformations.
In practise, we can pullback these fields onto the fluid worldvolume into objects invariant under the global symmetries
\begin{align}
    \bbg_{1,2\alpha\beta}
    &= g_{1,2\mu\nu}(X_{1,2})\, 
    \dow_\alpha X^\mu_{1,2}\,
    \dow_\beta X^\nu_{1,2}, \nn\\
    \bbA_{1,2\alpha}
    &= A_{1,2\mu} (X_{1,2})\, 
    \dow_\alpha X^\mu_{1,2}
    + \dow_\alpha \varphi_{1,2}.
    \label{eq:global-invariants-hydro}
\end{align}
We also equip the fluid worldvolume with a fixed thermal vector $\bbbeta^\alpha = \delta^\alpha_0/T_0$, which characterise the equilibrium thermal state around which the hydrodynamic theory is being setup.

To model the additional gapped modes present in the MIS theory, we will also introduce a pair of tensor fields $\bbkappa_{1,2\alpha\beta}(\sigma)$ and a pair vector fields $\bbupsilon_{1,2\alpha}(\sigma)$  on the fluid worldvolume. Since we only intend to introduce new degrees of freedom corresponding to stresses and fluxes and not for energy-momentum and charge densities, we will choose these new fields to be transverse to the thermal vector, i.e.
\begin{equation}
    \bbbeta^\alpha \bbkappa_{1,2\alpha\beta}
    = 0, \qquad    
    \bbbeta^\alpha \bbupsilon_{1,2\alpha}
    = 0.
\end{equation}
Both these fields are taken to be invariant under the global spacetime and U(1) symmetries. 

We can define the average-difference basis $f_r = (f_1+f_2)/2$, $f_a = (f_1-f_2)/\hbar$ for various quantities, which will be useful later. Due to the non-linear nature of the theory, $\bbg_{r,a\alpha\beta}$ and $\bbA_{r,a\mu}$ are non-trivially related to $X_{r,a}^\mu$, $\varphi_{r,a}$, $g_{r,a\mu\nu}$, $A_{r,a\mu}$. For example, in the classical limit we find
\begin{align}
    \bbg_{r\alpha\beta}
    &= g_{r\mu\nu}(X_r)\, 
    \dow_\alpha X^\mu_r\,
    \dow_\beta X^\nu_r
    + \cO(\hbar), \nn\\
    \bbg_{a\alpha\beta}
    &= \Big( g_{a\mu\nu}(X_r)\, 
    + \lie_{X_a} g_{r\mu\nu}(X_r)\Big)
    \dow_\alpha X^\mu_{r}\, \dow_\beta X^\nu_{r}
    + \cO(\hbar), \nn\\
    \bbA_{r\alpha}
    &= A_{r\mu} (X_{r})\, 
    \dow_\alpha X^\mu_{r}
    + \dow_\alpha \varphi_{r}
    + \cO(\hbar), \nn\\
    \bbA_{a\alpha}
    &= \Big( A_{a\mu} (X_{r}) + \lie_{X_a}A_{r\mu}(X_r) \Big)
    \dow_\alpha X^\mu_r
    + \dow_\alpha \varphi_{a} + \cO(\hbar),
\end{align}
up to quantum corrections. The Schwinger-Keldysh effective action of the theory can be expressed as $S[{\mathbb\Phi}_r,{\mathbb\Phi}_a;\bbbeta^\alpha]$, in terms of the global symmetry invariants $\mathbb\Phi_{r,a} = (\half\bbg_{r,a\alpha\beta},\bbA_{r,a\alpha},\half\bbkappa_{r,a\alpha\beta},\bbupsilon_{r,a\alpha})$ and the fixed thermal vector $\bbbeta^\alpha$, integrated over the fluid worldvolume coordinates $\sigma^\alpha$.

\paragraph*{Fluid worldvolume symmetries:} 

We also impose local diffeomorphisms $\sigma'^\alpha(\sigma)$ and U(1) gauge transformations $\lambda(\sigma)$ of the fluid worldvolume acting on the dynamical fields as
\begin{align}
    X^\mu_{1,2}(\sigma) 
    &\to X'^\mu_{1,2}(\sigma'(\sigma))
    = X^\mu_{1,2}(\sigma), \nn\\
    \phi_{1,2}(\sigma) 
    &\to \phi'_{1,2}(\sigma'(\sigma)) 
    = \phi_{1,2}(\sigma) + \lambda(\sigma).
\end{align}
The MIS fields $\bbkappa_{1,2\alpha\beta}$ and $\bbupsilon_{1,2\alpha}$ are taken to invariant under worldvolume U(1) gauge transformations and transform as appropriately ranked tensors under worldvolume diffeomorphisms.
These transformations are taken to be time-independent, satisfying
\begin{equation}
    \bbbeta^\beta\dow_\beta\sigma'^\alpha(\sigma) 
    = \bbbeta^\beta, \qquad 
    \bbbeta^\alpha\dow_\alpha\lambda(\sigma) = 0.
\end{equation}
One consequence of this is that all global symmetry invariants in \cref{eq:global-invariants-hydro} are invariant under worldvolume gauge transformations, except the spatial components of $\bbbeta^\alpha\bbA_{r\alpha}$, similar to the diffusion model. Worldvolume diffeomorphisms act on all global symmetry invariants as usual.

\paragraph*{Physical spacetime formulation:} The fluid worldvolume picture of the hydrodynamic effective field theory with two copies of spacetimes is theoretically neat and appealing. However, for practical purposes, it is more transparent to move to a single physical spacetime formulation, defined via $x^\mu = X_r^\mu(\sigma)$. We can use pullbacks with respect to this map to define objects that are invariant under the fluid worldvolume diffeomorphisms
\begin{align}
    G_{r\mu\nu} 
    &= \frac{\dow\sigma^\alpha}{\dow x^\mu}
    \frac{\dow\sigma^\beta}{\dow x^\nu}
    \bbg_{r\alpha\beta}
    = g_{r\mu\nu} + \cO(\hbar), \nn\\
    G_{a\mu\nu} 
    &= \frac{\dow\sigma^\alpha}{\dow x^\mu}
    \frac{\dow\sigma^\beta}{\dow x^\nu}
    \bbg_{a\alpha\beta}
    = g_{a\mu\nu} + \lie_{X_a} g_{r\mu\nu}
    + \cO(\hbar), \nn\\
    B_{r\mu}
    &= \frac{\dow\sigma^\alpha}{\dow x^\mu}
    \bbA_{r\alpha}
    = A_{r\mu} + \dow_\mu\varphi_r + \cO(\hbar), \nn\\
    B_{a\mu}
    &= \frac{\dow\sigma^\alpha}{\dow x^\mu}
    \bbA_{a\alpha}
    = A_{a\mu} + \dow_\mu\varphi_a
    + \lie_{X_a} A_{r\mu} + \cO(\hbar),
\end{align}
together with the MIS fields
\begin{equation}
    \kappa_{r,a\mu\nu} 
    = \frac{\dow\sigma^\alpha}{\dow x^\mu}
    \frac{\dow\sigma^\beta}{\dow x^\nu}
    \bbkappa_{r\alpha\beta}, \qquad
    \upsilon_{r,a\mu}
    = \frac{\dow\sigma^\alpha}{\dow x^\mu}
    \bbupsilon_{r,a\alpha}.
\end{equation}
Similarly, we can define the physical spacetime thermal vector by pushing forward the fluid worldvolume thermal vector
\begin{align}
    \beta^\mu(x) &= \bbbeta^\alpha \frac{\dow x^\mu}{\dow \sigma^\alpha(x)}
    = \frac{1}{T_0} \frac{\dow x^\mu}{\dow \sigma^0(x)},
    \label{eq:hydro-beta}
\end{align}
which can be used to define the fluid velocity and temperature as $\beta^\mu = u^\mu/T$. The fluid velocity is normalised as $u^\mu u^\nu G_{r\mu\nu} = -1$. The fluid worldvolume gauge transformations become the diagonal shift symmetry on the physical spacetime, acting on $B_{r\mu}$ as
\begin{equation}
    B_{r\mu} \to B_{r\mu} + \dow_\mu\lambda, \qquad 
    \beta^\mu \dow_\mu\lambda = 0.
\end{equation}
The temporal components of $B_{r\mu}$ can be used to define the chemical potential via
\begin{equation}
    \frac{\mu}{T} = \beta^\mu B_{r\mu} + \frac{\mu_0}{T_0},
\end{equation}
which are invariant under the diagonal shift symmetry.
The compromise for going to the physical spacetime formulation is that a diagonal part of the global physical spacetime symmetries becomes non-manifest
\begin{equation}
    x^\mu \to x'^\mu(x),
\end{equation}
which acts on various fields on the physical spacetime as expected according to their tensor structure. In the physical spacetime formulation, the Schwinger-Keldysh effective action can be expressed as $S[{\Phi}_r,{\Phi}_a;\beta^\mu]$, in terms of the global symmetry invariants $\Phi_{r,a} = (\half G_{r,a\mu\nu},B_{r,a\mu},\half\kappa_{r,a\mu\nu},\upsilon_{r,a\mu})$ and the thermal vector $\beta^\mu$. Note that, unlike the diffusion model, the thermal vector $\beta^\mu$ here is no longer a constant.

\paragraph*{Schwinger-Keldysh generating functional:}

The Schwinger-Keldysh generating functional for hydrodynamics can be defined using the effective action as
\begin{equation}
    {\cal Z}[\phi_r,\phi_a]
    = \int\cD\psi_r\,\cD\psi_a
    \exp(iS[\Phi_{r},\Phi_{a};\beta^\mu]),
\end{equation}
where the path integral is performed over the set of dynamical fields $\psi_{r,a} = (X_{r,a}^\mu,\varphi_{r,a},\bbkappa_{r,a\mu\nu},\bbupsilon_{r,a\mu})$ and depends on the background fields $\phi_{r,a}=(\half g_{r,a\mu\nu},A_{r,a\mu})$. The generating functional is required to satisfy the conditions in \cref{eq:SK-conditions-pf}, which translate in terms of the effective action as
\begin{equation}
    S[\Phi_r,\Phi_a = 0;\beta^\mu] = 0, \qquad 
    S[\Phi_r,-\Phi_a;\beta^\mu] 
    = -S^*[\Phi_r,\Phi_a;\beta^\mu], \qquad 
    \Im S[\Phi_r,\Phi_a;\beta^\mu] \geq 0.
\end{equation}

\paragraph*{Dynamical KMS symmetry:}

The Schwinger-Keldysh generating functional is required to satisfy the dynamical KMS symmetry given by its action on the background fields in \cref{eq:KMS-full}, or the one in \cref{eq:KMS-back} in the classical limit. The action of KMS symmetry on the dynamical fields is naturally defined on the the fluid worldvolume as
\begin{gather}
    X_1^\mu(\sigma) \to 
    \Theta X_1^\mu(\sigma), \qquad
    X_2^\mu(\sigma) \to 
    \Theta X_2^\mu(\sigma + i\hbar\,\Theta\bbbeta)
    - i\hbar\, \Theta\beta_0^\mu, \nn\\
  \varphi_1(\sigma) \to 
  \Theta\varphi_1(\sigma), \qquad
  \varphi_2(\sigma) \to 
  \Theta \varphi_2(\sigma + i\hbar\,\Theta\bbbeta), \nn\\
  \bbkappa_{1\alpha\beta}(\sigma) \to  
    \Theta\bbkappa_{1\alpha\beta}(\sigma), \qquad
    \bbkappa_{2\alpha\beta}(\sigma) \to 
    \Theta\bbkappa_{2\alpha\beta}(\sigma + i\hbar\,\Theta\bbbeta), \nn\\
    \bbupsilon_{1\alpha}(\sigma) \to  
    \Theta\bbupsilon J_{1\alpha}(\sigma), \qquad
    \bbupsilon_{2\alpha}(\sigma) \to 
    \Theta\bbupsilon_{2\alpha}(\sigma + i\hbar\,\Theta\bbbeta).
\end{gather}
The extra constant contribution in the transformation $X_2^\mu$ is taken so that KMS symmetry preserve the equilibrium configuration $X_{1,2}^\mu(\sigma) = \delta^\mu_\alpha \sigma^\alpha$, $\varphi_{1,2}(\sigma) = 0$. In the physical spacetime formulation in the classical limit, these give rise to
\begin{gather}
    \beta^\mu \to \Theta\beta^\mu, \qquad 
    X^\mu_a \to  \Theta\lb X^\mu_a
    + i(\beta^\mu - \beta^\mu_0) \rb, \nn\\
    \varphi_r \to \Theta\varphi_r, \qquad 
    \varphi_a \to \Theta
    \lb \varphi_a + i\lie_\beta\varphi_r \rb, \nn\\
    \kappa_{r\mu\nu} \to \Theta\kappa_{r\mu\nu}, \qquad 
    \kappa_{a\mu\nu} \to \Theta
    \lb \kappa_{a\mu\nu} + i\lie_\beta\kappa_{r\mu\nu} \rb, \nn\\
    \upsilon_{r\mu} \to \Theta\upsilon_{r\mu}, \qquad 
    \upsilon_{a\mu} \to \Theta\lb 
    \upsilon_{a\mu} + i\lie_\beta\upsilon_{r\mu} \rb.
    \label{eq:KMS-classical-hydro}
\end{gather}
These transformation properties induce the following dynamical KMS transformation on the building blocks of the effective action
\begin{equation}
    \Phi_r \to \Theta\Phi_r, \qquad 
    \Phi_a \to \Theta\lb\Phi_a
    + i\lie_\beta\Phi_r\rb, \qquad 
    \beta^\mu \to \Theta\beta^\mu,
    \label{eq:KMS-Phira_hydro}
\end{equation}
which is the appropriate generalisation of the transformations \eqref{eq:compound-diff} from the diffusion model.

\paragraph*{Hydrodynamic effective action:}

Truncating the theory to at most quadratic order in $\Phi_{a}$ fields and dropping any explicit spatial derivatives, the most general hydrodynamic effective action for MIS-hydrodynamics model is given as
\begin{align}
    S &= \int\df^{d+1}x\sqrt{-g_r}
    \bigg[
    \half G_{a\mu\nu} \lb \epsilon\, u^\mu u^\nu 
    - \cF\Delta^{\mu\nu} \rb
    + B_{a\mu} n u^\mu \nn\\
    &\hspace{5em}
    - \half\kappa_{a\mu\nu}\!
    \lb \frac{\chi^\ssfS_\kappa}{2}\Delta^{\mu\nu}\tr\kappa_r
    + \chi^\ssfT_\kappa \kappa_r^{\langle\mu\nu\rangle}
    \rb
    - \chi_\upsilon\upsilon_{a\mu}\upsilon^\mu_{r}
    \nn\\
    &\hspace{5em}
    + \half G_{a\mu\nu} \lb 
    \frac{\chi^\ssfS_\kappa}{2}  
    \kappa^{\mu\nu}_r\tr\kappa_r
    + \chi^\ssfT_\kappa
    \lb \kappa^{\mu\rho}_r\kappa^{\nu}_{r\rho}
    - \frac1d \kappa_r^{\mu\nu}\tr\kappa_r \rb
    + \chi_\upsilon\upsilon^\mu_r\upsilon^\nu_r \rb \nn\\
    &\hspace{5em}
    + \frac{iT}{4}\zeta\,\Delta^{\mu\nu} \Delta^{\rho\sigma}
    \lb G_{a\mu\nu} 
    + \frac{\chi_\kappa^\ssfS}{\alpha_\kappa^\ssfS}
    \kappa_{a\mu\nu} \rb\!\!
    \lb 
    G_{a\rho\sigma}
    + i\lie_\beta g_{r\rho\sigma}
    + \frac{\chi_\kappa^\ssfS}{\alpha_\kappa^\ssfS}
    \lb 
    \kappa_{a\rho\sigma}
    + i\lie_\beta\kappa_{r\rho\sigma} \rb
    \rb  \nn\\
    &\hspace{5em}   
    + \frac{iT}{2}\eta\,
    \Delta^{\rho\langle\mu}
    \Delta^{\nu\rangle\sigma}
    \lb G_{a\mu\nu} 
    + \frac{\chi_\kappa^\ssfT}{\alpha_\kappa^\ssfT}
    \kappa_{a\mu\nu} \rb\!\!
    \lb G_{a\rho\sigma}
    + i\lie_\beta g_{r\rho\sigma}
    + \frac{\chi_\kappa^\ssfT}{\alpha_\kappa^\ssfT}
    \lb \kappa_{a\rho\sigma}
    + i\lie_\beta\kappa_{r\rho\sigma} \rb
    \rb \nn\\
    &\hspace{5em}
    + i T\sigma\, \Delta^{\mu\nu} 
    \lb B_{a\mu}
    + \frac{\chi_\upsilon}{\alpha_\upsilon} 
    \upsilon_{a\mu} \rb\!\!
    \lb B_{a\nu} + i\delta_\scB A_\nu 
    + \frac{\chi_\upsilon}{\alpha_\upsilon}
    \lb \upsilon_{a\nu} + i\lie_\beta\upsilon_{r\nu} \rb \rb
    \bigg],
    \label{eq:action-hydro}
\end{align}
where we have already imposed the hydrodynamic frame conditions as given in \cref{eq:hydro-reparametrisation} for clarity. The action is manifestly invariant under the spacetime global symmetries and worldvolume gauge symmetries. It also trivially obeys the first two Schwinger-Keldysh conditions, while the third one requires
\begin{equation}
    \zeta \geq0, \qquad \eta \geq 0, \qquad 
    \sigma \geq 0.
\end{equation}
Using $\Theta = \rmT$ or PT, the terms in the first three lines map to themselves (up to a coordinate flip in the integral $x^\mu \to \Theta x^\mu$), with the residual terms that sum to a total derivative and drop out from the effective action, i.e.
\begin{align}
    &i\sqrt{-g_r}\Bigg[
    \half \lb \epsilon\, u^\mu u^\nu 
    - \cF\Delta^{\mu\nu} \rb \lie_\beta G_{r\mu\nu}
    + n u^\mu \lie_\beta B_{r\mu} 
    - \half \lb \frac{\chi^\ssfS_\kappa}{2}\Delta^{\mu\nu}\tr\kappa
    + \chi^\ssfT_\kappa \kappa^{\langle\mu\nu\rangle}
    \rb
    \lie_\beta\kappa_{\mu\nu}
    - \chi_\upsilon \upsilon^\mu
    \lie_\beta\upsilon_{\mu}
    \nn\\
    &\hspace{5em}
    + \half \lb 
    \frac{\chi^\ssfS_\kappa}{2}  
    \kappa^{\mu\nu}\tr\kappa
    + \chi^\ssfT_\kappa
    \lb \kappa^{\mu\rho}\kappa^{\nu}_\rho
    - \frac1d \kappa^{\mu\nu}\tr\kappa \rb
    + \chi_\upsilon\upsilon^\mu\upsilon^\nu \rb
    \lie_\beta G_{r\mu\nu}
    \Bigg] \nn\\
    &= i\sqrt{-g_r} \bigg(
    - \cF \nabla_\mu \beta^\mu
    + s \lie_\beta T
    + n \lie_\beta \mu
    - \frac{\chi_\kappa^\ssfS}{8} \lie_\beta (\tr\kappa)^2
    - \frac{\chi_\kappa^\ssfT}{4}
    \lie_\beta\!\lb \kappa^2 - \frac1d (\tr\kappa)^2 \rb
    - \frac{\chi_\upsilon}{2} \lie_\beta\upsilon^2 \Bigg), \nn\\
    &= -\dow_\mu\!\lb \sqrt{-g_r}\, i\beta^\mu\,\cF \rb.
    \label{eq:total-der-hydro}
\end{align}
We have identified $\kappa_{\mu\nu} \equiv \kappa_{r\mu\nu}$ and $\upsilon_{\mu}\equiv \upsilon_{r\mu}$ and used the thermodynamic relations \eqref{eq:thermo-hydro-Pi}. On account of \cref{eq:KMS-Phira_hydro}, the last three lines in the effective action are individually KMS-invariant. If $\Theta={\rm CPT}$ or CT, we will also need to require $n$ to be an odd function of $\mu$, while all other coefficients are even functions of $\mu$. 

This action is precisely the one we derived in \cref{eq:action-hydro-raw}, with the KMS conditions \eqref{eq:KMS-conditions} already implemented, as we can see by performing the following field redefinition
\begin{gather}
    \kappa_{\mu\nu} = \kappa_{r\mu\nu}, \qquad 
    \upsilon_\mu = \upsilon_{r\mu}, \nn\\
    W_{a\mu\nu}
    = \Delta_\mu^\rho\Delta_\nu^\sigma G_{a\rho\sigma}
    + \lb \frac1d \frac{\chi_\kappa^\ssfS}{\alpha_\kappa^\ssfS} \Delta_{\mu\nu}
    \Delta^{\rho\sigma}
    + \frac{\chi_\kappa^\ssfT}{\alpha_\kappa^\ssfT}
    \Delta_{\langle\mu}^\rho\Delta_{\nu\rangle}^\sigma
    \rb \kappa_{a\rho\sigma}, \qquad
    V_{a\mu}
    = \Delta_\mu^\nu B_{a\nu}
    + \frac{\chi_\upsilon}{\alpha_\upsilon}\upsilon_{a\mu}.
\end{gather}
While making the identification, it is also useful to note that $\dow_\mu(\mu/T) + \beta^\lambda F_{r\lambda\mu} = \lie_\beta B_{r\mu}$. This results in the standard prescription of KMS symmetry given in \cref{eq:KMS-hydro-standard}. To realise the alternate prescriptions in \cref{eq:KMS-hydro-alt}, we need to modify the SK-EFT by imposing a diagonal shift symmetries between the vector and tensor degrees of freedom respectively; see \cref{app:alt}.

Note that the dynamical KMS transformation of $\Phi_a = (\half G_{r,a\mu\nu},B_{r,a\mu},\half\kappa_{r,a\mu\nu},\upsilon_{r,a\mu})$ fields in \cref{eq:KMS-Phira_hydro} involves a Lie derivative $\lie_\beta$ along the thermal vector $\beta^\mu$, which forces the occurrences of $\lie_\beta$ in the dissipative terms in the last three lines of the effective action \eqref{eq:total-der-hydro}. In particular, if $\lie_\beta\kappa_{r\mu\nu}$ and $\lie_\beta\upsilon_{r\mu}$ were replaced with $\beta^\lambda\nabla_\lambda\kappa_{r\mu\nu}$ and $\beta^\lambda\nabla_\lambda\upsilon_{r\mu}$, these terms would no longer be non-linearly compatible with the dynamical KMS symmetry. This ties back to our comments towards the ends of \cref{sec:MIS-hydro,sec:MSR-hydro}, that the right choice of definitions of $\kappa_{\mu\nu}$ and $\upsilon_\mu$ in the classical hydrodynamic equations is crucial for compatibility of the EFT with the dynamical KMS symmetry and thus guarantee the agreement with all higher-point FDT requirements.

\subsection{Conformal hydrodynamics}
\label{sec:conformal}

The M\"uller-Israel-Stewart framework of relativistic hydrodynamics hugely simplifies in the conformal limit and was considered in the holographic model of~\cite{Baier:2007ix}. Conformal symmetry is defined as the invariance of the hydrodynamic equations under a conformal rescaling of background metric, i.e.
\begin{equation}
    g_{\mu\nu} \to \Omega^2 g_{\mu\nu}, \qquad 
    A_\mu \to A_\mu,
\end{equation}
for arbitrary function $\Omega(x)$. The hydrodynamic conservation equations \eqref{eq:hydro-conservation} are left invariant under this transformation, provided that the conserved currents scale as
\begin{equation}
    T^{\mu\nu} \to \Omega^{-d-3}\, T^{\mu\nu}, \qquad 
    J^\mu \to \Omega^{-d-1} J^\mu,
\end{equation}
and the energy-momentum tensor is traceless
\begin{equation}
    T^\mu_{~\mu} = 0.
\end{equation}
Conformal transformations are realised on the hydrodynamic fields as
\begin{gather}
    u^\mu \to \Omega^{-1}u^\mu, \qquad 
    T \to \Omega^{-1} T, \qquad 
    \mu \to \Omega^{-1}\mu, \qquad
    \kappa_{\mu\nu} \to \kappa_{\mu\nu}, \qquad 
    \upsilon_\mu \to \upsilon_\mu.
\end{gather}
Note that we are free to choose the conformal weight of the fields $\kappa_{\mu\nu}$ and $\upsilon_\mu$. However, had we rescaled these fields to fix $\alpha^\ssfS_\kappa$, $\alpha^\ssfT_\kappa$, and $\alpha_\upsilon$ introduced in \cref{eq:frame-choice-diff,eq:hydro-reparametrisation} to a particular value, the conformal weight of these fields would also be fixed by the conformal weights of the conserved currents. By scaling arguments, we can obtain the conformal weights of various coefficients in the hydrodynamic model
\begin{gather}
    \epsilon \to \Omega^{-d-1}\epsilon, \qquad 
    p \to \Omega^{-d-1}p, \qquad 
    n \to \Omega^{-d}n, \qquad 
    s \to \Omega^{-d}s, \nn\\
    \chi_\kappa^\ssfS \to \Omega^{-d+3}\chi_\kappa^\ssfS, \qquad 
    \chi_\kappa^\ssfT \to \Omega^{-d+3}\chi_\kappa^\ssfT, \qquad 
    \chi_\upsilon \to \Omega^{-d+1}\chi_\upsilon, \nn\\
    \alpha_\kappa^\ssfS \to \Omega^{-d+1}
    \alpha_\kappa^\ssfS, \qquad 
    \alpha_\kappa^\ssfT \to \Omega^{-d+1}\alpha_\kappa^\ssfT, \qquad 
    \alpha_\upsilon \to \Omega^{-d+1}\alpha_\upsilon, \nn\\
    \zeta \to \Omega^{-d}\zeta, \qquad 
    \eta \to \Omega^{-d}\eta, \qquad 
    \sigma \to \Omega^{-d+2}\sigma.
\end{gather}
Plugging these into the thermodynamic relation \eqref{eq:thermo-hydro-Pi}, we are led to the conformality constraints between thermodynamic coefficients
\begin{equation}
    \epsilon
    = d\, p, \qquad 
    d\,\cP + \frac{\chi_\kappa^\ssfS}{2} (\tr\kappa)^2
    + \chi_\kappa^\ssfT
    \lb \kappa^2 - \frac1d (\tr\kappa)^2 \rb
    + \chi_\upsilon \upsilon^2
    = 0.
\end{equation}
Furthermore, requiring the constitutive relations in \cref{eq:mis-hydro-consti} to transform homogeneously under non-constant conformal transformations, forces us to fix
\begin{equation}
    \zeta = 0 \qquad\implies\qquad \tr\kappa = 0.
\end{equation}
These conditions together ensure that $T^\mu_{~\mu} = 0$. The final constitutive relations are given as
\begin{align}
    T^{\mu\nu} 
    &= \epsilon \lb u^\mu u^\nu + \frac1d \Delta^{\mu\nu} \rb + \Pi^{\mu\nu}, \nn\\
    J^\mu 
    &= n\,u^\mu + \cJ^\mu,
\end{align}
together with the non-linear relations between $\Pi^{\mu\nu}$, $\cJ^\mu$ and $\kappa_{\mu\nu}$, $\upsilon_\mu$, i.e.
\begin{align}
    \Pi^{\mu\nu} 
    &= \alpha_\kappa^\ssfT \kappa^{\mu\nu}
    + \chi_\kappa^\ssfT
    \lb \kappa^{\mu\rho}\kappa^{\nu}_{~\rho}
    - \frac1d \Delta^{\mu\nu} \kappa^2
    \rb
    + \chi_\upsilon 
    \lb \upsilon^{\mu}\upsilon^{\nu}
    - \frac1d \Delta^{\mu\nu} \upsilon^2 \rb
    , \nn\\
    \cJ^\mu 
    &= \alpha_\upsilon \upsilon^\mu.
\end{align}
The dynamical equations for $\kappa_{\mu\nu}$ and $\upsilon_\mu$ take the form
\begin{align}
    \alpha_\kappa^\ssfT \kappa^{\mu\nu}
    &= 
    - 2\eta\,\Delta^{\rho\langle\mu}
    \Delta^{\nu\rangle\sigma}
    \lb \nabla_\rho u_\sigma
    + \frac{T\chi^\ssfT_\kappa}{2\alpha_\kappa^\ssfT}
    \lie_\beta \kappa_{\rho\sigma}
    \rb, \nn\\
    \alpha_\upsilon\upsilon^\mu
    &= -\sigma\,\Delta^{\mu\nu}
    \lb T \nabla_\nu\frac{\mu}{T}
    + u^\lambda F_{\lambda\nu} 
    + \frac{T\chi_\upsilon}{\alpha_\upsilon} 
    \lie_\beta\upsilon_\nu
    \rb.
\end{align}

We can also obtain the relaxation equations for the dissipative stress tensor and flux. To this end, let us take a simple ansatz for $\alpha_\kappa^\ssfT = c_\kappa^\ssfT \epsilon^{(d-1)/(d+1)}$ and $\alpha_\upsilon = c_\upsilon \epsilon^{(d-1)/(d+1)}$, for constants $c_\kappa^\ssfT$ and $c_\upsilon$. With this choice, we find
\begin{align}
    \Pi^{\mu\nu}
    &= 
    - 2\eta\,\Delta^{\rho\langle\mu}
    \Delta^{\nu\rangle\sigma} \nabla_\rho u_\sigma
    - \tau_\Pi^\ssfT\,\Delta^{\rho\langle\mu}
    \Delta^{\nu\rangle\sigma}
    \lb T\lie_\beta \Pi_{\rho\sigma}
    + \frac{d-1}{d} \Pi_{\rho\sigma}\nabla_\lambda u^\lambda
    \rb\nn\\
    &\qquad 
    + \frac{\tau_\Pi^\ssfT}{\eta}
    \lb \Pi^{\mu\rho}\Pi^{\nu}_{~\rho}
    - \frac1d \Delta^{\mu\nu} \Pi^2
    \rb
    + \frac{\tau_\cJ}{\sigma}
    \lb \cJ^{\mu}\cJ^{\nu}
    - \frac1d \Delta^{\mu\nu} \cJ^2 \rb
    + \cO(\dow^3), \nn\\
    \cJ^\mu
    &= -\sigma\,\Delta^{\mu\nu}
    \lb T \nabla_\nu\frac{\mu}{T}
    + u^\lambda F_{\lambda\nu} \rb
    - \tau_\cJ\,\Delta^{\mu\nu}
    \lb T\lie_\beta\cJ_\nu 
    + \frac{d-1}{d}\cJ_\nu \nabla_\lambda u^\lambda \rb 
    + \cO(\dow^3),
\end{align}
where we have ignored 3- and higher-derivative corrections. We have also used the leading order energy conservation equation $\lie_\beta \epsilon + (\epsilon+p)/T\,\nabla_\mu u^\mu = 0$. These equations can be directly compared with the ones derived in~\cite{Baier:2007ix}. However, unlike~\cite{Baier:2007ix}, our hydrodynamic equations are not designed to be exhaustive at second-order in derivatives. In particular, we have ignored any dependence of the constitutive relations on the fluid vorticity that is present in~\cite{Baier:2007ix}.

The effective action for MIS-hydrodynamics in the conformal limit simplifies to
\begin{align}
    S &= \int\df^{d+1}x\sqrt{-g_r}
    \bigg[
    \half G_{a\mu\nu} \lb u^\mu u^\nu 
    + \frac1d\Delta^{\mu\nu} \rb \epsilon
    + B_{a\mu} n u^\mu 
    - \frac{\chi^\ssfT_\kappa}{2}\kappa_{a\mu\nu} \kappa^{\mu\nu}_r
    - \chi_\upsilon\upsilon_{a\mu}\upsilon^\mu_{r}
    \nn\\
    &\hspace{5em}
    + \half G_{a\langle\mu\nu\rangle} \lb 
    \chi^\ssfT_\kappa
    \kappa^{\mu\rho}_r\kappa^{\nu}_{r\rho}
    + \chi_\upsilon\upsilon^\mu_r\upsilon^\nu_r \rb \nn\\
    &\hspace{5em}   
    + \frac{iT}{2}\eta\,
    \Delta^{\rho\langle\mu}
    \Delta^{\nu\rangle\sigma}
    \lb G_{a\mu\nu} 
    + \frac{\chi_\kappa^\ssfT}{\alpha_\kappa^\ssfT}
    \kappa_{a\mu\nu} \rb\!\!
    \lb G_{a\rho\sigma}
    + i\lie_\beta g_{r\rho\sigma}
    + \frac{\chi_\kappa^\ssfT}{\alpha_\kappa^\ssfT}
    \lb \kappa_{a\rho\sigma}
    + i\lie_\beta\kappa_{r\rho\sigma} \rb
    \rb \nn\\
    &\hspace{5em}
    + i T\sigma\, \Delta^{\mu\nu} 
    \lb B_{a\mu}
    + \frac{\chi_\upsilon}{\alpha_\upsilon} 
    \upsilon_{a\mu} \rb\!\!
    \lb B_{a\nu} + i\delta_\scB A_\nu 
    + \frac{\chi_\upsilon}{\alpha_\upsilon}
    \lb \upsilon_{a\nu} + i\lie_\beta\upsilon_{r\nu} \rb \rb
    \bigg],
\end{align}
which is better suited for potential holographic applications.

\section{Discussion}
\label{sec:discussion}

In this work, we have constructed UV-regularised stable and causal Schwinger-Keldysh effective field theories for relativistic diffusion and hydrodynamics. The SK-EFTs are appropriately coupled to double-copy background sources and non-linearly realise the dynamical KMS symmetry, and thus can be used to compute $n$-point time-ordered correlation functions, including perturbative stochastic loop corrections, consistent with FDT, stability, and causality requirements. 
We derived the respective SK-EFTs using both the MSR and SK formalisms, although the latter is more systematic and general and can potentially be used to include non-Gaussian stochastic interactions into the framework. 

We also find that it is possible to construct multiple models in the SK formalism that, up to certain field redefinitions, ultimately result in the same UV-regularised SK effective action; see \cref{app:alt}. As a consequence, the SK-EFTs discussed in this work simultaneously realise multiple dynamical KMS symmetries that act differently on the UV degrees of freedom. 
This is not really surprising because the UV sectors of these models are not constrained by the global symmetries in the hydrodynamic regime.

To construct these causal and stable SK-EFTs, we used hydrodynamic models inspired from the Maxwell-Cattaneo model of relativistic diffusion and the M\"uller-Israel-Stewart model of relativistic hydrodynamics, wherein one promotes the charge flux and spatial stress tensor to independent dynamical degrees of freedom with respective characteristic relaxation times. However, it is interesting to note that the classical evolution equations of the models we constructed, given in \cref{eq:calJ-consti,eq:mis-hydro-consti} respectively, are non-linearly different from the conventional treatments of MC-diffusion and MIS-hydrodynamics respectively. The precise form of the equations depends on the particular choice of variables for the relaxed degrees of freedom and we found that using charge flux or spatial stress tensor directly for this purpose does not generically non-linearly agree with the dynamical KMS symmetry. We are forced to work in the so-called ``thermodynamic frame'', where the choice of variables is enforced upon us by the grand canonical free energy density of the system under consideration. In retrospect, this is not surprising since SK-EFTs are designed to describe thermal fluctuations, while the very notion of relativistic temperature is not treated consistently in the traditional Landau frame; see e.g.~\cite{Kovtun:2022vas}. Thus, extensions of relativistic hydrodynamics based on Landau frame, such as \cite{Baier:2007ix}, do not lend themselves naturally to producing EFTs that respect FDT for $n$-point functions. As a future direction, it will be interesting to explore this connection between the choice of variables and the SK formalism.

Generally speaking, relaxation times are second-order transport coefficients in the hydrodynamic derivative expansion. In this work, our goal was to develop the general framework for extending relativistic SK-EFTs to include UV sectors that are consistent with stability, causality, and FDT requirements. To this end, we have analysed minimal models that fit the bill and have not attempted to be complete or rigorous at second order in derivatives, which would be an interesting avenue to pursue in the future. Care must be taken, however, because certain second order terms might end up worsening the causality and stability issues. A simple example of this is allowing an arbitrary $\sigma_\upsilon$ coefficient in the MC-diffusion model as given in \cref{eq:non-frame-upsilon-eq}, not fixed by the frame condition in \cref{eq:extra-coeff-frame}. This modifies the dispersion relations in the denominator of \cref{eq:retarded-J} by a term $+iD_\upsilon \omega k^2$, where the coefficient $D_\upsilon = \lb\sigma_\upsilon-\sigma\chi^2/\alpha_\upsilon^2\rb D_n/\chi_\upsilon$ is non-negative due to the constraints in \cref{eq:SK-ineq}. In a boosted frame of reference with small boost speed $v_0\ll 1$, this yields an unstable mode that goes as $\omega = i\tau_\cJ/(v_0^2 D_\upsilon) + \ldots$. To fix these issues, we will need to raise the powers of $\omega$ in the dispersion relations further by introducing higher-derivative terms in the effective theory and thus continuing the vicious cycle.

We should also emphasise that these SK-EFTs are not meant to describe physics near the relaxation timescales with any level of generality. They should only be viewed as introducing a consistent UV-regularisation into the hydrodynamic framework, while all physical predictions are still only universally valid at the longest spacetime scales within the hydrodynamic regime. If we wish to reliably describe physics near the relaxation timescales, we will need more information regarding the relaxed degrees of freedom and the relevant symmetries applicable at these scales, generically leading to qualitatively distinct results. For example, as we discussed in \cref{foot:sf-relax}, the relaxed vector modes in MC-diffusion model can come from momentum density relaxed due to impurities or superfluid velocity relaxed due to vortices. Near the relaxation timescales, the former realises an approximate translation symmetry while the latter realises an approximate higher-form symmetry; see e.g.~\cite{Armas:2021vku, Armas:2023tyx}, both of which will lead to drastically distinct physical signatures. Similarly, recalling \cref{foot:plastic}, the tensor modes in MIS-hydrodynamics might arise from a plastic crystal, which also realises an approximate higher-form symmetry near the relaxation timescales; see e.g.~\cite{Fukuma:2011pr, Armas:2022vpf}. Without the knowledge of the underlying small(er) scale physics, the predictability of these models is only limited to the universal hydrodynamic sector.

As we mentioned in the introduction, the recently proposed BDNK formalism also provides an algorithm to bypass stability and causality issues in relativistic hydrodynamics by invoking the freedom to arbitrarily redefine hydrodynamic fields. However, this formalism, as it stands, is not well-suited for being lifted to a SK-EFT and describing stochastic fluctuations. Relegating details to \cref{sec:bdnk}, one finds that tuning the transport coefficients in the BDNK formalism to ensure stability and causality at the classical level leads to potentially negative symmetric 2-point correlation functions and thus violations of unitarity. 
That being said, BDNK-hydrodynamics provides a certain advantage over MIS-hydrodynamics within the context of classical evolution: it is not straightforward to demonstrate causality and stability of the fully non-linear equations of MIS-hydrodynamics~\cite{Bemfica:2020xym}, while it is relatively simple to prove this for BDNK-hydrodynamics~\cite{Bemfica:2019knx, Hoult:2020eho}. It will be interesting to explore whether one can adjust the BDNK formalism so as to yield a consistent SK-EFT and yet preserve the inherent simplicity of the non-linear hydrodynamic equations. On the other hand, it will also be interesting to explore whether the SK-EFT framework for MIS-hydrodynamics developed in this work, and appropriate extensions to other MIS-inspired models in the literature~\cite{Geroch:1990bw, Denicol:2012cn, Gavassino:2023odx, Gavassino:2023qwl}, can help us analyse the stability and causality of these models at the full non-linear level.

The primary practical objective of SK-EFTs is to provide a systematic platform for studying statistical fluctuation corrections to classical hydrodynamic observables such as the retarded correlation functions of conserved densities and the corresponding fluxes. This entails using the linearised SK-EFT with interactions, like the one presented in \cref{sec:3pt-functions}, to compute stochastic loop corrections to various correlation functions order-by-order in the loop-expansion. This analysis for non-UV-regularised models appeared in~\cite{Chen-Lin:2018kfl, Jain:2020zhu, Jain:2020hcu} and more recently for a UV-regularised model of diffusion in~\cite{Abbasi:2022aao}. 
We comment on the model of~\cite{Abbasi:2022aao} in the second half of \cref{sec:bdnk}.
It will be interesting to undertake the computation of stochastic loop corrections using the KMS-compatible SK-EFTs developed in our work, which we leave for future explorations.

It will also be interesting to extend our results to more intricate dissipative relativistic hydrodynamic theories such as relativistic superfluid hydrodynamics~\cite{landau1959fluid, Bhattacharya:2011tra, Bhattacharyya:2012xi} or relativistic magnetohydrodynamics~\cite{Grozdanov:2016tdf, Hernandez:2017mch, Armas:2018atq, Armas:2018zbe, Glorioso:2018kcp, Armas:2022wvb}, which are all plagued by similar stability and causality issues. It might also be interesting to extend these ideas of UV-regularisation to non-relativistic SK-EFTs, such as the ones discussed in~\cite{Jain:2020vgc, Armas:2020mpr}. While an additional UV sector is not necessitated by any stability or causality requirements in non-relativistic contexts, the loop integrals in the respective SK-EFTs still need to be regulated with a KMS-compatible prescription and the results of this work might provide a viable and simple resolution to this end.

This work is closely related to the recent work of~\cite{Mullins:2023tjg}, where authors developed a general mechanism based on information current for introducing stochastic interactions into causal and stable models of linear relativistic hydrodynamics, like linear MC-diffusion and MIS-hydrodynamics. In a follow-up work that appears on arXiv on the same day as this paper, the authors have also utilised the information current perspective together with the MSR formalism to construct free EFTs for these linear hydrodynamic models consistent with 2-point FDT~\cite{Mullins:new}. In particular, the authors have identified a modified-KMS symmetry that is responsible for implementing 2-point FDT in their formalism. As we discussed in \cref{foot:pseudo-kms}, this is nothing but an alternate prescription of KMS transformations admissible in UV-regularised SK-EFTs. By contrast, the SK-EFTs constructed here correspond to the fully non-linear theories inspired from MC-diffusion and MIS-hydrodynamics, and include arbitrarily non-linear interactions consistent with all $n$-point FDT requirements. It will be interesting to further explore the connections with the work of~\cite{Mullins:2023tjg}, in particular with the information current identified by the authors and how it is realised in the fully non-linear EFT.

\acknowledgements

We would like to thank N.\,Abbasi, G.\,Denicol, M.\,Hippert, M.\,Kaminski, and J.\,Noronha for various useful discussions. We are also thankful to the authors of~\cite{Mullins:new} for sharing a draft of their paper prior to submission.
AJ is funded by the European Union’s
Horizon 2020 research and innovation programme under
the Marie Skłodowska-Curie grant agreement NonEqbSK
No. 101027527. AJ are partly supported by the Netherlands Organization
for Scientific Research (NWO) and by the Dutch Institute
for Emergent Phenomena (DIEP) cluster at the University
of Amsterdam. PK is supported in part by the NSERC of Canada.
This project was initiated during ``The Many Faces of Relativistic Fluid Dynamics'' program at KITP, Santa Barbara, supported by the National Science Foundation under Grant No. NSF PHY-1748958.

\appendix 

\section{Comparison with BDNK formalism}
\label{sec:bdnk}

In the introduction, we mentioned an alternate formalism for stabilising the gapped modes in a boosted frame of reference in relativistic hydrodynamic theories, without the need for introducing additional relaxed degrees of freedom, known as the Bemfica-Disconzi-Noronha-Kovtun (BDNK) formalism~\cite{Bemfica:2017wps, Kovtun:2019hdm, Bemfica:2019knx, Hoult:2020eho}. Let us look at the diffusion model for concreteness. The key idea is to untie the physical density $-J^\mu u_\mu^0$ from the thermodynamic density $n$ that appears in the thermodynamic relations \eqref{eq:firstlaw} and, just like the flux $\cJ^\mu$, allow for arbitrary constitutive relations for $J^\mu u^0_\mu$ as well. To this end, let us split the constitutive relations into the ideal and dissipative parts
\begin{equation}
    J^\mu = \lb n + \cN\rb u^\mu_0 + \cJ^\mu.
\end{equation}
The expression for entropy production modifies from \cref{eq:second-law} to 
\begin{equation}
    \dow_\mu S^\mu
    = - \frac{1}{T_0}\cN u_0^\mu\dow_\mu\mu 
    - \frac{1}{T_0} \cJ^\mu \Big( \dow_\mu\mu 
    + u_0^\rho F_{\rho\nu} \Big),
\end{equation}
where $S^\mu = (s - \mu/T_0\,\cN) u^\mu_0 -\mu/T_0\,\cJ^\mu$. This allows us to read out the constitutive relations
\begin{align}
    \cN &= \lambda\, u_0^\mu\dow_\mu\mu, \nn\\
    \cJ^\mu &= - \sigma\,\Delta^{\mu\nu}
    \lb \dow_\nu \mu + u_0^\rho F_{\rho\nu} \rb,
    \label{eq:bdnk-consti}
\end{align}
for some arbitrary dissipative transport coefficients $\lambda$ and $\sigma$. Looking at the expression for entropy production, i.e.
\begin{equation}
    \dow_\mu S^\mu
    = - \frac{1}{T_0\lambda}\cN^2
    + \frac{1}{T_0\sigma} \cJ_\mu\cJ^\mu,
    \label{eq:dS-bdnk}
\end{equation} we might be tempted to claim that $\sigma$ is non-negative while $\lambda$ is non-positive.
However, note that entropy production is only required to be positive onshell. Therefore, before imposing the positivity of entropy production, we should replace
\begin{equation}
    \cN
    = \frac{\lambda}{\chi} u_0^\mu \dow_\mu n 
    = -\frac{\lambda}{\chi} \lb u_0^\mu \dow_\mu\cN 
    + \dow_\mu \cJ^\mu \rb.
    \label{eq:higher-der-entropy}
\end{equation}
In particular, we see that $\cN$ is a 2-derivative onshell and thus the contribution from $\lambda$ to the entropy production is two orders suppressed relative to the contribution from $\sigma$. It was observed in~\cite{Bhattacharyya:2013lha,Bhattacharyya:2014bha} that the second law does not impose any constraints beyond the leading order in derivatives; we can always introduce even further derivative order corrections to the constitutive relations so that the the positivity of entropy production is ensured with only the leading order dissipative coefficients being non-negative. As a consequence, the positivity of entropy production production only imposes $\sigma\geq 0$, while $\lambda$ is left unconstrained.

We can compute the retarded correlation functions for this theory using the variational formulae in \cref{eq:retarded-var}. We find
\begin{align}
    G^R_{J^tJ^t}(\omega,k)
    &= \frac{-(1-i\omega\tau)\sigma k^2}
    {i\omega(1-i\omega\tau) - Dk^2}, \nn\\
    G^R_{J^i_\perp J^i_\perp}(\omega,k)
    &= i\omega\sigma\,k^{ij},
    \label{eq:retarded-bdnk}
\end{align}
where $\tau = \lambda/\chi$. Note that these correlation functions are considerably different from the ones obtained in the Maxwell-Cattaneo theory of diffusion in \cref{eq:retarded-J}. Nonetheless, they get the job done: in arbitrary boosted frame of reference, we find a diffusive and a gapped mode
\begin{equation}
    \omega = kv_0\cos\theta
    - i D k^2 \sqrt{1-v_0^2}
    \lb 1 - v_0^2\cos^2\theta \rb + \ldots, \qquad 
    \omega = -i\frac{\sqrt{1-v_0^2}}{\tau - v^2_0 D}
    + \ldots,
\end{equation}
and provided that we take $\tau\geq D \geq 0$, they are stable for any boost parameter. Given that $\chi\geq 0$, this essentially implies that stability requires $\lambda \geq \sigma \geq 0$. As discussed near \cref{eq:higher-der-entropy}, despite the positive $\lambda$, these constraints are not at odds with the positivity of entropy production.

The positivity of $\lambda$ does pose a different problem. We can revisit our discussion of stochastic fluctuations from \cref{sec:MSR-diff} for the BDNK model. We can introduce noise $\theta_\cN$ and $\theta^\mu_\cJ$ in the constitutive relations \eqref{eq:bdnk-consti} and set-up a path integral over noise configurations similar to \cref{eq:bdnk-consti}, but with the new weight factor
\begin{equation}
    \exp(-\frac14 \int \df^{d+1}x \lB
     \left\langle\frac{-1}{T_0\tilde\lambda}
     \right\rangle_{\!\!\theta}
    \theta_\cN^2
    + \left\langle
    \frac{\Delta_{\mu\nu}}{T_0\tilde\sigma}
    \right\rangle_{\!\!\theta}
    \theta^\mu_\cJ \theta^\nu_\cJ \rB).
    \label{eq:bdnk-weight}
\end{equation}
Following through the procedure in \cref{sec:MSR-diff}, we can obtain the symmetric 2-point Green's functions
\begin{align}
    G^S_{J^tJ^t}(\omega,k)
    &= 2T_0\frac{
    (1+\omega^2\tau^2)\tilde\sigma k^2
    - \tilde\lambda D^2 k^4}
    {|i\omega(1-i\omega\tau) - D k^2|^2}, \nn\\
    G^S_{J^i_\perp J^i_\perp}(\omega,k)
    &= 2T_0\tilde\sigma\,k^{ij}.
    \label{eq:symmetric-bdnk}
\end{align}
Requiring the symmetric and retarded Green's functions to satisfy the fluctuation-dissipation theorem will force us to take 
\begin{equation}
    \tilde\sigma = \sigma, \qquad 
    \tilde\lambda = \lambda,
\end{equation}
meaning that $\tilde\lambda$ is also positive. This, however, means that the $\theta_\cN^2$ term in the noise weight factor \eqref{eq:bdnk-weight} has the wrong sign: configurations with larger stochastic noise contribute more to the classical expectation values of operators. Also, the symmetric correlation function $G^S_{J^tJ^t}(\omega,k)$ is nothing but the variance of a random stochastic variable $J^t$ and thus is not allowed to be negative for any $\omega,k_i$. Ignoring this problem for the moment, we will be led to the effective action for BDNK-diffusion
\begin{equation}
    S
    = \int \df^{d+1}x
    \bigg[
    B_{a\mu} n u^\mu_0
    + iT_0\lb -\lambda\, u^\mu_0 u^\nu_0 
    + \sigma\,\Delta^{\mu\nu}
    \rb B_{a\mu}
    \lb B_{a\nu} 
    + i\beta_0^\lambda \dow_\lambda B_{r\nu} \rb
    \bigg].~~
    \label{eq:action-SK-bdnk}   
\end{equation}
In fact, we already derived this action using the Schwinger-Keldysh formalism in \cref{app:SK-MIS}. It follows from the general Schwinger-Keldysh effective action in \cref{eq:action-SK-raw} upon dropping all dependence on the gapped Maxwell-Cattaneo fields and skipping imposing any hydrodynamic frame requirement. However, the third SK-condition in \cref{eq:SK-conditions} is violated for positive $\lambda$, causing higher-frequency configurations of the noise field $\varphi_a$ to contribute arbitrarily more to the path integral. In fact, in the MSR formalism, we needed to perform a Gaussian integral over noise fields $\theta_\cN$, $\theta^\mu_\cJ$ to derive this effective action, similar to our discussion preceding \cref{eq:action-MIS}, which will be ill-defined with the wrong sign of $\theta_\cN^2$ term. These problems can be avoided by taking $\lambda$ to be negative, however this will cause the mode spectrum to be unstable and compromise the entire purpose of introducing gapped modes in the first place. 

It is worth pointing out that the wrong sign of $\lambda$ in the path integral smells similar to the apparent reduction of entropy in \cref{eq:dS-bdnk} for positive $\lambda$. However, there, we could brush the problem under the rug by noting that we can always find some higher-derivative corrections so that entropy production remains positive semi-definite while leaving $\lambda$ unconstrained; see the discussion in~\cite{Bhattacharyya:2013lha, Bhattacharyya:2014bha}. It is, in principle, possible that a higher-derivative generalisation of BDNK-diffusion will circumvent the problems with the associated Schwinger-Keldysh path integral as well. However, in its current form, BDNK-diffusion model cannot be used to reliably compute  UV-regulated Green's functions and stochastic fluctuations. 

The arguments presented above also qualitatively apply to the full theory of relativistic BDNK-hydrodynamics~\cite{Bemfica:2017wps, Kovtun:2019hdm, Bemfica:2019knx, Hoult:2020eho}. We start by untying the physical energy current $-T^{\mu\nu}u_\nu$ and charge density $-J^\mu u_\mu$ from the respective thermodynamic quantities $\epsilon u^\mu$ and $n$, allowing for arbitrary derivative corrections $\cE u^\mu + \cQ^\mu$ and $\cN$ relating the two. These corrections generically admit $\lambda$-like transport coefficients described above, which are required to be positive by the stability and causality and negative by the consistency of the EFT formalism, thus leading to a contradiction.

\paragraph*{From MC-diffusion to BDNK-diffusion:}

Despite our comments above, there is still something to be understood. Note that, taking $\tau$ and $\alpha_\upsilon$ to be constant, we can take a gradient of the constitutive relations of MC-diffusion in \cref{eq:consti-cJ-relax} and precisely land on the equations of BDNK-diffusion
\begin{equation}
    \dow_\mu\bigg( 
    \!\lb n
    + \lambda\,u_0^\nu\dow_\nu\mu \rb u^\mu_0
    - \sigma\, \Delta^{\mu\nu}\Big(
    \dow_\nu\mu - F_{\nu\rho}u^\rho_0
    \Big)
    \bigg)
    = 0,
    \label{eq:bdnk-like-diff}
\end{equation}
where we have identified $\lambda = \chi\tau$.
So, at least for constant $\tau$ and $\alpha_\upsilon$, the classical equations of motion of the two theories are precisely the same. The difference between the two is really the coupling to background fields: e.g. while $A_\mu u^\mu_0$ couples to $n$ in MC-diffusion, it couples to $n-\lambda\dow_t\mu$ in BDNK-diffusion. Given so, it should be possible to derive a SK-EFT for BDNK-diffusion for constant $\tau$ by using its MC counterpart. To this end, let us start with the effective action for MC-diffusion in the form \eqref{eq:action-MIS} and assume that all coefficients appearing therein are just functions of $n$ and not $\upsilon^2$. If so, $\upsilon_\mu$ becomes a Lagrange multiplier to set
\begin{equation}
    V_{a\mu} 
    - \frac{1}{\alpha_\upsilon}
    u_0^\lambda \dow_\lambda\!\lb \alpha_\upsilon\tau V_{a\mu} \rb
    = B_{a\mu}.
\end{equation}
Provided that we take $\tau$ and $\alpha_\upsilon$ to be constants, this can be explicitly solved in terms of 
non-local relation $V_{a\mu} = (1-\tau u_0^\lambda\dow_\lambda)^{-1}B_{a\mu}$. Plugging this back in and performing a field redefinition
\begin{equation}
    \varphi_a \to \hat\varphi_a - \tau
    u_0^\mu \dow_\mu \hat\varphi_a,
\end{equation}
we are led to the effective action
\begin{align}
    S
    &= 
    \int \df^{d+1}x
    \bigg[ \hat B_{a\mu} n u^\mu_0 
    + T_0 \lambda u^\mu_0 u_0^\nu 
    \,\hat B_{a\mu}\, 
    \beta_0^\lambda \dow_\lambda B_{r\nu}
    + iT_0\sigma\,\Delta^{\mu\nu} \hat B_{a\mu} \lb 
    \hat B_{a\nu}
    + i\beta_0^\lambda\dow_\lambda B_{r\nu}
    \rb
    \bigg],
    \label{eq:action-bdnk-new}
\end{align}
where we have defined
\begin{equation}
    \hat B_{a\mu} 
    = \dow_\mu\hat\varphi_a 
    + \frac{1}{(1-\tau u_0^\lambda\dow_\lambda)} A_{a\mu}.
\end{equation}
It should also be noted that taking $\mu(n)$ to not depend on $\upsilon^2$, the thermodynamic relation \eqref{eq:thermo-J} forces us to also take $\chi_\upsilon = \alpha_\upsilon^2\tau/\sigma$ to be constant; see our previous discussion around \cref{eq:MIS-diff-simple}. Together with the requirement that $\tau$ and $\alpha_\upsilon$ are constants, this forces us to take the conductivity $\sigma$ to also be a constant as well. Even though the two effective actions \eqref{eq:action-SK-bdnk} and \eqref{eq:action-bdnk-new} lead to the same classical equations of motion, modulo the constraints on coefficients outlined above, they are quite different from each other. Firstly, the $\lambda$ term does not appear in the imaginary part of the action at all, so the whole problem concerning the wrong sign of the path integral is avoided. Secondly, the coupling to background fields is different, so the correlation functions we obtain from this effective action are those given in \cref{eq:retarded-J,eq:symmetric-J} and not the ones from \cref{eq:retarded-bdnk,eq:symmetric-bdnk}. The downside, however, is that there is no simple realisation of the dynamical KMS symmetry in the new effective action \eqref{eq:action-bdnk-new}, even though it was derived from a KMS-invariant effective action \eqref{eq:action-MIS}. 
We refer to this new EFT as BDNK$^*$-diffusion to distinguish it from the original (and inconsistent) EFT of BDNK-diffusion in \eqref{eq:action-SK-bdnk}.
The effective action \eqref{eq:action-bdnk-new} of BDNK$^*$-diffusion becomes simpler if we choose the local rest frame $u_0^\mu = \delta^\mu_t$ and only turn on the time-components of the background fields $A_{r,at}$. We get
\begin{align}
    S
    &= 
    \int \df^{d+1}x
    \bigg[ \dow_t\hat\varphi_a\! 
    \lb n + \tau \dow_t n \rb 
    - D\,\dow_i\hat\varphi_a \dow^i n
    + iT_0\sigma\,\dow_i\hat\varphi_a
    \dow^i\hat\varphi_a
    + A_{at} n
    + \sigma\,\dow_i\hat\varphi_a \dow^i A_{rt}
    \bigg],
    \label{eq:action-bdnk-new-noback}
\end{align}
where $\sigma$ and $\tau$ are constants, while $D$ can be an arbitrary function of $n$.

The EFT for BDNK$^*$-diffusion in \cref{eq:action-bdnk-new-noback} was recently analysed in the work of~\cite{Abbasi:2022aao}, however the authors allowed the conductivity $\sigma$ to not be constant and depend arbitrarily on $n$. As we discussed above, $\sigma$ and $\tau$ need to be constant for this effective action to be derived from the KMS-invariant effective action in \cref{eq:action-MIS}. Without this constraint, the tree-level 3- and higher-point functions derived using the EFT in \cref{eq:action-bdnk-new-noback} will not satisfy FDTs; we look at these in detail in \cref{sec:3pt-functions}. This further implies that stochastic loop corrections to 2-point correlation functions computed using \cref{eq:action-bdnk-new-noback} with non-constant $\sigma$ and $\tau$ will also violate the FDT in \cref{eq:fdt}, however it would be interesting to verify this explicitly. In the context of~\cite{Abbasi:2022aao}, the coefficients $\lambda_\sigma$ and $\lambda'_\sigma$ in their work must be switched off for the results to be compatible with FDTs at full non-linear level. We note that the authors in~\cite{Abbasi:2022aao} only computed the loop-corrected symmetric 2-point correlation function of density using the EFT in \cref{eq:action-bdnk-new-noback} and assumed the 2-point FDT to define the associated retarded correlation function. Hence, the constraints arising from non-linear FDTs was invisible in their computation. If we wish to probe the effects of non-constant $\sigma$ and $\tau$ in the EFT, we will need to invoke the effective action for Maxwell-Cattaneo model of diffusion given in \cref{eq:action-MIS} with relaxed vector degrees of freedom.

Our comments above are not intended as a criticism of the analysis of~\cite{Abbasi:2022aao}, but rather as a cautionary note when using the MSR formalism to study statistical interactions in stochastic thermal systems. Ordinarily in the MSR formalism, one only imposes the 2-point FDT at tree-level to fix the noise term in the action. In ordinary theories of diffusion and hydrodynamics (without UV-regularisation), this luckily also ensures conformity with higher-point FDTs at tree-level and one can faithfully use the resulting EFT to compute statistical loop corrections. However, this does not work as neatly with the EFT employed in~\cite{Abbasi:2022aao}. Although this EFT respects tree-level 2-point FDT, one finds that higher-point FDTs are violated by the correlation functions. Note that, if we had not used the correct thermodynamic frame for MIS-hydrodynamics in \cref{sec:full-hydro}, a similar problem would have occurred for us while constructing the respective EFT using the MSR formalism; see the comments at the end of \cref{sec:MSR-hydro}.

This construction has no obvious analogue for the full BDNK-hydrodynamics, even for a particular choice of transport coefficients. Because of the spacetime-dependent velocity field $u^\mu$, the classical dynamical equations of MIS-hydrodynamics in \cref{eq:mis-hydro-consti} cannot be recast into a BDNK-like format, as was possible for the simple diffusion model in \cref{eq:bdnk-like-diff}. Therefore, the two theories are quite different even at the classical level and there is no simple way of deriving a consistent effective action for BDNK-hydrodynamics, similar to the one in \cref{eq:action-bdnk-new}, utilising its MIS counterpart.

\section{Linearised effective field theory and correlation functions}
\label{sec:3pt-functions}

Given the Schwinger-Keldysh generating functional $\cZ[\phi_r,\phi_a]$, we can compute various time-ordered $n$-point correlation functions as
\begin{equation}
    G_{r\ldots a\ldots}
    = \lb \frac{-i\delta}{\delta \phi_a}\ldots \rb
    \lb \frac{\delta}{\delta \phi_r}\ldots \rb
    \ln\cZ[\phi_r,\phi_a].
\end{equation}
For example, $G^R = G_{ra}$ and $G^S = G_{rr}$ are the retarded and symmetric 2-point correlation functions, while $G_{raa}$, $G_{rra}$, and $G_{rrr}$ are the retarded, partially-retarded, and completely symmetric 3-point correlation functions. When computed in a thermal state with global constant temperature $T_0$ and local rest frame velocity $u^\mu_0 = \delta^\mu_0$, the Fourier-space 2- and 3-point functions are required to satisfy the fluctuation-dissipation theorems
\begin{subequations}
\begin{align}
    G_{rr}(p_1,p_2)
    &= \frac{T_0}{i\omega}
    \Big( G_{ra}(p_1;p_2) - G_{ra}^*(p_1;p_2) \Big), \\
    G_{rra}(p_1,p_2;p_3)
    &= -\frac{T_0}{i\omega_2} \Big( 
    G_{raa}(p_1;p_2,p_3)
    - G^*_{raa}(p_3;p_2,p_1)
    \Big) 
    + (1\leftrightarrow 2), \\
    G_{rrr}(p_1,p_2,p_3)
    &= -\frac{T_0^2}{\omega_2\omega_3}
    \Big( G_{raa}(p_1;p_2,p_3)
    + G^*_{raa}(p_1;p_2,p_3) \Big)
    + (1\leftrightarrow 2)
    + (1\leftrightarrow 3),
\end{align}
\label{eq:fdt-3pt}%
\end{subequations}
where $\sum_{n}p^\mu_n = 0$, and so on for higher-point functions~\cite{Wang:1998wg}.\footnote{The convention for correlation functions in~\cite{Wang:1998wg} is related to ours as $G^{\text{WH}}_{r\ldots a\ldots} = i/2(-1)^{n_a}(-2i)^{n_r}G_{r\ldots a\ldots}$, where $n_{r}$ and $n_a$ are numbers of ``r'' and ``a'' type fields in the correlation function.} These theorems are guaranteed by dynamical KMS symmetry of the Schwinger-Keldysh effective field theory.

\paragraph*{BDNK$^*$-diffusion:}

As an illustration, let us consider the effective action for BDNK$^*$-diffusion given in \cref{eq:action-bdnk-new-noback}. Let us also allow the relaxation time $\tau$ and conductivity $\sigma$ in this action to depend on $n$, to see how this causes violations of higher-point FDTs. The discussion here is a straight-forward generalisation of the work in~\cite{Jain:2020zhu, Jain:2020hcu}. Denoting $n$ by solid lines and $\hat\varphi_a$ by wavy lines, we find the nonzero bare propagators
\begin{equation}
    \raisebox{-13pt}{
        \tikz[thick]{
        \draw [right] (-16mm,0)--(-8mm,0);
        \draw [aux] (-8mm,0)--(0,0);
        \node at (-8mm,-3mm) {$p$};
    }} \quad= \frac{1}{F(p)}, \qquad 
    \raisebox{-13pt}{
        \tikz[thick]{
        \draw [right] (-16mm,0)--(-8mm,0);
        \draw [] (-8mm,0)--(0,0);
        \node at (-8mm,-3mm) {$p$};
    }} \quad = \frac{2T_0\sigma k^2}{|F(p)|^2},
    \label{eq:propagators-bdnk}
\end{equation}
where $F(p) = \omega (1 - i\tau\omega) + iD_n k^2$. There are also two 3-point interaction vertices
\begin{equation}
    \raisebox{-20pt}{
        \tikz[thick]{
        \node at (-3mm,3mm) {$\lambda,\lambda_\tau$};
        \draw [right] (-8.1mm,0)--(-8mm,0);
        \draw [aux] (-8mm,0)--(0,0);
        \node at (-8mm,-3mm) {$p_1$};
        \draw (0mm,0)--(6mm,6mm);
        \draw [left] (6mm,6mm)--(6.1mm,6.1mm);
        \node at (8mm,4mm) {$p_2$};
        \draw [] (0mm,0)--(6mm,-6mm);
        \draw [left] (6mm,-6mm)--(6.1mm,-6.1mm);
        \node at (8mm,-4mm) {$p_3$};
    }}
    \quad = -\frac{i}{2} 
    \lb \lambda k_1^2 - \lambda_\tau \omega_1^2 \rb, \qquad
    \raisebox{-20pt}{
        \tikz[thick]{
        \node at (-1mm,3mm) {$\tilde\lambda$};
        \draw [right] (-8.1mm,0)--(-8mm,0);
        \draw [] (-8mm,0)--(0,0);
        \node at (-8mm,-3mm) {$p_1$};
        \draw [aux] (0mm,0)--(6mm,6mm);
        \draw [left] (6mm,6mm)--(6.1mm,6.1mm);
        \node at (8mm,4mm) {$p_2$};
        \draw [aux] (0mm,0)--(6mm,-6mm);
        \draw [left] (6mm,-6mm)--(6.1mm,-6.1mm);
        \node at (8mm,-4mm) {$p_3$};
    }}
    \quad = T_0\chi\tilde\lambda (k_2\cdot k_3),
\end{equation}
where $\lambda = \dow D/\dow n$, $\tilde\lambda = \chi^{-1}\dow\sigma/\dow n$, and $\lambda_\tau = \dow \tau/\dow n$. We can similarly work out higher-point interaction vertices as needed. Finally, since the background field $A_{rt}$ couples non-trivially to the dynamical fields $n$ and $\varphi_a$, we have background coupling vertices
\begin{equation}
    \raisebox{-4pt}{
        \tikz[thick]{
        \draw [->>] (-8.1mm,0)--(-8mm,0);
        \draw [right,aux] (-8mm,0)--(0,0);
        \node at (-12.5mm,0mm) {$p$};
    }} \quad = i\sigma k^2, \qquad 
    \raisebox{-20pt}{
        \tikz[thick]{
        \node at (-1mm,4mm) {$\tilde\lambda$};
        \draw [->>] (-0.1mm,0)--(-0mm,0);
        \node at (-4.5mm,0mm) {$p_1$};
        \draw [aux] (0mm,0)--(6mm,6mm);
        \draw [left] (6mm,6mm)--(6.1mm,6.1mm);
        \node at (8mm,4mm) {$p_2$};
        \draw [] (0mm,0)--(6mm,-6mm);
        \draw [left] (6mm,-6mm)--(6.1mm,-6.1mm);
        \node at (8mm,-4mm) {$p_3$};
    }} \quad = -i\chi\tilde\lambda (k_1\cdot k_2),
\end{equation}
where we have denoted an $A_{at}$ insertion with double arrow. We have not included a diagrammatic notation for $A_{at}$ because it couples trivially to $n$.\footnote{The choice of variables here is slightly different from our previous work~\cite{Jain:2020zhu, Jain:2020hcu}: here we are working with fluctuations in density $n(\mu) - n_0$, whereas in~\cite{Jain:2020zhu, Jain:2020hcu} we worked with $\delta n = n(\mu - A_{rt}) - n_0$, resulting in different background coupling vertices. One consequence of this being that there are non-trivial interactions vertices among $A_{at}$ and $A_{rt}$ in~\cite{Jain:2020zhu, Jain:2020hcu}. Since this is just a choice of variables, the final answers for the correlation functions remain unaffected.}
Blending these ingredients together, the two-point symmetric and retarded functions at tree-level are given as
\begin{equation}
    G_{ra}(p) =\quad
    \raisebox{-13pt}{
        \tikz[thick]{
        \draw [right] (-16mm,0)--(-8mm,0);
        \draw [aux] (-8mm,0)--(0,0);
        \draw [->>] (0mm,0)--(2mm,0);
        \node at (-8mm,-3mm) {$p$};
    }} \quad= \frac{i\sigma k^2}{F(p)}, \qquad 
    G_{rr}(p) = \quad
    \raisebox{-13pt}{
        \tikz[thick]{
        \draw [right] (-16mm,0)--(-8mm,0);
        \draw [] (-8mm,0)--(0,0);
        \node at (-8mm,-3mm) {$p$};
    }} \quad = \frac{2T_0\sigma k^2}{|F(p)|^2},
    \label{eq:2pt-bdnk}
\end{equation}
It is easy to check that they satisfy the fluctuation-dissipation theorem in \cref{eq:fdt-3pt}.
We can also compute the tree-level 3-point correlation functions: retarded
\begin{subequations}
\begin{align}
    G_{raa}(p_1,p_2,p_3) 
    &= 
    \raisebox{-23pt}{
        \tikz[thick]{
        \node at (-3mm,3mm) {$\lambda,\lambda_\tau$};
        \draw [right] (-10mm,0)--(-5mm,0);
        \draw [aux] (-5mm,0)--(0,0);
        \node at (-5mm,-3mm) {$p_1$};
        \draw [] (0mm,0)--(3mm,3mm);
        \draw [left] (2.7mm,2.7mm)--(3mm,3mm);
        \draw [aux] (3mm,3mm)--(6.5mm,6.5mm);
        \draw [-<<] (6.5mm,6.5mm)--(8mm,8mm);
        \node at (8mm,3mm) {$p_2$};
        \draw [] (0mm,0)--(3mm,-3mm);
        \draw [left] (2.7mm,-2.7mm)--(3mm,-3mm);
        \draw [aux] (3mm,-3mm)--(6.5mm,-6.5mm);
        \draw [-<<] (6.5mm,-6.5mm)--(8mm,-8mm);
        \node at (8mm,-3mm) {$p_3$};
    }}
    + \lb \raisebox{-13.5pt}{
        \tikz[thick]{
        \node at (-1mm,3mm) {$\tilde\lambda$};
        \draw [right] (-10mm,0)--(-5mm,0);
        \draw [aux] (-5mm,0)--(0,0);
        \node at (-5mm,-3mm) {$p_1$};
        \draw [] (0mm,0)--(3mm,3mm);
        \draw [left] (2.7mm,2.7mm)--(3mm,3mm);
        \draw [aux] (3mm,3mm)--(6.5mm,6.5mm);
        \draw [-<<] (6.5mm,6.5mm)--(8mm,8mm);
        \node at (8mm,3mm) {$p_2$};
        \draw [-<<] (0mm,0mm)--(2mm,-2mm);
        \node at (4mm,-4mm) {$p_3$};
    }} + (2\leftrightarrow 3) \rb
    \nn\\
    &=
    \frac{i\sigma^2k_2^2k_3^2
    \lb\lambda k_1^2 - \lambda_\tau \omega_1^2\rb}{F(p_1)F(p_2)^*F(p_3)^*}
    - \chi\sigma\tilde\lambda \lb \frac{ k_2^2 (k_1\cdot k_3)}{F(p_1)F(p_2)^*}
    + (2\leftrightarrow 3) \rb,
\end{align}
partially-retarded
\begin{align}
    G_{rra}(p_1,p_2,p_3)
    &= 
    \lb \raisebox{-27.5pt}{
        \tikz[thick]{
        \node at (-1mm,3mm) {$\lambda$};
        \draw [right] (-10mm,0)--(-5mm,0);
        \draw [] (-5mm,0)--(0,0);
        \node at (-5mm,-3mm) {$p_1$};
        \draw [aux] (0mm,0)--(4mm,4mm);
        \draw [left] (4mm,4mm)--(8mm,8mm);
        \node at (8mm,3mm) {$p_2$};
        \draw [] (0mm,0)--(3mm,-3mm);
        \draw [left] (2.7mm,-2.7mm)--(3mm,-3mm);
        \draw [aux] (3mm,-3mm)--(6.5mm,-6.5mm);
        \draw [-<<] (6.5mm,-6.5mm)--(8mm,-8mm);
        \node at (8mm,-3mm) {$p_3$};
    }} + (1\leftrightarrow 2) \rb
    + \raisebox{-27.5pt}{
        \tikz[thick]{
        \node at (-1mm,3mm) {$\tilde\lambda$};
        \draw [right] (-10mm,0)--(-5mm,0);
        \draw [aux] (-5mm,0)--(0,0);
        \node at (-5mm,-3mm) {$p_1$};
        \draw [aux] (0mm,0)--(4mm,4mm);
        \draw [left] (4mm,4mm)--(8mm,8mm);
        \node at (8mm,3mm) {$p_2$};
        \draw [] (0mm,0)--(3mm,-3mm);
        \draw [left] (2.7mm,-2.7mm)--(3mm,-3mm);
        \draw [aux] (3mm,-3mm)--(6.5mm,-6.5mm);
        \draw [-<<] (6.5mm,-6.5mm)--(8mm,-8mm);
        \node at (8mm,-3mm) {$p_3$};
    }}
    + \lb \raisebox{-13.5pt}{
        \tikz[thick]{
        \node at (-1mm,3mm) {$\tilde\lambda$};
        \draw [right] (-10mm,0)--(-5mm,0);
        \draw [aux] (-5mm,0)--(0,0);
        \node at (-5mm,-3mm) {$p_1$};
        \draw (0mm,0)--(4mm,4mm);
        \draw [left] (4mm,4mm)--(8mm,8mm);
        \node at (8mm,3mm) {$p_2$};
        \draw [-<<] (0mm,0mm)--(2mm,-2mm);
        \node at (4mm,-4mm) {$p_3$};
    }} + (1\leftrightarrow 2) \rb
    \nn\\
    &= - 2T_0\sigma^2 \lb \frac{k_1^2 k_3^2
    \lb\lambda k_2^2 - \lambda_\tau \omega_2^2 \rb}
    {|F(p_1)|^2F(p_2)F(p_3)^*}
    + (1\leftrightarrow 2) \rb \nn\\
    &\qquad 
    - 2iT_0\chi\sigma\tilde\lambda
    \lb \frac{k_3^2 (k_1\cdot k_2)}{F(p_1)F(p_2)F(p_3)^*}
    +  \lb \frac{k_2^2(k_1\cdot k_3)}{F(p_1)|F(p_2)|^2}
    + (1\leftrightarrow 2)
    \rb\rb,
\end{align}
and symmetric
\begin{align}
    G_{rrr}(p_1,p_2,p_3)
    &=
    \lb \raisebox{-20pt}{
        \tikz[thick]{
        \node at (-1mm,3mm) {$\lambda$};
        \draw [right] (-10mm,0)--(-5mm,0);
        \draw [] (-5mm,0)--(0,0);
        \node at (-5mm,-3mm) {$p_1$};
        \draw [] (0mm,0)--(4mm,4mm);
        \draw [left] (4mm,4mm)--(8mm,8mm);
        \node at (8mm,3mm) {$p_2$};
        \draw [aux] (0mm,0)--(4mm,-4mm);
        \draw [left] (4mm,-4mm)--(8mm,-8mm);
        \node at (8mm,-3mm) {$p_3$};
    }} + (1\leftrightarrow 3)
    + (2\leftrightarrow 3) \rb
    + \lb \raisebox{-20pt}{
        \tikz[thick]{
        \node at (-1mm,3mm) {$\tilde\lambda$};
        \draw [right] (-10mm,0)--(-5mm,0);
        \draw [aux] (-5mm,0)--(0,0);
        \node at (-5mm,-3mm) {$p_1$};
        \draw [aux] (0mm,0)--(4mm,4mm);
        \draw [left] (4mm,4mm)--(8mm,8mm);
        \node at (8mm,3mm) {$p_2$};
        \draw [] (0mm,0)--(4mm,-4mm);
        \draw [left] (4mm,-4mm)--(8mm,-8mm);
        \node at (8mm,-3mm) {$p_3$};
    }} 
    + (1\leftrightarrow 3)
    + (2\leftrightarrow 3)
    \rb \nn\\
    &= -4iT_0^2\sigma^2
    \lb \frac{k_1^2k_2^2
    \lb \lambda k_3^2 - \lambda_\tau \omega_3^2 \rb }{|F(p_1)|^2|F(p_2)|^2 F(p_3)}
    + (1\leftrightarrow 3)
    + (2\leftrightarrow 3)
    \rb \nn\\
    &\qquad 
    + 4T_0^2\chi\sigma\tilde\lambda
    \lb \frac{k_3^2(k_1\cdot k_2)}{F(p_1)F(p_2)|F(p_3)|^2}
    + (1\leftrightarrow 3)
    + (2\leftrightarrow 3)
    \rb.
\end{align}
\end{subequations}
We find that these correlation functions do not satisfy the fluctuation-dissipation theorems in \cref{eq:fdt-3pt} for non-constant $\sigma$ and $\tau$, i.e. nonzero $\tilde\lambda$ and $\lambda_\tau$. The $G_{rra}$ and $G_{rrr}$ fluctuation-dissipation theorems have residual contributions
\begin{subequations}
\begin{align}
    &\lb 
    \frac{\lambda_\tau}{\omega_1\omega_2\omega_3}
    \lb
    \omega_1^3 k_2^2 k_3^2
    + \omega_2^3 k_1^2 k_3^2
    + \omega_3^3 k_1^2k_2^2 \rb
    - \frac{\tau\chi\tilde\lambda}{\sigma}
    \Big( k_1^2 (k_2\cdot k_3)
    + k_2^2(k_1\cdot k_3)
    + k_3^2(k_1\cdot k_2) \Big) \rb \nn\\
    &\qquad\times
    \frac{T_0\sigma^2\omega_3 }{F(p_1)F(p_2)F(p_3)^*},
\end{align}
and
\begin{align}
& \lb \frac{\lambda_\tau}{\omega_1\omega_2\omega_3}
    \lb \omega_1^3k_2^2k_3^2
    + \omega_3^3k_1^2k_2^2
    + \omega_2^3k_1^2k_3^2
    \rb
    - \frac{\tau\chi\tilde\lambda}{\sigma}
    \Big( k_1^2(k_2\cdot k_3)
    + k_2^2(k_1\cdot k_3)
    + k_3^2(k_1\cdot k_2) \Big)
    \rb \times  \nn\\
    &\qquad\times 2iT_0^2\sigma^2
    \lb \frac{-1}{F(p_1)F(p_2)F(p_3)}
    + \lb \frac{\omega_1}{|F(p_1)|^2F(p_2)F(p_3)}
    + (1\leftrightarrow 2)
    + (1\leftrightarrow 3)
    \rb 
    \rb,
\end{align}
\end{subequations}
on the right-hand sides respectively. For non-zero relaxation time $\tau\neq 0$, the FDTs are only satisfied $\chi\tilde\lambda = \dow\sigma/\dow n = 0$ and $\lambda_\tau = \dow\tau/\dow n =0$, i.e. $\sigma$ and $\tau$ are taken to be a constants. Note that $\sigma$ is allowed to be non-constant in the ordinary diffusion model when $\tau = 0$. Since higher-point interaction vertices also contribute to 2-point correlation functions via stochastic interactions, the loop-corrected 2-point correlation functions will also violate the respective FDT in \cref{eq:fdt-3pt} for non-constant $\sigma$ or $\tau$, when $\tau\neq 0$.

\paragraph*{MC-diffusion:}

To include the effects of non-constant $\sigma$ and $\tau$ while respecting FDTs, we will need to invoke the full EFT for MC-diffusion derived in \cref{eq:action-MIS} with additional vector degrees of freedom. We could, of course, work with the form of the effective action in \cref{eq:final-action-diff} derived using the SK formalism, but the interactions are simpler in the MSR incarnation of the effective action in \cref{eq:action-MIS}. Recall that the two are related via field redefinitions in \cref{eq:field-redef-msr}. For simplicity, we will assume $\mu$ to just depend on $n$ and choose $\alpha_\upsilon=1$, which, owing to our discussion around \cref{eq:MIS-diff-simple}, will imply that $\tau/\sigma$ is still a constant. Nonetheless, we only really care about the generality of the ``hydrodynamic'' transport coefficient $\sigma$; the additional relaxed sector included for stability and causality reasons is anyway not universal, so we might as well work with the simplest such extension. For completeness, let us rewrite the effective action \eqref{eq:MIS-diff-simple} in the local rest frame $u^\mu_0 = \delta^\mu_t$ and with only the time-components of the background gauge fields $A_{r,at}$ turned on, leading to
\begin{equation}
    S
    = \int \df^{d+1}x
    \bigg[ \dow_t\varphi_a n
    + \dow_i\varphi_a\upsilon^i
    - V_a^i\!\lb \upsilon_i
    + \tau \dow_t \upsilon_i \rb
    - D_n V_a^i \dow_in
    + iT_0\sigma\,V_{a}^i V_{a i}
    + A_{at} n 
    + \sigma\, V_a^i \dow_i A_{rt}
    \bigg].
\end{equation}
As opposed to the simple propagators in \cref{eq:propagators-bdnk}, we have a more intricate set of propagators in this case due the additional vector degrees of freedom. Denoting $n$ and $\varphi_a$ still by solid and wavy lines, and $\upsilon^i$ and $V_{a}^i$ by the respective thicker versions, we find
\begin{gather}
    \raisebox{-13pt}{
        \tikz[thick]{
        \draw [right] (-16mm,0)--(-8mm,0);
        \draw [aux] (-8mm,0)--(0,0);
        \node at (-8mm,-3mm) {$p$};
    }}\quad 
    = \frac{1 - i\tau\omega}{F(p)}, 
    \qquad 
    \raisebox{-13pt}{
        \tikz[thick]{
        \draw [right] (-16mm,0)--(-8mm,0);
        \draw [] (-8mm,0)--(0,0);
        \node at (-8mm,-3mm) {$p$};
    }}
    \quad = \frac{2T_0\sigma k^2}{|F(p)|^2}, \nn\\[0.5em]
    \raisebox{-13pt}{
        \tikz[thick]{
        \draw [ultra thick] (-16mm,0)--(-8mm,0);
        \draw [right] (-8.1mm,0)--(-8mm,0);
        \draw [aux, ultra thick] (-8mm,0)--(0,0);
        \node at (-8mm,-3mm) {$p$};
    }}\quad 
    = \frac{-i\omega}{F(p)}\frac{k^ik^j}{k^2}
    - \frac{i}{1-i\tau\omega} k^{ij}, \qquad
    \raisebox{-13pt}{
        \tikz[thick]{
        \draw [ultra thick] (-16mm,0)--(-8mm,0);
        \draw [right] (-8.1mm,0)--(-8mm,0);
        \draw [ultra thick] (-8mm,0)--(0,0);
        \node at (-8mm,-3mm) {$p$};
    }}\quad 
    = \frac{2T_0\sigma \omega^2}{|F(p)|^2} 
    \frac{k^ik^j}{k^2}
    + \frac{2 T_0\sigma}{|1-i\tau\omega|^2} k^{ij}, \nn\\[0.5em]
        \raisebox{-13pt}{
        \tikz[thick]{
        \draw [right] (-16mm,0)--(-8mm,0);
        \draw [aux, ultra thick] (-8mm,0)--(0,0);
        \node at (-8mm,-3mm) {$p$};
    }}\quad 
    = \frac{-ik^i}{F(p)}, \qquad 
    \raisebox{-13pt}{
        \tikz[thick]{
        \draw [ultra thick] (-16mm,0)--(-8mm,0);
        \draw [right] (-8.1mm,0)--(-8mm,0);
        \draw [aux] (-8mm,0)--(0,0);
        \node at (-8mm,-3mm) {$p$};
    }}\quad 
    = \frac{-iD_n k^i}{F(p)}, \qquad 
    \raisebox{-13pt}{
        \tikz[thick]{
        \draw [right] (-16mm,0)--(-8mm,0);
        \draw [ultra thick] (-8mm,0)--(0,0);
        \node at (-8mm,-3mm) {$p$};
    }}\quad 
    = \frac{2T_0\sigma \omega\,k^i}{|F(p)|^2}.
\end{gather}
We have three 3-point interaction vertices
\begin{equation}
    \raisebox{-20pt}{
        \tikz[thick]{
        \node at (-1mm,3mm) {$\lambda$};
        \draw [right] (-8.1mm,0)--(-8mm,0);
        \draw [aux, ultra thick] (-8mm,0)--(0,0);
        \node at (-8mm,-3mm) {$p_1$};
        \draw (0mm,0)--(6mm,6mm);
        \draw [left] (6mm,6mm)--(6.1mm,6.1mm);
        \node at (8mm,4mm) {$p_2$};
        \draw [] (0mm,0)--(6mm,-6mm);
        \draw [left] (6mm,-6mm)--(6.1mm,-6.1mm);
        \node at (8mm,-4mm) {$p_3$};
    }}
    \quad = \frac{\lambda}{2} k_1^i, \qquad
    \raisebox{-20pt}{
        \tikz[thick]{
        \node at (-2mm,3mm) {$\lambda_\tau$};
        \draw [right] (-8.1mm,0)--(-8mm,0);
        \draw [aux, ultra thick] (-8mm,0)--(0,0);
        \node at (-8mm,-3mm) {$p_1$};
        \draw [ultra thick] (0mm,0)--(6mm,6mm);
        \draw [left] (6mm,6mm)--(6.1mm,6.1mm);
        \node at (8mm,4mm) {$p_2$};
        \draw [] (0mm,0)--(6mm,-6mm);
        \draw [left] (6mm,-6mm)--(6.1mm,-6.1mm);
        \node at (8mm,-4mm) {$p_3$};
    }}
    \quad = \lambda_\tau \omega_2 \delta^{ij}, \qquad
    \raisebox{-20pt}{
        \tikz[thick]{
        \node at (-1.5mm,3mm) {$\tilde\lambda$};
        \draw [right] (-8.1mm,0)--(-8mm,0);
        \draw [] (-8mm,0)--(0,0);
        \node at (-8mm,-3mm) {$p_1$};
        \draw [aux, ultra thick] (0mm,0)--(6mm,6mm);
        \draw [left] (6mm,6mm)--(6.1mm,6.1mm);
        \node at (8.5mm,4mm) {$p_2$};
        \draw [aux, ultra thick] (0mm,0)--(6mm,-6mm);
        \draw [left] (6mm,-6mm)--(6.1mm,-6.1mm);
        \node at (8.5mm,-4mm) {$p_3$};
    }}
    \quad = -T_0\chi\tilde\lambda \delta^{ij}.
\end{equation}
Note that there are no interactions involving the noise field $\varphi_a$. We can similarly work out higher-point interaction vertices as needed. Finally, we have the background coupling vertices
\begin{equation}
    \raisebox{-4pt}{
        \tikz[thick]{
        \draw [->>] (-8.1mm,0)--(-8mm,0);
        \draw [ultra thick, aux] (-8mm,0)--(0,0);
        \draw [right] (-0.1mm,0)--(0,0);
        \node at (-12.5mm,0mm) {$p$};
    }} \quad = \sigma k^i, \qquad 
    \raisebox{-20pt}{
        \tikz[thick]{
        \node at (-1mm,4mm) {$\tilde\lambda$};
        \draw [->>] (-0.1mm,0)--(-0mm,0);
        \node at (-4.5mm,0mm) {$p_1$};
        \draw [aux, ultra thick] (0mm,0)--(6mm,6mm);
        \draw [left] (6mm,6mm)--(6.1mm,6.1mm);
        \node at (8mm,4mm) {$p_2$};
        \draw [] (0mm,0)--(6mm,-6mm);
        \draw [left] (6mm,-6mm)--(6.1mm,-6.1mm);
        \node at (8mm,-4mm) {$p_3$};
    }} \quad = \chi\tilde\lambda k_1^i,
\end{equation}
where we have still denoted the insertion of $A_{rt}$ with a double arrow.
Using this EFT, the 2-point correlation functions of density at tree-level can be computed simply as
\begin{equation}
    G_{ra}(p) =\quad
    \raisebox{-13pt}{
        \tikz[thick]{
        \draw [right] (-16mm,0)--(-8mm,0);
        \draw [aux, ultra thick] (-8mm,0)--(0,0);
        \draw [->>] (0mm,0)--(2mm,0);
        \node at (-8mm,-3mm) {$p$};
    }} \quad= \frac{i\sigma k^2}{F(p)}, \qquad 
    G_{rr}(p) = \quad
    \raisebox{-13pt}{
        \tikz[thick]{
        \draw [right] (-16mm,0)--(-8mm,0);
        \draw [] (-8mm,0)--(0,0);
        \node at (-8mm,-3mm) {$p$};
    }} \quad = \frac{2T_0\sigma k^2}{|F(p)|^2},
\end{equation}
and yield the same results as in \cref{eq:2pt-bdnk}, consistent with FDT. The computation for tree-level 3-point correlation functions is a little more involved and we find: retarded
\begin{subequations}
\begin{align}
    G_{raa}(p_1,p_2,p_3) 
    &= 
    \raisebox{-23pt}{
        \tikz[thick]{
        \node at (-1mm,3mm) {$\lambda$};
        \draw [right] (-10mm,0)--(-5mm,0);
        \draw [aux,ultra thick] (-5mm,0)--(0,0);
        \node at (-5mm,-3mm) {$p_1$};
        \draw [] (0mm,0)--(3mm,3mm);
        \draw [left] (2.7mm,2.7mm)--(3mm,3mm);
        \draw [aux,ultra thick] (3mm,3mm)--(6.5mm,6.5mm);
        \draw [-<<] (6.5mm,6.5mm)--(8mm,8mm);
        \node at (8mm,3mm) {$p_2$};
        \draw [] (0mm,0)--(3mm,-3mm);
        \draw [left] (2.7mm,-2.7mm)--(3mm,-3mm);
        \draw [aux,ultra thick] (3mm,-3mm)--(6.5mm,-6.5mm);
        \draw [-<<] (6.5mm,-6.5mm)--(8mm,-8mm);
        \node at (8mm,-3mm) {$p_3$};
    }}
    + \lb \raisebox{-23pt}{
        \tikz[thick]{
        \node at (-1mm,3mm) {$\lambda_\tau$};
        \draw [right] (-10mm,0)--(-5mm,0);
        \draw [aux,ultra thick] (-5mm,0)--(0,0);
        \node at (-5mm,-3mm) {$p_1$};
        \draw [ultra thick] (0mm,0)--(3mm,3mm);
        \draw [left] (2.7mm,2.7mm)--(3mm,3mm);
        \draw [aux,ultra thick] (3mm,3mm)--(6.5mm,6.5mm);
        \draw [-<<] (6.5mm,6.5mm)--(8mm,8mm);
        \node at (8mm,3mm) {$p_2$};
        \draw [] (0mm,0)--(3mm,-3mm);
        \draw [left] (2.7mm,-2.7mm)--(3mm,-3mm);
        \draw [aux,ultra thick] (3mm,-3mm)--(6.5mm,-6.5mm);
        \draw [-<<] (6.5mm,-6.5mm)--(8mm,-8mm);
        \node at (8mm,-3mm) {$p_3$};
    }} +  (2\leftrightarrow 3) \rb
    + \lb \raisebox{-13.5pt}{
        \tikz[thick]{
        \node at (-1mm,3mm) {$\tilde\lambda$};
        \draw [right] (-10mm,0)--(-5mm,0);
        \draw [aux, ultra thick] (-5mm,0)--(0,0);
        \node at (-5mm,-3mm) {$p_1$};
        \draw [] (0mm,0)--(3mm,3mm);
        \draw [left] (2.7mm,2.7mm)--(3mm,3mm);
        \draw [aux,ultra thick] (3mm,3mm)--(6.5mm,6.5mm);
        \draw [-<<] (6.5mm,6.5mm)--(8mm,8mm);
        \node at (8mm,3mm) {$p_2$};
        \draw [-<<] (0mm,0mm)--(2mm,-2mm);
        \node at (4mm,-4mm) {$p_3$};
    }} + (2\leftrightarrow 3) \rb
    \nn\\
    &=
    \frac{i\sigma^2\lambda k_1^2 k_2^2 k_3^2}{F(p_1)F(p_2)^*F(p_3)^*}
    + i\sigma^2\lambda_\tau \frac{\omega_2^2 k_3^2 (k_1\cdot k_2)
    + \omega_3^2 k_2^2 (k_1\cdot k_3)}{F(p_1)F(p_2)^*F(p_3)^*} \nn\\
    &\qquad 
    - \chi\sigma\tilde\lambda \lb 
    \frac{k_2^2(k_1\cdot k_3)}{F(p_1)F(p_2)^*}
    + (2\leftrightarrow 3) \rb,
\end{align}
partially-retarded
\begin{align}
    G_{rra}(p_1,p_2,p_3)
    &= 
    \lb \raisebox{-23pt}{
        \tikz[thick]{
        \node at (-1mm,3mm) {$\lambda$};
        \draw [right] (-10mm,0)--(-5mm,0);
        \draw [] (-5mm,0)--(0,0);
        \node at (-5mm,-3mm) {$p_1$};
        \draw [aux,ultra thick] (0mm,0)--(4mm,4mm);
        \draw [left] (4mm,4mm)--(8mm,8mm);
        \node at (8mm,3mm) {$p_2$};
        \draw [] (0mm,0)--(3mm,-3mm);
        \draw [left] (2.7mm,-2.7mm)--(3mm,-3mm);
        \draw [aux,ultra thick] (3mm,-3mm)--(6.5mm,-6.5mm);
        \draw [-<<] (6.5mm,-6.5mm)--(8mm,-8mm);
        \node at (8mm,-3mm) {$p_3$};
    }} + (1\leftrightarrow 2) \rb
    + \lb \raisebox{-23pt}{
        \tikz[thick]{
        \node at (-2mm,3mm) {$\lambda_\tau$};
        \draw [right] (-10mm,0)--(-5mm,0);
        \draw [ultra thick] (-5mm,0)--(0,0);
        \node at (-5mm,-3mm) {$p_1$};
        \draw [aux,ultra thick] (0mm,0)--(4mm,4mm);
        \draw [left] (4mm,4mm)--(8mm,8mm);
        \node at (8mm,3mm) {$p_2$};
        \draw [] (0mm,0)--(3mm,-3mm);
        \draw [left] (2.7mm,-2.7mm)--(3mm,-3mm);
        \draw [aux,ultra thick] (3mm,-3mm)--(6.5mm,-6.5mm);
        \draw [-<<] (6.5mm,-6.5mm)--(8mm,-8mm);
        \node at (8mm,-3mm) {$p_3$};
    }}
    + \raisebox{-23pt}{
        \tikz[thick]{
        \node at (-2mm,3mm) {$\lambda_\tau$};
        \draw [right] (-10mm,0)--(-5mm,0);
        \draw [] (-5mm,0)--(0,0);
        \node at (-5mm,-3mm) {$p_1$};
        \draw [aux,ultra thick] (0mm,0)--(4mm,4mm);
        \draw [left] (4mm,4mm)--(8mm,8mm);
        \node at (8mm,3mm) {$p_2$};
        \draw [ultra thick] (0mm,0)--(3mm,-3mm);
        \draw [left] (2.7mm,-2.7mm)--(3mm,-3mm);
        \draw [aux,ultra thick] (3mm,-3mm)--(6.5mm,-6.5mm);
        \draw [-<<] (6.5mm,-6.5mm)--(8mm,-8mm);
        \node at (8mm,-3mm) {$p_3$};
    }}
    + (1\leftrightarrow 2) \rb \nn\\
    &\qquad 
    + \raisebox{-23pt}{
        \tikz[thick]{
        \node at (-1mm,3mm) {$\tilde\lambda$};
        \draw [right] (-10mm,0)--(-5mm,0);
        \draw [aux, ultra thick] (-5mm,0)--(0,0);
        \node at (-5mm,-3mm) {$p_1$};
        \draw [aux, ultra thick] (0mm,0)--(4mm,4mm);
        \draw [left] (4mm,4mm)--(8mm,8mm);
        \node at (8mm,3mm) {$p_2$};
        \draw [] (0mm,0)--(3mm,-3mm);
        \draw [left] (2.7mm,-2.7mm)--(3mm,-3mm);
        \draw [aux,ultra thick] (3mm,-3mm)--(6.5mm,-6.5mm);
        \draw [-<<] (6.5mm,-6.5mm)--(8mm,-8mm);
        \node at (8mm,-3mm) {$p_3$};
    }}
    + \lb \raisebox{-13.5pt}{
        \tikz[thick]{
        \node at (-1mm,3mm) {$\tilde\lambda$};
        \draw [right] (-10mm,0)--(-5mm,0);
        \draw [aux,ultra thick] (-5mm,0)--(0,0);
        \node at (-5mm,-3mm) {$p_1$};
        \draw (0mm,0)--(4mm,4mm);
        \draw [left] (4mm,4mm)--(8mm,8mm);
        \node at (8mm,3mm) {$p_2$};
        \draw [-<<] (0,0)--(2mm,-2mm);
        \node at (4mm,-4mm) {$p_3$};
    }} + (1\leftrightarrow 2) \rb
    \nn\\
    &= - 2T_0\sigma^2 \lambda  \lb \frac{k_1^2 k_2^2 k_3^2}
    {|F(p_1)|^2F(p_2)F(p_3)^*}
    + (1\leftrightarrow 2) \rb \nn\\
    &\qquad 
    - 2T_0\sigma^2\lambda_\tau
    \lb \frac{\omega_1^2k_3^2 (k_1\cdot k_2)
    + k_1^2\omega_3^2 (k_2\cdot k_3)}{|F(p_1)|^2F(p_2)F(p_3)^*}
    + (1\leftrightarrow 2)
    \rb
    \nn\\
    &\qquad 
    - 2iT_0\chi\sigma\tilde\lambda
    \lb \frac{k_3^2 (k_1\cdot k_2)}{F(p_1)F(p_2)F(p_3)^*}
    +  \lb \frac{k_2^2(k_1\cdot k_3)}{F(p_1)|F(p_2)|^2}
    + (1\leftrightarrow 2)
    \rb\rb,
\end{align}
and symmetric
\begin{align}
    G_{rrr}(p_1,p_2,p_3)
    &=
    \lb \raisebox{-20pt}{
        \tikz[thick]{
        \node at (-1mm,3mm) {$\lambda$};
        \draw [right] (-10mm,0)--(-5mm,0);
        \draw [] (-5mm,0)--(0,0);
        \node at (-5mm,-3mm) {$p_1$};
        \draw [] (0mm,0)--(4mm,4mm);
        \draw [left] (4mm,4mm)--(8mm,8mm);
        \node at (8mm,3mm) {$p_2$};
        \draw [aux, ultra thick] (0mm,0)--(4mm,-4mm);
        \draw [left] (4mm,-4mm)--(8mm,-8mm);
        \node at (8mm,-3mm) {$p_3$};
    }} + (1\leftrightarrow 3)
    + (2\leftrightarrow 3) \rb
    + \lb \raisebox{-20pt}{
        \tikz[thick]{
        \node at (-2mm,3mm) {$\lambda_\tau$};
        \draw [right] (-10mm,0)--(-5mm,0);
        \draw [] (-5mm,0)--(0,0);
        \node at (-5mm,-3mm) {$p_1$};
        \draw [ultra thick] (0mm,0)--(4mm,4mm);
        \draw [left] (4mm,4mm)--(8mm,8mm);
        \node at (8mm,3mm) {$p_2$};
        \draw [aux, ultra thick] (0mm,0)--(4mm,-4mm);
        \draw [left] (4mm,-4mm)--(8mm,-8mm);
        \node at (8mm,-3mm) {$p_3$};
    }} + (1\leftrightarrow 2\leftrightarrow 3) \rb \nn\\
    &\qquad 
    + \lb \raisebox{-20pt}{
        \tikz[thick]{
        \node at (-1mm,3mm) {$\tilde\lambda$};
        \draw [right] (-10mm,0)--(-5mm,0);
        \draw [aux, ultra thick] (-5mm,0)--(0,0);
        \node at (-5mm,-3mm) {$p_1$};
        \draw [aux, ultra thick] (0mm,0)--(4mm,4mm);
        \draw [left] (4mm,4mm)--(8mm,8mm);
        \node at (8mm,3mm) {$p_2$};
        \draw [] (0mm,0)--(4mm,-4mm);
        \draw [left] (4mm,-4mm)--(8mm,-8mm);
        \node at (8mm,-3mm) {$p_3$};
    }} 
    + (1\leftrightarrow 3)
    + (2\leftrightarrow 3)
    \rb \nn\\
    &= -4iT_0^2\sigma^2\lambda 
    \lb \frac{k_1^2k_2^2k_3^2}{|F(p_1)|^2|F(p_2)|^2 F(p_3)}
    + (1\leftrightarrow 3)
    + (2\leftrightarrow 3)
    \rb \nn\\
    &\qquad 
    - 4iT_0^2\sigma^2\lambda_\tau \lb 
    \frac{\omega_2^2k_1^2(k_2\cdot k_3)
    + \omega_1^2k_2^2(k_1\cdot k_3)}
    {|F(p_1)|^2|F(p_2)|^2F(p_3)}
    + (1\leftrightarrow 3)
    + (2\leftrightarrow 3) \rb\nn\\
    &\qquad 
    + 4T_0^2\chi\sigma\tilde\lambda
    \lb \frac{k_3^2(k_1\cdot k_2)}{F(p_1)F(p_2)|F(p_3)|^2}
    + (1\leftrightarrow 3)
    + (2\leftrightarrow 3)
    \rb.
\end{align}
\end{subequations}
Plugging these into the respective FDTs in \cref{eq:fdt-3pt}, we find that the residual terms on the right-hand side are given as
\begin{subequations}
\begin{align}
    &\lb \lambda_\tau - \frac{\chi \tau\tilde\lambda}{\sigma} \rb 
    T_0 \sigma^2 \omega_3 \lb 
    \frac{k_1^2 (k_2\cdot k_3)
    + k_2^2(k_1\cdot k_3)
    + k_3^2(k_1\cdot k_2)
    }{F(p_1)F(p_2)F(p_3)^*}
    \rb,
\end{align}
and
\begin{align}
    & 2iT_0^2\sigma^2 \lb \lambda_\tau - \frac{\chi \tau\tilde\lambda}{\sigma} \rb 
    \Big( k_1^2(k_2\cdot k_3)
    + k_2^2(k_1\cdot k_3)
    + k_3^2(k_1\cdot k_2) \Big) \times  \nn\\
    &\qquad 
    \lb \frac{-1}{F(p_1)F(p_2)F(p_3)}
    + \lb \frac{\omega_1}{|F(p_1)|^2F(p_2)F(p_3)}
    + (1\leftrightarrow 2)
    + (1\leftrightarrow 3)
    \rb 
    \rb,
\end{align}
\end{subequations}
respectively. As mentioned before, FDTs are only satisfied when $\lambda_\tau - \chi\tau \tilde\lambda/\sigma = \sigma\,\dow(\tau/\sigma)/\dow n = 0$, i.e. $\tau/\sigma$ is taken to be a constant. 

We can also use the diagrammatic representation and Feynman rules described above to compute loop corrections to 2-point correlation functions consistent with FDT requirements. We leave this exercise for future work. We also leave the generalisation of the discussion in this appendix to MIS-hydrodynamics for future explorations.

\section{Alternate KMS prescriptions}
\label{app:alt}

Since the additional relaxed degrees of freedom in MC-diffusion and MIS-hydrodynamics are not associated with any symmetries, we can actually construct different models in the SK formalism that ultimately give rise to the same effective action, but with different realisations of the dynamical KMS symmetry. As an example, let us consider the SK model for MC-diffusion from \cref{app:SK-MIS}, but with vector fields $\hat\upsilon_{1,2\mu}$ (normalised as $\beta_0^\mu\hat\upsilon_{1,2\mu} = 0$) instead of $\upsilon_{1,2\mu}$ that respect a ``spatial diagonal
shift symmetry'', i.e.
\begin{equation}
    \hat\upsilon_{1,2\mu} \to \hat\upsilon_{1,2\mu} + \lambda_\mu, \quad\text{such that}\quad 
    \beta^\mu_0 \lambda_\mu = 0, ~
    \beta_0^\lambda \dow_\lambda \lambda_\mu = 0.
\end{equation}
This means that all dependence on the average combination $\hat\upsilon_{r\mu}$ in the EFT must arise via its time-derivative $\upsilon_{\mu} \equiv u^\lambda_0\dow_\lambda\hat\upsilon_{r\mu}$, which we will use as the definition of the classical vector degree of freedom $\upsilon_\mu$ is MC-diffusion. We take the C,P,T eigenvalues of the spatial-components of the new fields $\upsilon_{1,2i}$, $\upsilon_{r,ai}$ to be $(-,-,+)$, while those of the time-components can be obtained using the normalisation conditions. With the identification of $\upsilon_\mu$ above, these lead to the correct C,P,T eigenvalues of $\upsilon_\mu$ as given in \cref{tab:CPT-rel}. The KMS transformation properties of $\hat\upsilon_{1,2\mu}$ and $\hat\upsilon_{r,a\mu}$ are given similar to \cref{eq:KMS-diff-def,eq:KMS-diff-classical}, which imply
\begin{equation}
    \upsilon_{\mu} \to \Theta\upsilon_{\mu}, \qquad 
    \hat\upsilon_{a\mu} \to \Theta\lb \hat\upsilon_{a\mu}
    + \frac{i}{T_0}\upsilon_{\mu} \rb.
\end{equation}
A simple EFT that realises this and all other requirements in \cref{app:SK-MIS} can be written down similar to \cref{eq:action-SK-raw} as
\begin{align}
    S
    &= \int \df^{d+1}x
    \bigg[
    B_{a\mu} n u_0^\mu
    - \chi_\upsilon \hat\upsilon^\mu_{a}
    u_0^\lambda \dow_\lambda\upsilon_\mu \nn\\
    &\hspace{8em} 
    + iT_0\lb -\hat\lambda\, u^\mu_0 u^\nu_0 
    + \hat\sigma\,\Delta^{\mu\nu}
    \rb B_{a\mu}
    \lb B_{a\nu} 
    + i\beta_0^\lambda \dow_\lambda B_{r\nu} \rb\nn\\
    &\hspace{8em} 
    + iT_0\hat\sigma_{\upsilon}\, \Delta^{\mu\nu}\hat\upsilon_{a\mu}
    \lb \hat\upsilon_{a\nu} 
    + i\beta_0\upsilon_{\nu} \rb \nn\\
    &\hspace{8em} 
    + iT_0\hat\gamma_\times\,\Delta^{\mu\nu} 
    \Big( B_{a\mu} \lb \hat\upsilon_{a\nu} 
    + i\beta_0\upsilon_{\nu} \rb
    + \hat\upsilon_{a\mu}
    \lb B_{a\nu} + i\beta_0^\lambda \dow_\lambda B_{r\nu} \rb
    \Big) \nn\\
    &\hspace{8em} 
    + iT_0\hat{\bar\gamma}_\times\,\Delta^{\mu\nu} 
    \Big( B_{a\mu} \lb \hat\upsilon_{a\nu} 
    + i\beta_0\upsilon_{\nu} \rb
    - \hat\upsilon_{a\mu}
    \lb B_{a\nu} + i\beta_0^\lambda \dow_\lambda B_{r\nu} \rb
    \Big)
    \bigg],
    \label{eq:action-SK-raw-alt}   
\end{align}
where all coefficients are functions of $\mu \equiv u_0^\mu B_{r\mu}$ and $\upsilon^2$, and satisfy the inequality constraints similar to \cref{eq:SK-ineq}. The first two terms satisfy the KMS condition up to a total-derivative term similar to \cref{eq:totalder-diff}. The remaining terms are manifestly KMS-invariant. However, since $\hat\upsilon_{r,a\mu}$ and $B_{r,a\mu}$ have opposite eigenvalues under T this time around, the ${\bar\gamma}_\times$ term is disallowed by $\Theta={\rm T,PT}$. Whereas for $\Theta={\rm CT,CPT}$, all terms are allowed, provided that $\bar\gamma_\times$ and $n$ is odd functions of $\mu$, while all other coefficients are even functions of $\mu$. Using this EFT, one can work out the vector equation of motion
\begin{equation}
    \chi_\upsilon \Delta^{\mu\nu} u_0^\lambda \dow_\lambda \upsilon_\nu
    = - (\hat\gamma_\times-\hat{\bar\gamma}_\times) 
    \Delta^{\mu\nu}\!\lb \dow_\nu \mu - F_{r\nu\rho}u^\rho_0 \rb
    - \hat\sigma_\upsilon\upsilon^{\mu},
\end{equation}
and the current
\begin{align}
    J^\mu_r 
    &= \lb n + \hat\lambda\, u_0^\lambda\dow_\lambda\mu \rb u_0^\mu
    - \hat\sigma\, \Delta^{\mu\nu} \lb \dow_\nu \mu - F_{r\nu\rho}u^\rho_0 \rb 
    - (\hat\gamma_\times+\hat{\bar\gamma}_\times) \upsilon^{\mu},
\end{align}
analogous to \cref{eq:non-frame-upsilon-eq,eq:general-frame}. Comparing these to the frame conditions in \cref{eq:param-J,eq:frame-choice-diff}, we are led to set
\begin{equation}
    \hat\lambda = 0, \qquad 
    \hat\sigma = 0, \qquad
    \hat\sigma_\upsilon = \frac{\alpha_\upsilon^2}{\sigma}, \qquad 
    \hat\gamma_\times 
    = 0, \qquad 
    \hat{\bar\gamma}_\times = -\alpha_\upsilon,
\end{equation}
instead of the analogous expressions~\eqref{eq:extra-coeff-frame} in the original SK model. The effective action, in this case, simplifies to
\begin{align}
    S
    &= \int \df^{d+1}x
    \bigg[
    B_{a\mu} n u_0^\mu
    - \chi_\upsilon \hat\upsilon^\mu_{a}
    u_0^\lambda \dow_\lambda\upsilon_\mu
    + \alpha_\upsilon\,\Delta^{\mu\nu} 
    \Big( B_{a\mu}\upsilon_{\nu}
    - \hat\upsilon_{a\mu}u_0^\lambda \dow_\lambda B_{r\nu}
    \Big) \nn\\
    &\hspace{18em} 
    + \frac{iT_0\alpha_\upsilon^2}{\sigma} \Delta^{\mu\nu}\hat\upsilon_{a\mu}
    \lb \hat\upsilon_{a\nu} 
    + i\beta_0\upsilon_{\nu} \rb
    \bigg].  
\end{align}
We again recover the effective action \eqref{eq:action-MIS} we derived using the MSR formalism, with the KMS condition \eqref{eq:sigma-KMS} already imposed, but with the field identifications
\begin{equation}
    \upsilon_\mu = u_0^\lambda\dow_\lambda\hat\upsilon_{r\mu}, \qquad
    V_{a\mu} = \frac{\alpha_\upsilon}{\sigma} \hat\upsilon_{a\mu}.
\end{equation}
Note that $\upsilon_\mu$ and $V_{a\mu}$ have opposite $\Theta$-eigenvalues under this prescription, precisely
leading to the alternate prescription of KMS symmetry given in \cref{eq:KMS-diff-alternate}.

In the full theory of MIS-hydrodynamics, we can actually write down three distinct SK models with different realisations of the dynamical KMS symmetry, depending on if we allow a spatial diagonal shift symmetry in the tensor degrees of freedom, in vector degrees of freedom, or both, taking the form
\begin{gather}
    \hat\kappa_{\mu\nu} \to \hat\kappa_{\mu\nu}
    + \lambda_{\mu\nu}, \quad\text{such that}\quad 
    \beta^\mu\lambda_{\mu\nu} = 0, \quad \lambda_{\mu\nu} = \lambda_{\nu\mu}, \quad 
    \lie_\beta\lambda_{\mu\nu} = 0, \nn\\
    \hat\upsilon_{\mu} \to \hat\upsilon_{\mu}
    + \lambda_{\mu}, \quad\text{such that}\quad 
    \beta^\mu\lambda_{\mu} = 0, \quad 
    \lie_\beta\lambda_{\mu} = 0.
\end{gather}
Following through the procedure outlined above, we find the same effective action as the MSR formalism in \cref{eq:action-hydro-raw}, with the KMS conditions \eqref{eq:KMS-conditions} already implemented, together with the field identifications of the tensor degrees of freedom
\begin{equation}
    \kappa_{\mu\nu} = T\lie_\beta\hat\kappa_{\mu\nu}, \qquad 
    W_{a\mu\nu}
    = \lb
    \frac1d\frac{\alpha_\kappa^\ssfS}{\zeta}
    \Delta_{\mu\nu}\Delta^{\rho\sigma}
    + \frac{\alpha_\kappa^\ssfT}{\eta} 
    \Delta_{\langle\mu}^\rho\Delta_{\nu\rangle}^\rho
    \rb
    \hat\kappa_{a\rho\sigma},
\end{equation}
and/or the vector degrees of freedom
\begin{equation}
    \upsilon_\mu = T\lie_\beta\hat\upsilon_\mu, \qquad 
    V_{a\mu} = \frac{\alpha_\upsilon}{\sigma} 
    \hat\upsilon_{a\mu}.
\end{equation}
As expected, the physical and auxiliary fields have opposite $\Theta$-eigenvalues under these identifications.
These precisely lead to the alternate prescriptions of KMS transformations mentioned in \cref{eq:KMS-hydro-alt}.

\addbibresources{new}
\makereferences

\end{document}